\documentstyle[12pt]{article} 
\newcommand{\sn}{{\rm sn}}
\newcommand{\cn}{{\rm cn}}
\newcommand{\dn}{{\rm dn}}

\begin{document} 

\vspace{15mm}  

\centerline{\Large \bf Jacobi's Inversion Problem}
\centerline{\Large \bf for Genus Two Hyperelliptic Integral}

\vspace{10mm}

\centerline{\large Kazuyasu Shigemoto\footnote{E-mail address:
shigemot@tezukayama-u.ac.jp}}

\centerline {{
\large Tezukayama University, Tezukayama 7, Nara 631, Japan
}}

\vspace{10mm}

\centerline{\bf Abstract} 
\noindent 
Hinted by the elliptic parameterization of the Ising model, the 
addition formula of the elliptic function forms to give the 
integrable $SU(2)$ group relation in the previous paper.
We then expect that the addition formula of the Abelian 
function with any genus will form to give some integrable 
Lie group structure. In this paper, we study Jacobi's 
inversion problem for hyperelliptic integral with genus two and we expect 
some $SU(2)$ structure for the addition formula of 
the hyperelliptic theta function with genus two.

\vspace{5mm}

\noindent
\begin{tabular}{|llll|} \hline 
Keywords: & Ising model  , & integrability condition , & Jacobi's inversion relation \\
   & addition formula, & hyperelliptic function,  & theta function\\
\hline
\end{tabular} 

\vspace{10mm}
\noindent {\bf Contents}\\
I Introduction\\
II Jacobi's inversion problem\\
III Theta identity of Riemann's type with two variables\\
IV Parameterization of the ratio of theta function with the symmetric 
function of $x_1$ and $x_2$\\
V Differential equation by using the addition formula of the theta function\\
VI Differential equation for $\tau_{12}=0$ case\\
VII Summary and discussion\\
Appendix A: Property of theta function\\
Appendix B: Various theta identity\\
Appendix C: Parameterization of the constant\\
Appendix D: Parameterization of the ratio of other theta function\\
Appendix E: Addition formula of theta function and differential formula\\


\setcounter{equation}{0}
\section{Introduction}

There are many two dimensional integrable statistical models \cite{Baxter}.
Let's consider the general Yang-Baxter equation of the spin model
in the form  
\begin{eqnarray}
  && U(x) V(x,y) U(y)=V(y)U(y,x) V(x).
\label{e1-1}
\end{eqnarray}   
This type of Yang-Baxter equation is called the integrability condition.
The meaning of the integrability condition is that the model has the group 
structure. As the Yang-Baxter equation 
of this type says that the product of three group actions for two different paths
gives the same group action, the group structure of the model is expected to 
be $SU(2)$.\\
Furthermore if the Yang-Baxter equation satisfies the difference property 
such as 
\begin{eqnarray}
  && U(x) V(x+y) U(y)=V(y)U(x+y) V(x),
\label{e1-2}
\end{eqnarray}  
we can understand this relation as some kind of addition formula, and 
we can solve the problem exactly by using only the addition formula.
\\
First and famous integrable and exactly solvable spin model is the Ising 
model\cite{Onsager}.
The structure of the Ising model is $SU(2)$\cite{Baxter,Onsager}. 
While the Boltzmann weight of the Ising model can be parameterized by 
the elliptic function\cite{Baxter}.   
Then we expect the correspondence between the $SU(2)$ group structure 
and the elliptic function. In other words, we expect that the symmetry 
of the elliptic function is $SU(2)$, and the origin of the 
addition formulae of the elliptic function comes from the $SU(2)$ group 
structure.
This is just the same as the addition formula of the trigonometric
 function, which is the Abelian function with genus zero.
The addition formula of the trigonometric function forms to give
the $U(1)$ group structure $\exp(i(x+y))=\exp(ix) \exp(iy)$ 
by parameterizing the circle with the 
trigonometric function through the Euler's relation 
$\exp(ix)=\cos(x)+i \sin(x)$.\\
For the elliptic function, which is the Abelian function with genus one, addition
formula closed as the rational function in the following complicated form 
\begin{eqnarray}
&& \sn(x+y)=\frac{\sn(x) \cn(y) \dn(y)+\sn(y)\cn(x)\dn(x)}{1-k^2 \sn^2(x) \sn^2(y)}, 
\label{e1-3}\\
&& \cn(x+y)=\frac{\cn(x) \cn(y)-\sn(x)\sn(y)\dn(x)\dn(y)}{1-k^2 \sn^2(x) \sn^2(y)}, 
\label{e1-4}\\
&& \dn(x+y)=\frac{\dn(x) \dn(y)-k^2 \sn(x)\sn(y)\cn(x)\cn(y)}{1-k^2 \sn^2(x) \sn^2(y)}.
\label{e1-5}
\end{eqnarray}\\
We expect the more precise structure than the closure of the rational function. 
As $x$ and $y$ are continuous variables, the closure of the algebraic 
function will be expected to give the Lie group structure. Then we expect that 
the addition
formula Eq.(\ref{e1-3}), Eq.(\ref{e1-4}) and Eq.(\ref{e1-5}) will be transformed into 
the Lie group structure.\\
In the previous papers, we have considered the surface of the sphere, 
which has $SU(2)$ symmetry, and we have parameterized the spherical 
trigonometry relations with the elliptic function.
Then the addition formula of the elliptic function forms to give 
the integrable $SU(2)$ Lie group structure of the Ising model 
in the form\cite{Shigemoto1,Shigemoto2} 
\begin{eqnarray}
&& U(x) V(x+y) U(y)= V(y) U(x+y) V(x),
\label{e1-6}\\
&&U(x)=\exp\{i{\rm am}(x,k) J_z\}, V(x)=\exp\{i{\rm am}(kx,1/k) J_x\}, 
\label{e1-7} 
\end{eqnarray} 
If we take $k\rightarrow 0$, Eq.(\ref{e1-6}) reduces to the addition formula of $U(1)$ 
in the form $\exp(ix J_z) \exp(iy J_z)=\exp(i(x+y)J_z)$.\\
For spin $1/2$ case, the above relation gives 
\begin{eqnarray}
&& U_{1/2}(x) V_{1/2}(x+y) U_{1/2}(y)= V_{1/2}(y) U_{1/2}(x+y) V_{1/2}(x), 
\label{e1-8}\\
&&U_{1/2}(x)=\left(\sqrt{\frac{1+\cn(x,k)}{2}} +i \sigma_z \sqrt{\frac{1-\cn(x,k)}{2}}\right),
\nonumber\\
&&V_{1/2}(x)=
\left(\sqrt{\frac{1+\cn(kx,1/k)}{2}} +i \sigma_x \sqrt{\frac{1-\cn(kx,1/k)}{2}}\right)
\nonumber\\
&&=\left(\sqrt{\frac{1+\dn(x,k)}{2}} +i \sigma_x \sqrt{\frac{1-\dn(x,k)}{2}}\right) ,
\label{e1-9}\\ 
&&\sigma_z=\left( \begin{array} {cc} 
  1 & 0\\
  0 & -1  \end{array} \right)  , 
\quad 
\sigma_x=\left( \begin{array} {cc} 
  0 & 1\\
  1 & 0 \end{array} \right) . \nonumber
\end{eqnarray} 
For spin $1$ case, the above relation gives 
\begin{eqnarray}
&& U_{1}(x) V_{1}(x+y) U_{1}(y)= V_{1}(y) U_{1}(x+y) V_{1}(x), 
\label{e1-10}\\
&&U_{1}(x)=\left(1-(1-\cn(x,k))J_z^2 +i J_z \sn(x,k) \right),
\nonumber\\
&&V_{1}(x)=\left( 1-(1-\cn(kx,1/k)) J_x^2+i J_x \sn(kx,1/k)\right)
\nonumber\\
&&=\left( 1-(1-\dn(x,k)) J_x^2+i J_x k \sn(x,k)\right) , 
\label{e1-11} \\
&&J_z=\left( \begin{array} {ccc} 
  0 & -i & 0 \\
  i & 0 & 0 \\
  0 & 0 & 0  \end{array} \right)  , 
\quad 
J_x=\left( \begin{array} {ccc} 
  0 & 0 & 0 \\
  0 & 0 & -i \\
  0 & i & 0  \end{array} \right) . \nonumber
\end{eqnarray}
In this way, the addition formula of the elliptic function forms to give 
the integrable $SU(2)$ group structure.\\
For the Abelian function with genus two, we have 15 hyperelliptic functions
$f_i(x,y)$ $(i=1, \cdots, 15)$ and the addition formula becomes\cite{Kossak} 
\begin{eqnarray}
&&\hskip -20 mm f_i(x_1+x_2,y_1+y_2)=\frac{F(f_1(x_1,y_1),f_1(x_2,y_2)\cdots 
f_{15}(x_1,y_1),f_{15}(x_2,y_2))}
{G(f_1(x_1,y_1),f_1(x_2,y_2)\cdots 
f_{15}(x_1,y_1),f_{15}(x_2,y_2))},
(i=1, \cdots, 15)
\label{e1-12}
\end{eqnarray}
where $F$ and $G$ are polynomial of the argument. We expect that this 
hyperelliptic addition formula will form some integrable 
Lie group structure. 
To have the hint of the structure of the addition formula of the hyperelliptic
theta function, we first revisit  
the Jacobi's inversion problem of the hyperelliptic integral with genus two. 
This Jacobi's inversion problem is solved 
independently by G\"{o}pel\cite{Goepel1,Goepel2} and 
Rosenhain\cite{Rosenhain1,Rosenhain2,Rosenhain3}. Rosenhain's paper is more 
precise so that we follow according to Rosenhain's paper. 

\vspace{10mm}

\setcounter{equation}{0}
\section{Jacobi's inversion problem}

\subsection{Abelian integral with genus one case}
We take the $4$th order polynomial of the form
$f_4(x)=(1-x^2)(1-k^2x^2)$ and consider the Abelian integral
\begin{eqnarray}
du=\frac{dx}{\sqrt{f_4(x)}} .
\label{e2-1}
\end{eqnarray}
The Abelian function in this case are $x$, $\sqrt{1-x^2}$, $\sqrt{1-k^2x^2}$.
Abelian function $x$ is given as the inverse function of $u$ from 
Eq.(\ref{e2-1}), which is given as one of the Jacobi's elliptic function
$x=\sn(u,k)$. Furthermore,  Abelian function
$x=\sn(u,k)$, $\sqrt{1-x^2}=\cn(u,k)$, $\sqrt{1-k^2x^2}=\dn(u,k)$ 
are expressed as the ratio of the theta function. For $x=\sn(u,k)$ case , we have 
\begin{eqnarray}
\hskip -5mm x=\sn(u,k)=\sn(u;\tau)=
-\vartheta\left[\begin{array}{c} 0\\ 0\\ \end{array}\right](0;\tau)
\vartheta\left[\begin{array}{c} 1\\ 1\\ \end{array}\right](z;\tau)
/\vartheta\left[\begin{array}{c} 1\\ 0\\ \end{array}\right](0;\tau)
\vartheta\left[\begin{array}{c} 0\\ 1\\ \end{array}\right](z;\tau) ,
\label{e2-2}
\end{eqnarray}
where
\begin{eqnarray}
&&K=\int_0^1 \frac{dx}{\sqrt{f_4(x)}}, \quad 
K'=\sqrt{-1} \int_{1/k}^1 \frac{dx}{\sqrt{f_4(x)}} ,
\quad \tau=\sqrt{-1}K'/K , 
\label{e2-3}\\
&&k=
\vartheta^2\left[\begin{array}{c} 1\\ 0\\ \end{array}\right](0;\tau)/
\vartheta^2\left[\begin{array}{c} 0\\ 0\\ \end{array}\right](0;\tau) ,
\quad 
u=2K z, 
\label{e2-4}
\end{eqnarray}
which connects $\tau$ with $k$, and connects $u$ with $z$.

\subsection{Abelian integral with genus two case}
We take the $5$th order polynomial of the form
$f_5(x)=x(1-x)(1-k_0^2 x)(1-k_1^2 x)(1-k_2^2 x)$ 
and consider the Abelian integral \\
\begin{eqnarray}
&&du=\frac{(P+Q x_1)dx_1}{\sqrt{f_5(x_1)}}+\frac{(P+Q x_2)dx_2}{\sqrt{f_5(x_2)}} , 
\label{e2-5}\\
&&dv=\frac{(R+S x_1)dx_1}{\sqrt{f_5(x_1)}}+\frac{(R+S x_2)dx_2}{\sqrt{f_5(x_2)}} .
\label{e2-6}
\end{eqnarray}
The Jacobi's inversion problem for genus two case is given in the followings:
the single-valued function is the symmetric combination of $x_1$ and $x_2$, 
and such single valued function is expressed as the ratio of the hyperelliptic theta 
function with the above two variables $u$ and $v$.\\
For example, $x_1 x_2$ and $(1-x_1)(1-x_2)$, which are some of the symmetric 
combination of $x_1$ and $x_2$,
are given by  \\
\begin{eqnarray}
&&x_1 x_2=c_1 
\vartheta^2 \left[\begin{array}{cc} 1 \ 0 \\ 1 \ 1 \\
\end{array}\right](u,v)/
\vartheta^2 \left[\begin{array}{cc} 0 \ 0 \\ 1 \ 1 \\
\end{array}\right](u,v)  , 
\label{e2-7}\\
&&(1-x_1)(1-x_2)=c_2 
\vartheta^2 \left[\begin{array}{cc} 1 \ 0 \\ 0 \ 1 \\
\end{array}\right](u,v)/
\vartheta^2 \left[\begin{array}{cc} 0 \ 0 \\ 1 \ 1 \\
\end{array}\right](u,v)  , 
\label{e2-8}
\end{eqnarray} \\
Similarly, $(1-k_0^2 x_1)(1-k_0^2 x_2)$,  $(1-k_1^2 x_1)(1-k_1^2 x_2)$,
 $(1-k_2^2 x_1)(1-k_2^2 x_2)$ are expressed as the ratio of the theta function
with two variable $u$ and $v$.\\
We will show the above statement by first parameterize the symmetric combination of
$x_1$ and $x_2$ by using the theta identity, and next we derive the differential equation
by using the addition formula of the theta function, which gives the differential 
equation in the form of Eq.(\ref{e2-5}) and Eq.(\ref{e2-6}). 

\vspace{10mm}
\setcounter{equation}{0}
\section{Riemann theta identity with two variables}

The theta function with two variables is defined by  
\begin{eqnarray}
&&\vartheta\left[\begin{array}{cc} a \ c \\ b \ d \\ 
\end{array}\right](u,v;\tau_1, \tau_2, \tau_{12})
\nonumber\\
&&=\sum_{m,n\in Z} \exp 
\left\{ \pi i  
\Bigl( \tau_1 (m+\frac{a}{2})^2+\tau_2 (n+\frac{c}{2})^2+
2 \tau_{12} (m+\frac{a}{2}) (n+\frac{c}{2}) \Bigr) \right.
\nonumber\\
&& \left. +2\pi i \ 
\Bigl( (m+\frac{a}{2})(u+\frac{b}{2})+(n+\frac{c}{2})(v+\frac{d}{2})
\Bigr)
\right\}
\label{e3-1}\\
&&=\sum_{m,n\in Z} \exp 
\left\{ \frac{\pi i}{4}  
\Bigl( \tau_1 (2m+a)^2 +\tau_2 (2n+c)^2 +
 2\tau_{12} (2m+a) (2n+c) \Bigr. \right.
\nonumber\\
&& \left. \Bigl. +2(2m+a)(2u+b) +2(2n+c)(2v+d) \Bigr) \right\}  . 
\label{e3-2}
\end{eqnarray}\\
Using the above, we define \\
\begin{eqnarray}
M(u, v)=\prod_{i=1}^{4} \vartheta\left[\begin{array}{cc} 0 \ 0 \\ 0 \ 0 \\
\end{array}\right](u_i,v_i)
+\prod_{i=1}^{4} \vartheta\left[\begin{array}{cc} 0 \ 1 \\ 0 \ 0 \\
\end{array}\right](u_i,v_i)  , 
\label{e3-3}\\
M'(u, v)=\prod_{i=1}^{4} \vartheta\left[\begin{array}{cc} 1 \ 0 \\ 0 \ 0 \\
\end{array}\right](u_i,v_i)  
+\prod_{i=1}^{4} \vartheta\left[\begin{array}{cc} 1 \ 1 \\ 0 \ 0 \\
\end{array}\right](u_i,v_i)  , 
\label{e3-4}\\
M''(u, v)=\prod_{i=1}^{4} \vartheta\left[\begin{array}{cc} 1 \ 0 \\ 1 \ 0 \\
\end{array}\right](u_i,v_i)
+\prod_{i=1}^{4} \vartheta\left[\begin{array}{cc} 1 \ 1 \\ 1 \ 0 \\
\end{array}\right](u_i,v_i)   , 
\label{e3-5}\\
M'''(u, v)=\prod_{i=1}^{4} \vartheta\left[\begin{array}{cc} 0 \ 0 \\ 1 \ 0 \\
\end{array}\right](u_i,v_i)
+\prod_{i=1}^{4} \vartheta\left[\begin{array}{cc} 0 \ 1 \\ 1 \ 0 \\
\end{array}\right](u_i,v_i) .
\label{e3-6}
\end{eqnarray}\\
Next we take the following combination
\begin{eqnarray}
&&M(u, v)+M'(u, v)
\nonumber\\
&&=\prod_{i=1}^{4} 
\vartheta\left[\begin{array}{cc} 0 \ 0 \\ 0 \ 0 \\
\end{array}\right](u_i,v_i)
+\prod_{i=1}^{4} \vartheta\left[\begin{array}{cc} 0 \ 1 \\ 0 \ 0 \\
\end{array}\right](u_i,v_i)
+\prod_{i=1}^{4} \vartheta\left[\begin{array}{cc} 1 \ 0 \\ 0 \ 0 \\
\end{array}\right](u_i,v_i)
+\prod_{i=1}^{4} \vartheta\left[\begin{array}{cc} 1 \ 1 \\ 0 \ 0 \\
\end{array}\right](u_i,v_i)
\nonumber\\
&&=\sum_{m'_i,n'_i \in Z' } \exp 
\left\{ \frac{\pi i}{4}  
\sum_{i=1}^4 \Bigl( \tau_1 {m'_i}^2 +\tau_2 {n'_i}^2 +
 2\tau_{12} m'_i n'_i 
+4 m'_i u_i+ 4 n'_i v_i \Bigr) \right\} , 
\label{e3-7}
\end{eqnarray}
where $m'_i,n'_i \in Z'$ means that $(m'_1,m'_2,m'_3,m'_4)$ all take even integer 
or all take odd integer, and $(n_1,n'_2,n'_3,n'_4)$ all take even integer 
or all take odd integer.\\
We define the Riemann matrix $A$
\begin{eqnarray}
A=\frac{1}{2}\left(\begin{array}{cccc}
1 & 1 & 1 & 1\\
1 & 1 & -1 & -1\\
1 & -1 & 1 & -1\\
1 & -1 & -1 & 1
\end{array}\right)  , 
\label{e3-8}
\end{eqnarray}
which is the orthogonal matrix with 
$A^{T}=A^{-1}$. \\
Using this Riemann matrix, we transform $m'_i,n'_i,u_i,v_i$ in the form \\
\begin{eqnarray}
&&\left(\begin{array}{c}
\tilde{m}_1\\
\tilde{m}_2\\
\tilde{m}_3\\
\tilde{m}_4\\
\end{array}\right)
=A\left(\begin{array}{c}
m'_1\\
m'_2\\
m'_3\\
m'_4\\
\end{array}\right)
=\left(\begin{array}{c}
(m'_1+m'_2+m'_3+m'_4)/2\\
(m'_1+m'_2-m'_3-m'_4)/2\\
(m'_1-m'_2+m'_3-m'_4)/2\\
(m'_1-m'_2-m'_3+m'_4)/2\\
\end{array}\right) ,
\label{e3-9}\\
&&\left(\begin{array}{c}
\tilde{n}_1\\
\tilde{n}_2\\
\tilde{n}_3\\
\tilde{n}_4\\
\end{array}\right)
=A\left(\begin{array}{c}
n'_1\\
n'_2\\
n'_3\\
n'_4\\
\end{array}\right)
=\left(\begin{array}{c}
(n'_1+n'_2+n'_3+n'_4)/2\\
(n'_1+n'_2-n'_3-n'_4)/2\\
(n'_1-n'_2+n'_3-n'_4)/2\\
(n'_1-n'_2-n'_3+n'_4)/2\\
\end{array}\right) ,
\label{e3-10}\\
&&\left(\begin{array}{c}
\tilde{u}_1\\
\tilde{u}_2\\
\tilde{u}_3\\
\tilde{u}_4\\
\end{array}\right)
=A\left(\begin{array}{c}
u_1\\
u_2\\
u_3\\
u_4\\
\end{array}\right)
=\left(\begin{array}{c}
(u_1+u_2+u_3+u_4)/2\\
(u_1+u_2-u_3-u_4)/2\\
(u_1-u_2+u_3-u_4)/2\\
(u_1-u_2-u_3+u_4)/2\\
\end{array}\right) ,
\label{e2-11}\\
&&\left(\begin{array}{c}
\tilde{v}_1\\
\tilde{v}_2\\
\tilde{v}_3\\
\tilde{v}_4\\
\end{array}\right)
=A\left(\begin{array}{c}
v_1\\
v_2\\
v_3\\
v_4\\
\end{array}\right)
=\left(\begin{array}{c}
(v_1+v_2+v_3+v_4)/2\\
(v_1+v_2-v_3-v_4)/2\\
(v_1-v_2+v_3-v_4)/2\\
(v_1-v_2-v_3+v_4)/2\\
\end{array}\right) .
\label{e3-12}
\end{eqnarray}\\
Then we have 
\begin{eqnarray}
&&M(u, v)+M'(u, v)
\nonumber\\
&&=\sum_{\tilde{m}_i,\tilde{n}_i \in Z' } \exp 
\left\{ \frac{\pi i}{4}  
\sum_{i=1}^4 \Bigl( \tau_1 {\tilde{m}_i}^2 +\tau_2 {\tilde{n}_i}^2 +
 2\tau_{12} \tilde{m}_i \tilde{n}_i 
+4 \tilde{m}_i \tilde{u}_i+ 4 \tilde{n}_i \tilde{v}_i \Bigr) \right\}
\nonumber\\ 
&&=M(\tilde{u}, \tilde{v})+M'(\tilde{u}, \tilde{v})  . 
\label{e3-13} 
\end{eqnarray}\\
The reason why the above relation is satisfied is the followings:
if we consider $\tilde{m_1}=(m'_1+m'_2+m'_3+m'_4)/2$, then as $\{m'_1,m'_2,m'_3,m'_4\}$
are all even integer or are all odd integer, $\tilde{m_1}$ becomes integer. Furthermore,
if we consider $\tilde{m_2}=(m'_1+m'_2-m'_3-m'_4)/2$, then $\tilde{m_2}$ also becomes 
integer, and $\tilde{m}_1-\tilde{m}_2=m'_3+m'_4=({\rm even\ integer})$, so that we have 
$\{ \tilde{m}_1,\tilde{m}_2,\tilde{m}_3,\tilde{m}_4 \}$ are all even integer or are all 
odd integer.\\
We simply denote $M=M(u, v)$, $M'=M'(u, v)$, $M''=M''(u,v)$, $M'''=M'''(u, v)$ and
 $\tilde{M}=M(\tilde{u}, \tilde{v})$, $\tilde{M}'=M'(\tilde{u}, \tilde{v})$, 
$\tilde{M}''=M''(\tilde{u}, \tilde{v})$, $\tilde{M}'''=M'''(\tilde{u}, \tilde{v})$.\\
Then we have the Riemann theta identity for two variables in the form 
\begin{eqnarray}
&&M+M'=\tilde{M}+\tilde{M}'  ,
\label{e3-14}\\
&&
{\rm that\ is}, \nonumber\\
&&\hskip -15mm
\prod_{i=1}^{4} \vartheta\left[\begin{array}{cc} 0 \ 0 \\ 0 \ 0 \\
\end{array}\right](u_i,v_i)
+\prod_{i=1}^{4} \vartheta\left[\begin{array}{cc} 0 \ 1 \\ 0 \ 0 \\
\end{array}\right](u_i,v_i)
+\prod_{i=1}^{4} \vartheta\left[\begin{array}{cc} 1 \ 0 \\ 0 \ 0 \\
\end{array}\right](u_i,v_i)
+\prod_{i=1}^{4} \vartheta\left[\begin{array}{cc} 1 \ 1 \\ 0 \ 0 \\
\end{array}\right](u_i,v_i)
\nonumber\\
&&\hskip-20mm
 =\prod_{i=1}^{4} \vartheta\left[\begin{array}{cc} 0 \ 0 \\ 0 \ 0 \\
\end{array}\right](\tilde{u}_i,\tilde{v}_i)
+\prod_{i=1}^{4} \vartheta\left[\begin{array}{cc} 0 \ 1 \\ 0 \ 0 \\
\end{array}\right](\tilde{u}_i,\tilde{v}_i)
+\prod_{i=1}^{4} \vartheta\left[\begin{array}{cc} 1 \ 0 \\ 0 \ 0 \\
\end{array}\right](\tilde{u}_i,\tilde{v}_i)
+\prod_{i=1}^{4} \vartheta\left[\begin{array}{cc} 1 \ 1 \\ 0 \ 0 \\
\end{array}\right](\tilde{u}_i,\tilde{v}_i) .
\label{e3-15}
\end{eqnarray}\\
Another Riemann theta identity is given by replacing the variable
in Eq.(\ref{e3-15}).
By replacing $u_1 \rightarrow u_1+1/2$, $u_2 \rightarrow u_2+1/2$,
$u_3\rightarrow u_3+1/2$, $u_4 \rightarrow u_4+1/2$, which gives 
$\tilde{u}_1\rightarrow \tilde{u}_1+1$, $\tilde{u}_2\rightarrow \tilde{u}_2$,
$\tilde{u}_3\rightarrow \tilde{u}_3$, $\tilde{u}_4\rightarrow \tilde{u}_4$,
and also $v_i$ are not changed.\\
This gives 
\begin{eqnarray}
&&M''+M'''=\tilde{M}-\tilde{M}'  ,
\label{e3-16}\\
&&
{\rm that\ is}, \nonumber\\
&&\hskip -15mm
\prod_{i=1}^{4} \vartheta\left[\begin{array}{cc} 0 \ 0 \\ 1 \ 0 \\
\end{array}\right](u_i,v_i)
+\prod_{i=1}^{4} \vartheta\left[\begin{array}{cc} 0 \ 1 \\ 1 \ 0 \\
\end{array}\right](u_i,v_i)
+\prod_{i=1}^{4} \vartheta\left[\begin{array}{cc} 1 \ 0 \\ 1 \ 0 \\
\end{array}\right](u_i,v_i)
+\prod_{i=1}^{4} \vartheta\left[\begin{array}{cc} 1 \ 1 \\ 1 \ 0 \\
\end{array}\right](u_i,v_i)
\nonumber\\
&&\hskip-20mm
 =\prod_{i=1}^{4} \vartheta\left[\begin{array}{cc} 0 \ 0 \\ 0 \ 0 \\
\end{array}\right](\tilde{u}_i,\tilde{v}_i)
+\prod_{i=1}^{4} \vartheta\left[\begin{array}{cc} 0 \ 1 \\ 0 \ 0 \\
\end{array}\right](\tilde{u}_i,\tilde{v}_i)
-\prod_{i=1}^{4} \vartheta\left[\begin{array}{cc} 1 \ 0 \\ 0 \ 0 \\
\end{array}\right](\tilde{u}_i,\tilde{v}_i)
-\prod_{i=1}^{4} \vartheta\left[\begin{array}{cc} 1 \ 1 \\ 0 \ 0 \\
\end{array}\right](\tilde{u}_i,\tilde{v}_i) , 
\label{e3-17}
\end{eqnarray}
where we use $\vartheta\left[\begin{array}{cc} 1 \ c \\ 0 \ 0 \\
\end{array}\right](\tilde{u}_i+1,\tilde{v}_i)
=-\vartheta\left[\begin{array}{cc} 1 \ c \\ 0 \ 0 \\
\end{array}\right](\tilde{u}_i,\tilde{v}_i)$.\\
Further, we replace  $u_1 \rightarrow u_1+1$, $u_2 \rightarrow u_2$,
$u_3 \rightarrow u_3$, $u_4 \rightarrow u_4$, which gives 
$\tilde{u}_1\rightarrow \tilde{u}_1+1/2$, 
$\tilde{u}_2\rightarrow \tilde{u}_2+1/2$,
$\tilde{u}_3\rightarrow \tilde{u}_3+1/2$,
$\tilde{u}_4\rightarrow \tilde{u}_4+1/2$,
in Eq.(\ref{e3-15}) and also $v_i$ are not changed.
This gives 
\begin{eqnarray}
&&M-M'=\tilde{M}''+\tilde{M}'''  ,
\label{e3-18}\\
&&
{\rm that\ is}, \nonumber\\
&&\hskip -15mm
\prod_{i=1}^{4} \vartheta\left[\begin{array}{cc} 0 \ 0 \\ 0 \ 0 \\
\end{array}\right](u_i,v_i)
+\prod_{i=1}^{4} \vartheta\left[\begin{array}{cc} 0 \ 1 \\ 0 \ 0 \\
\end{array}\right](u_i,v_i)
-\prod_{i=1}^{4} \vartheta\left[\begin{array}{cc} 1 \ 0 \\ 0 \ 0 \\
\end{array}\right](u_i,v_i)
-\prod_{i=1}^{4} \vartheta\left[\begin{array}{cc} 1 \ 1 \\ 0 \ 0 \\
\end{array}\right](u_i,v_i)
\nonumber\\
&&\hskip-20mm
 =\prod_{i=1}^{4} \vartheta\left[\begin{array}{cc} 0 \ 0 \\ 1 \ 0 \\
\end{array}\right](\tilde{u}_i,\tilde{v}_i)
+\prod_{i=1}^{4} \vartheta\left[\begin{array}{cc} 0 \ 1 \\ 1 \ 0 \\
\end{array}\right](\tilde{u}_i,\tilde{v}_i)
+\prod_{i=1}^{4} \vartheta\left[\begin{array}{cc} 1 \ 0 \\ 1 \ 0 \\
\end{array}\right](\tilde{u}_i,\tilde{v}_i)
+\prod_{i=1}^{4} \vartheta\left[\begin{array}{cc} 1 \ 1 \\ 1 \ 0 \\
\end{array}\right](\tilde{u}_i,\tilde{v}_i).
\label{e3-19}
\end{eqnarray}\\
Further, we replace  $u_1 \rightarrow u_1+1$, $u_2 \rightarrow u_2$,
$u_3 \rightarrow u_3$, $u_4 \rightarrow u_4$, which gives 
$\tilde{u}_1\rightarrow \tilde{u}_1+1/2$, 
$\tilde{u}_2\rightarrow \tilde{u}_2+1/2$,
$\tilde{u}_3\rightarrow \tilde{u}_3+1/2$,
$\tilde{u}_4\rightarrow \tilde{u}_4+1/2$,
in Eq.(\ref{e3-17}) and also $v_i$ are not changed.
This gives 
\begin{eqnarray}
&&-M''+M'''=-\tilde{M}''+\tilde{M}'''  ,
\label{e3-20}\\
&&
{\rm that\ is}, \nonumber\\
&&\hskip -15mm
\prod_{i=1}^{4} \vartheta\left[\begin{array}{cc} 0 \ 0 \\ 1 \ 0 \\
\end{array}\right](u_i,v_i)
+\prod_{i=1}^{4} \vartheta\left[\begin{array}{cc} 0 \ 1 \\ 1 \ 0 \\
\end{array}\right](u_i,v_i)
-\prod_{i=1}^{4} \vartheta\left[\begin{array}{cc} 1 \ 0 \\ 1 \ 0 \\
\end{array}\right](u_i,v_i)
-\prod_{i=1}^{4} \vartheta\left[\begin{array}{cc} 1 \ 1 \\ 1 \ 0 \\
\end{array}\right](u_i,v_i)
\nonumber\\
&&\hskip-20mm
 =\prod_{i=1}^{4} \vartheta\left[\begin{array}{cc} 0 \ 0 \\ 1 \ 0 \\
\end{array}\right](\tilde{u}_i,\tilde{v}_i)
+\prod_{i=1}^{4} \vartheta\left[\begin{array}{cc} 0 \ 1 \\ 1 \ 0 \\
\end{array}\right](\tilde{u}_i,\tilde{v}_i)
-\prod_{i=1}^{4} \vartheta\left[\begin{array}{cc} 1 \ 0 \\ 1 \ 0 \\
\end{array}\right](\tilde{u}_i,\tilde{v}_i)
-\prod_{i=1}^{4} \vartheta\left[\begin{array}{cc} 1 \ 1 \\ 1 \ 0 \\
\end{array}\right](\tilde{u}_i,\tilde{v}_i).
\label{e3-21}
\end{eqnarray}\\
In summary, we have 
\begin{eqnarray}
&&2M=\tilde{M}+\tilde{M}'+\tilde{M}''+\tilde{M}''',
\label{e3-22}\\
&&2M'=\tilde{M}+\tilde{M}'-\tilde{M}''-\tilde{M}''',
\label{e3-23}\\
&&2M''=\tilde{M}-\tilde{M}'+\tilde{M}''-\tilde{M}''',
\label{e3-24}\\
&&2M'''=\tilde{M}-\tilde{M}'-\tilde{M}''+\tilde{M}'''  .
\label{e3-25}
\end{eqnarray}\\
We inversely solve the above in the following form 
\begin{eqnarray}
&&2\tilde{M}=M+M'+M''+M''',
\label{e3-26}\\
&&2\tilde{M}'=M+M'-M''-M''',
\label{e3-27}\\
&&2\tilde{M}''=M-M'+M''-M''',
\label{e3-28}\\
&&2\tilde{M}'''=M-M'-M''+M'''  . 
\label{e3-29}
\end{eqnarray}\\
We can express the above relation in the form
\begin{eqnarray}
&&\left(\begin{array}{c}
M\\
M'\\
M''\\
M'''\\
\end{array}\right)
=A\left(\begin{array}{c}
\tilde{M}\\
\tilde{M}'\\
\tilde{M}''\\
\tilde{M}'''\\
\end{array}\right), \quad
\left(\begin{array}{c}
\tilde{M}\\
\tilde{M}'\\
\tilde{M}''\\
\tilde{M}'''\\
\end{array}\right)
=A\left(\begin{array}{c}
M\\
M'\\
M''\\
M'''\\
\end{array}\right) .
\label{e3-30}
\end{eqnarray}

\vskip 10mm
\setcounter{equation}{0}
\section{Parameterization of the ratio of theta function with the symmetric function of $x_1$ and $x_2$}

We parameterize the ratio of theta function with the symmetric function of 
$x_1$ and $x_2$.
The basis of the symmetric function of two variable $x_1$ and $x_2$ are 
$\{x_1 x_2, x_1+x_2\}$, so that the linear combination of three polynomial 
is dependent.\\
We take $x_1x_2$, $(1-x_1)(1-x_2)$, $(1-k_0^2 x_1)(1-k_0^2 x_2)$, we have
the following identity
\begin{eqnarray}
1=k_0^2 x_1x_2-\frac{k_0^2}{1-k_0^2}(1-x_1)(1-x_2)
+\frac{1}{1-k_0^2}(1-k_0^2x_1)(1-k_0^2x_2) . 
\label{e4-1}
\end{eqnarray}\\
While we have the following theta identity 
\begin{eqnarray}
&&1=
\vartheta^2 \left[\begin{array}{cc} 0 \ 0 \\ 0 \ 1 \\
\end{array}\right](0,0) \ 
\vartheta^2 \left[\begin{array}{cc} 1 \ 0 \\ 1 \ 1 \\
\end{array}\right](u,v)/
\vartheta^2 \left[\begin{array}{cc} 1 \ 0 \\ 0 \ 1 \\
\end{array}\right](0,0) \ 
\vartheta^2 \left[\begin{array}{cc} 0 \ 0 \\ 1 \ 1 \\
\end{array}\right](u,v)
\nonumber\\
&&+\vartheta^2 \left[\begin{array}{cc} 0 \ 0 \\ 1 \ 1 \\
\end{array}\right](0,0) \ 
\vartheta^2 \left[\begin{array}{cc} 1 \ 0 \\ 0 \ 1 \\
\end{array}\right](u,v)/
\vartheta^2 \left[\begin{array}{cc} 1 \ 0 \\ 0 \ 1 \\
\end{array}\right](0,0) \ 
\vartheta^2 \left[\begin{array}{cc} 0 \ 0 \\ 1 \ 1 \\
\end{array}\right](u,v)
\nonumber\\
&&-\vartheta^2 \left[\begin{array}{cc} 1 \ 1 \\ 1 \ 1 \\
\end{array}\right](0,0) \ 
\vartheta^2 \left[\begin{array}{cc} 0 \ 1 \\ 0 \ 1 \\
\end{array}\right](u,v)/
\vartheta^2 \left[\begin{array}{cc} 1 \ 0 \\ 0 \ 1 \\
\end{array}\right](0,0) \ 
\vartheta^2 \left[\begin{array}{cc} 0 \ 0 \\ 1 \ 1 \\
\end{array}\right](u,v)  ,
\label{e4-2}
\end{eqnarray}
which is given by Eq.(\ref{B1-3}) by replacing $u\rightarrow u+\tau/2+1/2$,
$v \rightarrow v+\tau_{12}/2+1/2$.\\
Then we have the parameterization
\begin{eqnarray}
&&\frac{\vartheta^2 \left[\begin{array}{cc} 0 \ 0 \\ 0 \ 1 \\
\end{array}\right](0,0) \ 
\vartheta^2 \left[\begin{array}{cc} 1 \ 0 \\ 1 \ 1 \\
\end{array}\right](u,v)}
{\vartheta^2 \left[\begin{array}{cc} 1 \ 0 \\ 0 \ 1 \\
\end{array}\right](0,0) \ 
\vartheta^2 \left[\begin{array}{cc} 0 \ 0 \\ 1 \ 1 \\
\end{array}\right](u,v)}
=k_0^2 x_1 x_2  , 
\label{e4-3}\\
&&\frac{\vartheta^2 \left[\begin{array}{cc} 0 \ 0 \\ 1 \ 1 \\
\end{array}\right](0,0) \ 
\vartheta^2 \left[\begin{array}{cc} 1 \ 0 \\ 0 \ 1 \\
\end{array}\right](u,v)}
{\vartheta^2 \left[\begin{array}{cc} 1 \ 0 \\ 0 \ 1 \\
\end{array}\right](0,0) \ 
\vartheta^2 \left[\begin{array}{cc} 0 \ 0 \\ 1 \ 1 \\
\end{array}\right](u,v)}
=-\frac{k_0^2 }{1-k_0^2}(1-x_1)(1-x_2)  , 
\label{e4-4}\\
&&\frac{\vartheta^2 \left[\begin{array}{cc} 1 \ 1 \\ 1 \ 1 \\
\end{array}\right](0,0) \ 
\vartheta^2 \left[\begin{array}{cc} 0 \ 1 \\ 0 \ 1 \\
\end{array}\right](u,v)}
{\vartheta^2 \left[\begin{array}{cc} 1 \ 0 \\ 0 \ 1 \\
\end{array}\right](0,0) \ 
\vartheta^2 \left[\begin{array}{cc} 0 \ 0 \\ 1 \ 1 \\
\end{array}\right](u,v)}
=-\frac{1}{1-k_0^2}(1-k_0^2x_1)(1-k_0^2x_2)  .
\label{e4-5}
\end{eqnarray} \\
Similarly we use the identity
\begin{eqnarray}
1=k_1^2 x_1x_2-\frac{k_1^2}{1-k_1^2}(1-x_1)(1-x_2)
+\frac{1}{1-k_1^2}(1-k_1^2x_1)(1-k_1^2x_2)  ,
\label{e4-6}
\end{eqnarray} \\
and
\begin{eqnarray}
&&1=
\vartheta^2 \left[\begin{array}{cc} 0 \ 0 \\ 0 \ 0 \\
\end{array}\right](0,0) \ 
\vartheta^2 \left[\begin{array}{cc} 1 \ 0 \\ 1 \ 1 \\
\end{array}\right](u,v)/
\vartheta^2 \left[\begin{array}{cc} 1 \ 0 \\ 0 \ 0 \\
\end{array}\right](0,0) \ 
\vartheta^2 \left[\begin{array}{cc} 0 \ 0 \\ 1 \ 1 \\
\end{array}\right](u,v)
\nonumber\\
&&+\vartheta^2 \left[\begin{array}{cc} 0 \ 0 \\ 1 \ 0 \\
\end{array}\right](0,0) \ 
\vartheta^2 \left[\begin{array}{cc} 1 \ 0 \\ 0 \ 1 \\
\end{array}\right](u,v)/
\vartheta^2 \left[\begin{array}{cc} 1 \ 0 \\ 0 \ 0 \\
\end{array}\right](0,0) \ 
\vartheta^2 \left[\begin{array}{cc} 0 \ 0 \\ 1 \ 1 \\
\end{array}\right](u,v)
\nonumber\\
&&+\vartheta^2 \left[\begin{array}{cc} 1 \ 1 \\ 1 \ 1 \\
\end{array}\right](0,0) \ 
\vartheta^2 \left[\begin{array}{cc} 0 \ 1 \\ 0 \ 0 \\
\end{array}\right](u,v)/
\vartheta^2 \left[\begin{array}{cc} 1 \ 0 \\ 0 \ 0 \\
\end{array}\right](0,0) \ 
\vartheta^2 \left[\begin{array}{cc} 0 \ 0 \\ 1 \ 1 \\
\end{array}\right](u,v)  ,
\label{e4-7}
\end{eqnarray}
which is derived from Eq.(\ref{B1-1}) by replacing 
$u\rightarrow u+\tau/2+1/2$,
$v \rightarrow v+\tau_{12}/2+1/2$.\\
Then we have the parameterization
\begin{eqnarray}
&&\frac{\vartheta^2 \left[\begin{array}{cc} 0 \ 0 \\ 0 \ 0 \\
\end{array}\right](0,0) \ 
\vartheta^2 \left[\begin{array}{cc} 1 \ 0 \\ 1 \ 1 \\
\end{array}\right](u,v)}
{\vartheta^2 \left[\begin{array}{cc} 1 \ 0 \\ 0 \ 0 \\
\end{array}\right](0,0) \ 
\vartheta^2 \left[\begin{array}{cc} 0 \ 0 \\ 1 \ 1 \\
\end{array}\right](u,v)}
=k_1^2 x_1 x_2 ,
\label{e4-8}\\
&&\frac{\vartheta^2 \left[\begin{array}{cc} 0 \ 0 \\ 1 \ 0 \\
\end{array}\right](0,0) \ 
\vartheta^2 \left[\begin{array}{cc} 1 \ 0 \\ 0 \ 1 \\
\end{array}\right](u,v)}
{\vartheta^2 \left[\begin{array}{cc} 1 \ 0 \\ 0 \ 0 \\
\end{array}\right](0,0) \ 
\vartheta^2 \left[\begin{array}{cc} 0 \ 0 \\ 1 \ 1 \\
\end{array}\right](u,v)}
=-\frac{k_1^2 }{1-k_1^2}(1-x_1)(1-x_2) ,
\label{e4-9}\\
&&\frac{\vartheta^2 \left[\begin{array}{cc} 1 \ 1 \\ 1 \ 1 \\
\end{array}\right](0,0) \ 
\vartheta^2 \left[\begin{array}{cc} 0 \ 1 \\ 0 \ 0 \\
\end{array}\right](u,v)}
{\vartheta^2 \left[\begin{array}{cc} 1 \ 0 \\ 0 \ 0 \\
\end{array}\right](0,0) \ 
\vartheta^2 \left[\begin{array}{cc} 0 \ 0 \\ 1 \ 1 \\
\end{array}\right](u,v)}
=\frac{1}{1-k_1^2}(1-k_1^2 x_1)(1-k_1^2 x_2) .
\label{e4-10}
\end{eqnarray} \\
Similarly we use the identity
\begin{eqnarray}
1=k_2^2 x_1x_2-\frac{k_2^2}{1-k_2^2}(1-x_1)(1-x_2)
+\frac{1}{1-k_2^2}(1-k_2^2x_1)(1-k_1^2x_2) ,
\label{e4-11}
\end{eqnarray}
and
\begin{eqnarray}
&&1=
\vartheta^2 \left[\begin{array}{cc} 0 \ 1 \\ 0 \ 0 \\
\end{array}\right](0,0) \ 
\vartheta^2 \left[\begin{array}{cc} 1 \ 0 \\ 1 \ 1 \\
\end{array}\right](u,v)/
\vartheta^2 \left[\begin{array}{cc} 1 \ 1 \\ 0 \ 0 \\
\end{array}\right](0,0) \ 
\vartheta^2 \left[\begin{array}{cc} 0 \ 0 \\ 1 \ 1 \\
\end{array}\right](u,v)
\nonumber\\
&&+\vartheta^2 \left[\begin{array}{cc} 0 \ 1 \\ 1 \ 0 \\
\end{array}\right](0,0) \ 
\vartheta^2 \left[\begin{array}{cc} 1 \ 0 \\ 0 \ 1 \\
\end{array}\right](u,v)/
\vartheta^2 \left[\begin{array}{cc} 1 \ 1 \\ 0 \ 0 \\
\end{array}\right](0,0) \ 
\vartheta^2 \left[\begin{array}{cc} 0 \ 0 \\ 1 \ 1 \\
\end{array}\right](u,v)
\nonumber\\
&&+\vartheta^2 \left[\begin{array}{cc} 1 \ 1 \\ 1 \ 1 \\
\end{array}\right](0,0) \ 
\vartheta^2 \left[\begin{array}{cc} 0 \ 0 \\ 0 \ 0 \\
\end{array}\right](u,v)/
\vartheta^2 \left[\begin{array}{cc} 1 \ 1 \\ 0 \ 0 \\
\end{array}\right](0,0) \ 
\vartheta^2 \left[\begin{array}{cc} 0 \ 0 \\ 1 \ 1 \\
\end{array}\right](u,v) , 
\label{e4-12}
\end{eqnarray}
which is derived from Eq.(\ref{B1-2}) by replacing 
$u\rightarrow u+\tau/2+1/2$,
$v \rightarrow v+\tau_{12}/2+1/2$.\\
Then we have the parameterization
\begin{eqnarray}
&&
\frac{\vartheta^2 \left[\begin{array}{cc} 0 \ 1 \\ 0 \ 0 \\
\end{array}\right](0,0) \ 
\vartheta^2 \left[\begin{array}{cc} 1 \ 0 \\ 1 \ 1 \\
\end{array}\right](u,v)}
{\vartheta^2 \left[\begin{array}{cc} 1 \ 1 \\ 0 \ 0 \\
\end{array}\right](0,0) \ 
\vartheta^2 \left[\begin{array}{cc} 0 \ 0 \\ 1 \ 1 \\
\end{array}\right](u,v)}
=k_2^2 x_1 x_2 ,
\label{e4-13}\\
&&\frac{\vartheta^2 \left[\begin{array}{cc} 0 \ 1 \\ 1 \ 0 \\
\end{array}\right](0,0) \ 
\vartheta^2 \left[\begin{array}{cc} 1 \ 0 \\ 0 \ 1 \\
\end{array}\right](u,v)}
{\vartheta^2 \left[\begin{array}{cc} 1 \ 1 \\ 0 \ 0 \\
\end{array}\right](0,0) \ 
\vartheta^2 \left[\begin{array}{cc} 0 \ 0 \\ 1 \ 1 \\
\end{array}\right](u,v)}
=-\frac{k_2^2 }{1-k_2^2}(1-x_1)(1-x_2) ,
\label{e4-14}\\
&&\frac{\vartheta^2 \left[\begin{array}{cc} 1 \ 1 \\ 1 \ 1 \\
\end{array}\right](0,0) \ 
\vartheta^2 \left[\begin{array}{cc} 0 \ 0 \\ 0 \ 0 \\
\end{array}\right](u,v)}
{\vartheta^2 \left[\begin{array}{cc} 1 \ 1 \\ 0 \ 0 \\
\end{array}\right](0,0) \ 
\vartheta^2 \left[\begin{array}{cc} 0 \ 0 \\ 1 \ 1 \\
\end{array}\right](u,v)}
=\frac{1}{1-k_2^2}(1-k_2^2 x_1)(1-k_2^2 x_2) .
\label{e4-15}
\end{eqnarray}\\
From this parameterization, the theta function with zero argument and various 
$\{k_0, k_1, \cdots, \}$ is connected. Some consistency conditions 
must be checked.
The detail of these connections and the consistency of the above 
relations is explained in the Appendix C.\\
The summary of Appendix C is given as follows 
\begin{eqnarray}
&&k_0^2=\frac{\vartheta^2 \left[\begin{array}{cc} 1 \ 0 \\ 0 \ 0 \\
\end{array}\right](0,0) \ 
\vartheta^2 \left[\begin{array}{cc} 1 \ 1 \\ 0 \ 0 \\
\end{array}\right](0,0)} 
{\vartheta^2 \left[\begin{array}{cc} 0 \ 0 \\ 0 \ 0 \\
\end{array}\right](0,0) \ 
\vartheta^2 \left[\begin{array}{cc} 0 \ 1 \\ 0 \ 0 \\
\end{array}\right](0,0)}, 
\label{e4-16}\\
&&k_1^2=\frac{\vartheta^2 \left[\begin{array}{cc} 1 \ 0 \\ 0 \ 1 \\
\end{array}\right](0,0) \ 
\vartheta^2 \left[\begin{array}{cc} 1 \ 1 \\ 0 \ 0 \\
\end{array}\right](0,0)} 
{\vartheta^2 \left[\begin{array}{cc} 0 \ 0 \\ 0 \ 1 \\
\end{array}\right](0,0) \ 
\vartheta^2 \left[\begin{array}{cc} 0 \ 1 \\ 0 \ 0 \\
\end{array}\right](0,0)},
\label{e4-17}\\
&&k_2^2=\frac{\vartheta^2 \left[\begin{array}{cc} 1 \ 0 \\ 0 \ 1 \\
\end{array}\right](0,0) \ 
\vartheta^2 \left[\begin{array}{cc} 1 \ 0 \\ 0 \ 0 \\
\end{array}\right](0,0)} 
{\vartheta^2 \left[\begin{array}{cc} 0 \ 0 \\ 0 \ 1 \\
\end{array}\right](0,0) \ 
\vartheta^2 \left[\begin{array}{cc} 0 \ 0 \\ 0 \ 0 \\
\end{array}\right](0,0)}, 
\label{e4-18}\\
&&{k'_{0}}^2=
\frac{\vartheta^2 \left[\begin{array}{cc} 0 \ 0 \\ 1 \ 0 \\
\end{array}\right](0,0) \ 
\vartheta^2 \left[\begin{array}{cc} 0 \ 1 \\ 1 \ 0 \\
\end{array}\right](0,0)} 
{\vartheta^2 \left[\begin{array}{cc} 0 \ 0 \\ 0 \ 0 \\
\end{array}\right](0,0) \ 
\vartheta^2 \left[\begin{array}{cc} 0 \ 1 \\ 0 \ 0 \\
\end{array}\right](0,0)}, 
\label{e4-19}\\
&&{k'_{1}}^2=
\frac{\vartheta^2 \left[\begin{array}{cc} 0 \ 0 \\ 1 \ 1 \\
\end{array}\right](0,0) \ 
\vartheta^2 \left[\begin{array}{cc} 0 \ 1 \\ 1 \ 0 \\
\end{array}\right](0,0)} 
{\vartheta^2 \left[\begin{array}{cc} 0 \ 0 \\ 0 \ 1 \\
\end{array}\right](0,0) \ 
\vartheta^2 \left[\begin{array}{cc} 0 \ 1 \\ 0 \ 0 \\
\end{array}\right](0,0)}, 
\label{e4-20}\\
&&{k'_{2}}^2=
\frac{\vartheta^2 \left[\begin{array}{cc} 0 \ 0 \\ 1 \ 1 \\
\end{array}\right](0,0) \ 
\vartheta^2 \left[\begin{array}{cc} 0 \ 0 \\ 1 \ 0 \\
\end{array}\right](0,0)} 
{\vartheta^2 \left[\begin{array}{cc} 0 \ 0 \\ 0 \ 1 \\
\end{array}\right](0,0) \ 
\vartheta^2 \left[\begin{array}{cc} 0 \ 0 \\ 0 \ 0 \\
\end{array}\right](0,0)}, 
\label{e4-21}\\
&&k_{01}^2=
\frac{\vartheta^2 \left[\begin{array}{cc} 1 \ 1 \\ 0 \ 0 \\
\end{array}\right](0,0) \ 
\vartheta^2 \left[\begin{array}{cc} 1 \ 1 \\ 1 \ 1 \\
\end{array}\right](0,0) \ 
\vartheta^2 \left[\begin{array}{cc} 0 \ 1 \\ 1 \ 0 \\
\end{array}\right](0,0) }
{\vartheta^2 \left[\begin{array}{cc} 0 \ 1 \\ 0 \ 0 \\
\end{array}\right](0,0) \ 
\vartheta^2 \left[\begin{array}{cc} 0 \ 0 \\ 0 \ 0 \\
\end{array}\right](0,0) \ 
\vartheta^2 \left[\begin{array}{cc} 0 \ 0 \\ 0 \ 1 \\
\end{array}\right](0,0) }, 
\label{e4-22}\\
&&k_{02}^2=
\frac{\vartheta^2 \left[\begin{array}{cc} 1 \ 0 \\ 0 \ 0 \\
\end{array}\right](0,0) \ 
\vartheta^2 \left[\begin{array}{cc} 1 \ 1 \\ 1 \ 1 \\
\end{array}\right](0,0) \ 
\vartheta^2 \left[\begin{array}{cc} 0 \ 0 \\ 1 \ 0 \\
\end{array}\right](0,0) }
{\vartheta^2 \left[\begin{array}{cc} 0 \ 0 \\ 0 \ 0 \\
\end{array}\right](0,0) \ 
\vartheta^2 \left[\begin{array}{cc} 0 \ 1 \\ 0 \ 0 \\
\end{array}\right](0,0) \ 
\vartheta^2 \left[\begin{array}{cc} 0 \ 0 \\ 0 \ 1 \\
\end{array}\right](0,0) }, 
\label{e4-23}\\
&&k_{12}^2=
\frac{\vartheta^2 \left[\begin{array}{cc} 1 \ 0 \\ 0 \ 1 \\
\end{array}\right](0,0) \ 
\vartheta^2 \left[\begin{array}{cc} 1 \ 1 \\ 1 \ 1 \\
\end{array}\right](0,0) \ 
\vartheta^2 \left[\begin{array}{cc} 0 \ 0 \\ 1 \ 1 \\
\end{array}\right](0,0) }
{\vartheta^2 \left[\begin{array}{cc} 0 \ 0 \\ 0 \ 1 \\
\end{array}\right](0,0) \ 
\vartheta^2 \left[\begin{array}{cc} 0 \ 1 \\ 0 \ 0 \\
\end{array}\right](0,0) \ 
\vartheta^2 \left[\begin{array}{cc} 0 \ 0 \\ 0 \ 0 \\
\end{array}\right](0,0) } .
\label{e4-24}
\end{eqnarray}\\
Inversely, we express the ratio of theta function with zero argument 
by $\{ k_0, k_1, \cdots\}$ in the form\\
\begin{eqnarray}
&&\frac{\vartheta^2 \left[\begin{array}{cc} 0 \ 0 \\ 1 \ 1 \\
\end{array}\right](0,0) \ } 
{\vartheta^2 \left[\begin{array}{cc} 0 \ 0 \\ 0 \ 0 \\
\end{array}\right](0,0) \ }=\frac{k_0 k'_2 k_{12}}{k_1 k_{02}},
\quad
\frac{\vartheta^2 \left[\begin{array}{cc} 0 \ 1 \\ 1 \ 0 \\
\end{array}\right](0,0) \ } 
{\vartheta^2 \left[\begin{array}{cc} 0 \ 0 \\ 0 \ 0 \\
\end{array}\right](0,0) \ }=\frac{k'_0 k_2 k_{01}}{k_1 k_{02}},
\label{e4-25}\\
&&\frac{\vartheta^2 \left[\begin{array}{cc} 0 \ 0 \\ 1 \ 0 \\
\end{array}\right](0,0) \ } 
{\vartheta^2 \left[\begin{array}{cc} 0 \ 0 \\ 0 \ 0 \\
\end{array}\right](0,0) \ }=\frac{k'_0 k'_2 }{k'_1 },
\quad
\frac{\vartheta^2 \left[\begin{array}{cc} 1 \ 1 \\ 0 \ 0 \\
\end{array}\right](0,0) \ } 
{\vartheta^2 \left[\begin{array}{cc} 0 \ 0 \\ 0 \ 0 \\
\end{array}\right](0,0) \ }=\frac{k_0 k'_2 k_{01}}{k'_1 k_{02}},
\label{e4-26}\\
&&\frac{\vartheta^2 \left[\begin{array}{cc} 1 \ 0 \\ 0 \ 1 \\
\end{array}\right](0,0) \ } 
{\vartheta^2 \left[\begin{array}{cc} 0 \ 0 \\ 0 \ 0 \\
\end{array}\right](0,0) \ }=\frac{k'_0 k_2 k_{12}}{k'_1 k_{02}},
\quad
\frac{\vartheta^2 \left[\begin{array}{cc} 1 \ 0 \\ 0 \ 0 \\
\end{array}\right](0,0) \ } 
{\vartheta^2 \left[\begin{array}{cc} 0 \ 0 \\ 0 \ 0 \\
\end{array}\right](0,0) \ }=\frac{k_0 k_2 }{k_1 },
\label{e4-27}\\
&&\frac{\vartheta^2 \left[\begin{array}{cc} 1 \ 1 \\ 1 \ 1 \\
\end{array}\right](0,0) \ } 
{\vartheta^2 \left[\begin{array}{cc} 0 \ 0 \\ 0 \ 0 \\
\end{array}\right](0,0) \ }=\frac{k_{01} k_{12}}{k_1 k'_1},
\quad
\frac{\vartheta^2 \left[\begin{array}{cc} 0 \ 0 \\ 0 \ 1 \\
\end{array}\right](0,0) \ } 
{\vartheta^2 \left[\begin{array}{cc} 0 \ 0 \\ 0 \ 0 \\
\end{array}\right](0,0) \ }=\frac{k_0 k'_0 k_{12}}{k_1 k'_1 k_{02}},
\label{e4-28}\\
&&\frac{\vartheta^2 \left[\begin{array}{cc} 0 \ 1 \\ 0 \ 0 \\
\end{array}\right](0,0) \ } 
{\vartheta^2 \left[\begin{array}{cc} 0 \ 0 \\ 0 \ 0 \\
\end{array}\right](0,0) \ }=\frac{k_2 k'_2 k_{01}}{k_1 k'_1 k_{02}} .
\label{e4-29}
\end{eqnarray}\\
Using these expressions, we have the parameterization of ratio of the theta function with 
the symmetric function of $x_1$ and $x_2$ in the form\\
\begin{eqnarray}
&&1) \quad \frac{\vartheta^2 \left[\begin{array}{cc} 1 \ 0 \\ 1 \ 1 \\
\end{array}\right](u,v) \ } 
{\vartheta^2 \left[\begin{array}{cc} 0 \ 0 \\ 1 \ 1 \\
\end{array}\right](u,v) \ }=k_0 k_1 k_2 \ x_1 x_2  ,
\label{e4-30}\\
&&2) \quad \frac{\vartheta^2 \left[\begin{array}{cc} 1 \ 0 \\ 0 \ 1 \\
\end{array}\right](u,v) \ } 
{\vartheta^2 \left[\begin{array}{cc} 0 \ 0 \\ 1 \ 1 \\
\end{array}\right](u,v) \ }=-\frac{k_0 k_1 k_2}{k'_0 k'_1 k'_2} \ (1-x_1) (1-x_2) , 
\label{e4-31}\\
&&3) \quad \frac{\vartheta^2 \left[\begin{array}{cc} 0 \ 1 \\ 0 \ 1 \\
\end{array}\right](u,v) \ } 
{\vartheta^2 \left[\begin{array}{cc} 0 \ 0 \\ 1 \ 1 \\
\end{array}\right](u,v) \ }=-\frac{k_1 k_2}{k'_0 k_{01} k_{02}} \ (1-k_0^2 x_1) (1-k_0^2 x_2),
\label{e4-32}\\
&&4) \quad \frac{\vartheta^2 \left[\begin{array}{cc} 0 \ 1 \\ 0 \ 0 \\
\end{array}\right](u,v) \ } 
{\vartheta^2 \left[\begin{array}{cc} 0 \ 0 \\ 1 \ 1 \\
\end{array}\right](u,v) \ }=\frac{k_0 k_2}{k'_1 k_{01} k_{12}} \ (1-k_1^2 x_1) (1-k_1^2 x_2),
\label{e4-33}\\
&&5) \quad \frac{\vartheta^2 \left[\begin{array}{cc} 0 \ 0 \\ 0 \ 0 \\
\end{array}\right](u,v) \ } 
{\vartheta^2 \left[\begin{array}{cc} 0 \ 0 \\ 1 \ 1 \\
\end{array}\right](u,v) \ }=\frac{k_0 k_1}{k'_2 k_{02} k_{12}} \ (1-k_2^2 x_1) (1-k_2^2 x_2) .
\label{e4-34}
\end{eqnarray}\\
The parameterization of other ratio of theta function is given in Appendix D, and we have 
\begin{eqnarray}
&&6) \quad \frac{\vartheta^2 \left[\begin{array}{cc} 0 \ 0 \\  0 \ 1 \\
\end{array}\right](u,v) }
{\vartheta^2 \left[\begin{array}{cc} 0 \ 0 \\  1 \ 1 \\
\end{array}\right](u,v)}
=-\frac{F_{01}(x_1) F_{01}(x_2) }{k'_0 k'_1 k'_2 (x_2-x_1)^2}
\left\{ \frac{\sqrt{f_5(x_1)}}{F_{01}(x_1)} - \frac{\sqrt{f_5(x_2)}}{F_{01}(x_2)}
\right\}^2  , 
\label{e4-35}\\
&&7) \quad \frac{\vartheta^2 \left[\begin{array}{cc} 0 \ 1 \\  1 \ 1 \\
\end{array}\right](u,v) }
{\vartheta^2 \left[\begin{array}{cc} 0 \ 0 \\  1 \ 1 \\
\end{array}\right](u,v)}
=\frac{k_1 k_2 F_{34}(x_1) F_{34}(x_2) }
{k'_1 k'_2 k_{01} k_{02} (x_2-x_1)^2}
\left\{ \frac{\sqrt{f_5(x_1)}}{F_{34}(x_1)} 
- \frac{\sqrt{f_5(x_2)}}{F_{34}(x_2)} \right\}^2  ,
\label{e4-36}\\
&&8) \quad \frac{\vartheta^2 \left[\begin{array}{cc} 0 \ 1 \\  1 \ 0 \\
\end{array}\right](u,v) }
{\vartheta^2 \left[\begin{array}{cc} 0 \ 0 \\  1 \ 1 \\
\end{array}\right](u,v)}
=-\frac{k_0 k_2 F_{24}(x_1) F_{24}(x_2) }
{k'_0 k'_2 k_{01} k_{12} (x_2-x_1)^2}
\left\{ \frac{\sqrt{f_5(x_1)}}{F_{24}(x_1)} 
- \frac{\sqrt{f_5(x_2)}}{F_{24}(x_2)} \right\}^2  , 
\label{e4-37}\\
&&9) \quad \frac{\vartheta^2 \left[\begin{array}{cc} 0 \ 0 \\  1 \ 0 \\
\end{array}\right](u,v) }
{\vartheta^2 \left[\begin{array}{cc} 0 \ 0 \\  1 \ 1 \\
\end{array}\right](u,v)}
=-\frac{k_0 k_1 F_{23}(x_1) F_{23}(x_2) }
{k'_0 k'_1 k_{02} k_{12} (x_2-x_1)^2}
\left\{ \frac{\sqrt{f_5(x_1)}}{F_{23}(x_1)} 
- \frac{\sqrt{f_5(x_2)}}{F_{23}(x_2)} \right\}^2   ,  
\label{e4-38}\\
&&10) \quad \frac{\vartheta^2 \left[\begin{array}{cc} 1 \ 1 \\  1 \ 1 \\
\end{array}\right](u,v) }
{\vartheta^2 \left[\begin{array}{cc} 0 \ 0 \\  1 \ 1 \\
\end{array}\right](u,v)}
=\frac{k_0 F_{12}(x_1) F_{12}(x_2) }
{k'_1 k'_2 k_{01} k_{02} (x_2-x_1)^2}
\left\{ \frac{\sqrt{f_5(x_1)}}{F_{12}(x_1)} 
- \frac{\sqrt{f_5(x_2)}}{F_{12}(x_2)} \right\}^2  ,  
\label{e4-39}\\
&&11) \quad \frac{\vartheta^2 \left[\begin{array}{cc} 1 \ 1 \\  1 \ 0 \\
\end{array}\right](u,v) }
{\vartheta^2 \left[\begin{array}{cc} 0 \ 0 \\  1 \ 1 \\
\end{array}\right](u,v)}
=-\frac{k_1 F_{13}(x_1) F_{13}(x_2) }
{k'_0 k'_2 k_{01} k_{12} (x_2-x_1)^2}
\left\{ \frac{\sqrt{f_5(x_1)}}{F_{13}(x_1)} 
- \frac{\sqrt{f_5(x_2)}}{F_{13}(x_2)} \right\}^2   ,  
\label{e4-40}\\
&&12) \quad \frac{\vartheta^2 \left[\begin{array}{cc} 1 \ 0 \\  1 \ 0 \\
\end{array}\right](u,v) }
{\vartheta^2 \left[\begin{array}{cc} 0 \ 0 \\  1 \ 1 \\
\end{array}\right](u,v)}
=-\frac{k_2 F_{14}(x_1) F_{14}(x_2) }
{k'_0 k'_1 k_{02} k_{12} (x_2-x_1)^2}
\left\{ \frac{\sqrt{f_5(x_1)}}{F_{14}(x_1)} 
- \frac{\sqrt{f_5(x_2)}}{F_{14}(x_2)} \right\}^2  ,  
\label{e4-41}\\
&&13) \quad \frac{\vartheta^2 \left[\begin{array}{cc} 1 \ 1 \\  0 \ 1 \\
\end{array}\right](u,v) }
{\vartheta^2 \left[\begin{array}{cc} 0 \ 0 \\  1 \ 1 \\
\end{array}\right](u,v)}
=-\frac{k_0 F_{02}(x_1) F_{02}(x_2) }
{k'_0 k_{01} k_{02}  (x_2-x_1)^2}
\left\{ \frac{\sqrt{f_5(x_1)}}{F_{02}(x_1)} 
- \frac{\sqrt{f_5(x_2)}}{F_{02}(x_2)} \right\}^2  ,  
\label{e4-42}\\
&&14) \quad \frac{\vartheta^2 \left[\begin{array}{cc} 1 \ 1 \\  0 \ 0 \\
\end{array}\right](u,v) }
{\vartheta^2 \left[\begin{array}{cc} 0 \ 0 \\  1 \ 1 \\
\end{array}\right](u,v)}
=\frac{k_1 F_{03}(x_1) F_{03}(x_2) }
{k'_1 k_{01} k_{12}  (x_2-x_1)^2}
\left\{ \frac{\sqrt{f_5(x_1)}}{F_{03}(x_1)} 
- \frac{\sqrt{f_5(x_2)}}{F_{03}(x_2)} \right\}^2  ,   
\label{e4-43}\\
&&15) \quad \frac{\vartheta^2 \left[\begin{array}{cc} 1 \ 0 \\  0 \ 0 \\
\end{array}\right](u,v) }
{\vartheta^2 \left[\begin{array}{cc} 0 \ 0 \\  1 \ 1 \\
\end{array}\right](u,v)}
=\frac{k_2 F_{04}(x_1) F_{04}(x_2) }
{k'_2 k_{02} k_{12}  (x_2-x_1)^2}
\left\{ \frac{\sqrt{f_5(x_1)}}{F_{04}(x_1)} 
- \frac{\sqrt{f_5(x_2)}}{F_{04}(x_2)} \right\}^2  ,
\label{e4-44}
\end{eqnarray}
where we use the following functions, 
$F_{01}(x)=x(1-x)$, $F_{02}(x)=x(1-k_0^2 x)$, $F_{03}(x)=x(1-k_1^2 x)$, 
$F_{04}(x)=x(1-k_2^2 x)$, $F_{12}(x)=(1-x)(1-k_0^2 x)$, $F_{13}(x)=(1-x)(1-k_1^2 x)$,
$F_{14}(x)=(1-x)(1-k_2^2 x)$, $F_{23}(x)=(1-k_0^2 x)(1-k_1^2 x)$, 
$F_{24}(x)=(1-k_0^2 x)(1-k_2^2 x)$, $F_{34}(x)=(1-k_1^2 x)(1-k_2^2 x)$.

\setcounter{equation}{0}
\section{Differential equation by using the addition formula of the theta function}

As we parameterize the ratio of all theta functions with the symmetric 
function of $x_1$ and $x_2$, we can connect $dx_1$ and $dx_2$ with $du$ and $dv$.
Here we will show that we have the following differential 
equation\\
\begin{eqnarray}
&& du =\frac{(P+Q x_1) \ dx_1}{\sqrt{f_5(x_1)}}
+\frac{(P+Q x_2) \ dx_2}{\sqrt{f_5(x_2)}}, 
\label{e5-1}\\      
&&dv =\frac{(R+S x_1) \ dx_1}{\sqrt{f_5(x_1)}}
+\frac{(R+S x_2) \ dx_2}{\sqrt{f_5(x_2)}}.     
\label{e5-2}
\end{eqnarray}\\
To show the above differential equation, we start from 
\begin{eqnarray}
&&\vartheta \left[\begin{array}{cc} 1 \ 0 \\  1 \ 1 \\
\end{array}\right](u,v) /
\vartheta \left[\begin{array}{cc} 0 \ 0 \\  1 \ 1 \\
\end{array}\right](u,v)=-\sqrt{k_0 k_1 k_2} \sqrt{x_1 x_2} , 
\label{e5-3}\\
&&\vartheta \left[\begin{array}{cc} 1 \ 0 \\  0 \ 1 \\
\end{array}\right](u,v) /
\vartheta \left[\begin{array}{cc} 0 \ 0 \\  1 \ 1 \\
\end{array}\right](u,v)=\sqrt{-1} \sqrt{\frac{k_0 k_1 k_2}{k'_0 k'_1 k'_2}}
 \sqrt{(1-x_1)(1- x_2)} , 
\label{e5-4}
\end{eqnarray}
and calculate $\partial_u \sqrt{x_1 x_2}$, $\partial_u \sqrt{(1-x_1)(1- x_2)}$, 
$\partial_v \sqrt{x_1 x_2}$, $\partial_v \sqrt{(1-x_1)(1- x_2)}$, that is, 
we calculate \\
\begin{eqnarray}
&&\frac{\partial}{\partial u}\left(
\vartheta \left[\begin{array}{cc} 1 \ 0 \\  1 \ 1 \\
\end{array}\right](u,v) /
\vartheta \left[\begin{array}{cc} 0 \ 0 \\  1 \ 1 \\
\end{array}\right](u,v)\right), \ 
\frac{\partial}{\partial u}\left(
\vartheta \left[\begin{array}{cc} 1 \ 0 \\  0 \ 1 \\
\end{array}\right](u,v) /
\vartheta \left[\begin{array}{cc} 0 \ 0 \\  1 \ 1 \\
\end{array}\right](u,v)\right), 
\label{e5-5}\\
&&\frac{\partial}{\partial v}\left(
\vartheta \left[\begin{array}{cc} 1 \ 0 \\  1 \ 1 \\
\end{array}\right](u,v) /
\vartheta \left[\begin{array}{cc} 0 \ 0 \\  1 \ 1 \\
\end{array}\right](u,v) \right), \ 
\frac{\partial}{\partial v}\left(
\vartheta \left[\begin{array}{cc} 1 \ 0 \\  0 \ 1 \\
\end{array}\right](u,v) /
\vartheta \left[\begin{array}{cc} 0 \ 0 \\  1 \ 1 \\
\end{array}\right](u,v)\right) , 
\label{e5-6}
\end{eqnarray}
by using the addition theorem of theta function.

\subsection{Differential equation}
Using Eq.(\ref{E-6}) in Appendix E, we have\\
\begin{eqnarray}
&& \frac{\partial}{\partial u} \sqrt{x_1 x_2}
=\frac{(-1)}{\sqrt{k_0 k_1 k_2}} 
\frac{\partial}{\partial u} 
\left(
\vartheta \left[\begin{array}{cc} 1 \ 0 \\  1 \ 1 \\
\end{array}\right](u,v) /
\vartheta \left[\begin{array}{cc} 0 \ 0 \\  1 \ 1 \\
\end{array}\right](u,v) \right)
\nonumber\\
&&=\frac{(-1)}{\sqrt{k_0 k_1 k_2}} 
\left\{
\frac{ 
\vartheta \left[\begin{array}{cc} 0 \ 0 \\  1 \ 0 \\
\end{array}\right](0,0) \ 
\partial_u \vartheta \left[\begin{array}{cc} 1 \ 0 \\  1 \ 0 \\
\end{array}\right](u,0)\Big|_0 }
{\vartheta \left[\begin{array}{cc} 1 \ 0 \\  0 \ 1 \\
\end{array}\right](0,0) \ 
\vartheta \left[\begin{array}{cc} 0 \ 0 \\  0 \ 1 \\
\end{array}\right](0,0) \ }
\frac{ 
\vartheta \left[\begin{array}{cc} 1 \ 0 \\  0 \ 0 \\
\end{array}\right](u,v) \ }
{\vartheta \left[\begin{array}{cc} 0 \ 0 \\  1 \ 1 \\
\end{array}\right](u,v) \ }
\frac{
\vartheta \left[\begin{array}{cc} 0 \ 0 \\  0 \ 0 \\
\end{array}\right](u,v)}
{\vartheta \left[\begin{array}{cc} 0 \ 0 \\  1 \ 1 \\
\end{array}\right](u,v) \ } \right.
\nonumber\\
&&\left. -\frac{ 
\vartheta \left[\begin{array}{cc} 0 \ 1 \\  1 \ 0 \\
\end{array}\right](0,0) \ 
\partial_u \vartheta \left[\begin{array}{cc} 1 \ 1 \\  1 \ 0 \\
\end{array}\right](u,0)\Big|_0 }
{\vartheta \left[\begin{array}{cc} 1 \ 0 \\  0 \ 1 \\
\end{array}\right](0,0) \ 
\vartheta \left[\begin{array}{cc} 0 \ 0 \\  0 \ 1 \\
\end{array}\right](0,0) \ }
\frac{ 
\vartheta \left[\begin{array}{cc} 1 \ 1 \\  0 \ 0 \\
\end{array}\right](u,v) \ }
{\vartheta \left[\begin{array}{cc} 0 \ 0 \\  1 \ 1 \\
\end{array}\right](u,v) \ }
\frac{
\vartheta \left[\begin{array}{cc} 0 \ 1 \\  0 \ 0 \\
\end{array}\right](u,v)}
{\vartheta \left[\begin{array}{cc} 0 \ 0 \\  1 \ 1 \\
\end{array}\right](u,v) \ } \right\} .
\nonumber\\
&&=-\frac{a}{2}\left(
\sqrt{\frac{x_2}{x_1}} \frac{(1-k_2^2x_2) \sqrt{f_5(x_1)}}{(x_2-x_1)}
-\sqrt{\frac{x_1}{x_2}}\frac{(1-k_2^2x_1) \sqrt{f_5(x_2)}}{(x_2-x_1)}\right)
\nonumber\\
&&+\frac{b}{2}\left(
\sqrt{\frac{x_2}{x_1}} \frac{(1-k_1^2 x_2) \sqrt{f_5(x_1)}}{(x_2-x_1)}
-\sqrt{\frac{x_1}{x_2}}\frac{(1-k_1^2 x_1) \sqrt{f_5(x_2)}}{(x_2-x_1)}\right)
\nonumber\\
&&=\frac{1}{2} \left(
\sqrt{\frac{x_2}{x_1}}\frac{(A+Bx_2) \sqrt{f_5(x_1)}}{(x_2-x_1)}
-\sqrt{\frac{x_1}{x_2}}\frac{(A+Bx_1) \sqrt{f_5(x_2)}}{(x_2-x_1)}\right) , 
\label{e5-7}
\end{eqnarray}
where we use Eq.(\ref{e4-34}), Eq.(\ref{e4-44}) and 
Eq.(\ref{e4-33}), Eq.(\ref{e4-43}).
The constants $a$, $b$, $A$, $B$ are given by\\
\begin{eqnarray}
&& a=\frac{2 }{k'_2 k_{02}k_{12}} 
\frac{\vartheta \left[\begin{array}{cc} 0 \ 0 \\  1 \ 0 \\
\end{array}\right](0,0) \ 
\partial_u \vartheta \left[\begin{array}{cc} 1 \ 0 \\  1 \ 0 \\
\end{array}\right](u,0)\Big|_{0}}
{\vartheta \left[\begin{array}{cc} 1 \ 0 \\  0 \ 1 \\
\end{array}\right](0,0) \ 
\vartheta \left[\begin{array}{cc} 0 \ 0 \\  0 \ 1 \\
\end{array}\right](0,0)} , 
\label{e5-8}\\
&&b=\frac{2 }{k'_1 k_{01} k_{12}} 
\frac{\vartheta \left[\begin{array}{cc} 0 \ 1 \\  1 \ 0 \\
\end{array}\right](0,0) \ 
\partial_u \vartheta \left[\begin{array}{cc} 1 \ 1 \\  1 \ 0 \\
\end{array}\right](u,0)\Big|_{0} }
{\vartheta \left[\begin{array}{cc} 1 \ 0 \\  0 \ 1 \\
\end{array}\right](0,0) \ 
\vartheta \left[\begin{array}{cc} 0 \ 0 \\  0 \ 1 \\
\end{array}\right](0,0) } ,
\label{e5-9}
\end{eqnarray}
and $A=-a+b$, $B=(ak_2^2-b k_1^2)$.\\
Similarly, using Eq.(\ref{E-10}), we have\\
\begin{eqnarray}
&& \frac{\partial}{\partial u} \sqrt{(1-x_1)(1- x_2)}
=\sqrt{-1} \sqrt{\frac{k'_0 k'_1 k'_2}{k_0 k_1 k_2}} \ 
\frac{\partial}{\partial u} 
\left(
\vartheta \left[\begin{array}{cc} 1 \ 0 \\  0 \ 1 \\
\end{array}\right](u,v) /
\vartheta \left[\begin{array}{cc} 0 \ 0 \\  1 \ 1 \\
\end{array}\right](u,v) \right)
\nonumber\\
&&=\sqrt{-1} \sqrt{\frac{k'_0 k'_1 k'_2}{k_0 k_1 k_2}} \
\left\{
-\frac{ 
\vartheta \left[\begin{array}{cc} 0 \ 0 \\  0 \ 0 \\
\end{array}\right](0,0) \ 
\partial_u \vartheta \left[\begin{array}{cc} 1 \ 0 \\  1 \ 0 \\
\end{array}\right](u,0)\Big|_0 }
{\vartheta \left[\begin{array}{cc} 1 \ 0 \\  0 \ 1 \\
\end{array}\right](0,0) \ 
\vartheta \left[\begin{array}{cc} 0 \ 0 \\  1 \ 1 \\
\end{array}\right](0,0) \ }
\frac{ 
\vartheta \left[\begin{array}{cc} 0 \ 0 \\  0 \ 0 \\
\end{array}\right](u,v) \ }
{\vartheta \left[\begin{array}{cc} 0 \ 0 \\  1 \ 1 \\
\end{array}\right](u,v) \ }
\frac{
\vartheta \left[\begin{array}{cc} 1 \ 0 \\  1 \ 0 \\
\end{array}\right](u,v)}
{\vartheta \left[\begin{array}{cc} 0 \ 0 \\  1 \ 1 \\
\end{array}\right](u,v) \ } \right.
\nonumber\\
&&\left. +\frac{ 
\vartheta \left[\begin{array}{cc} 0 \ 1 \\  0 \ 0 \\
\end{array}\right](0,0) \ 
\partial_u \vartheta \left[\begin{array}{cc} 1 \ 1 \\  1 \ 0 \\
\end{array}\right](u,0)\Big|_0 }
{\vartheta \left[\begin{array}{cc} 1 \ 0 \\  0 \ 1 \\
\end{array}\right](0,0) \ 
\vartheta \left[\begin{array}{cc} 0 \ 0 \\  1 \ 1 \\
\end{array}\right](0,0) \ }
\frac{ 
\vartheta \left[\begin{array}{cc} 0 \ 1 \\  0 \ 0 \\
\end{array}\right](u,v) \ }
{\vartheta \left[\begin{array}{cc} 0 \ 0 \\  1 \ 1 \\
\end{array}\right](u,v) \ }
\frac{
\vartheta \left[\begin{array}{cc} 1 \ 1 \\  1 \ 0 \\
\end{array}\right](u,v)}
{\vartheta \left[\begin{array}{cc} 0 \ 0 \\  1 \ 1 \\
\end{array}\right](u,v) \ } \right\}  .
\nonumber\\
&&=\frac{\tilde{a}}{2}\left(
\sqrt{\frac{1-x_2}{1-x_1}} \frac{(1-k_2^2x_2) \sqrt{f_5(x_1)}}{(x_2-x_1)}
-\sqrt{\frac{1-x_1}{1-x_2}}\frac{(1-k_2^2x_1) \sqrt{f_5(x_2)}}{(x_2-x_1)}\right)
\nonumber\\
&&-\frac{\tilde{b}}{2}\left(
\sqrt{\frac{1-x_2}{1-x_1}} \frac{(1-k_1^2 x_2) \sqrt{f_5(x_1)}}{(x_2-x_1)}
-\sqrt{\frac{x_1}{x_2}}\frac{(1-k_1^2 x_1) \sqrt{f_5(x_2)}}{(x_2-x_1)}\right)
\nonumber\\
&&=-\frac{1}{2} \left(
\sqrt{\frac{1-x_2}{1-x_1}}\frac{(\tilde{A}+\tilde{B}x_2) \sqrt{f_5(x_1)}}{(x_2-x_1)}
-\sqrt{\frac{1-x_1}{1-x_2}}\frac{(\tilde{A}+\tilde{B}x_1) \sqrt{f_5(x_2)}}{(x_2-x_1)}\right) , 
\label{e5-10}
\end{eqnarray}
where we use Eq.(\ref{e4-34}), Eq.(\ref{e4-41}) and Eq.(\ref{e4-33}), Eq.(\ref{e4-40}).
The constants $\tilde{a}$, $\tilde{b}$, $\tilde{A}$, $\tilde{B}$, are given by \
\begin{eqnarray}
&& \tilde{a}=\frac{2 }{k_{02}k_{12}} 
\frac{\vartheta \left[\begin{array}{cc} 0 \ 0 \\  0 \ 0 \\
\end{array}\right](0,0) \ 
\partial_u \vartheta \left[\begin{array}{cc} 1 \ 0 \\  1 \ 0 \\
\end{array}\right](u,0)\Big|_{0} }
{\vartheta \left[\begin{array}{cc} 1 \ 0 \\  0 \ 1 \\
\end{array}\right](0,0) \ 
\vartheta \left[\begin{array}{cc} 0 \ 0 \\  1 \ 1 \\
\end{array}\right](0,0) } , 
\label{e5-11}\\
&&\tilde{b}=\frac{2}{k_{01}k_{12}} 
\frac{\vartheta \left[\begin{array}{cc} 0 \ 1 \\  0 \ 0 \\
\end{array}\right](0,0) \ 
\partial_u \vartheta \left[\begin{array}{cc} 1 \ 1 \\  1 \ 0 \\
\end{array}\right](u,0)\Big|_{0} }
{\vartheta \left[\begin{array}{cc} 1 \ 0 \\  0 \ 1 \\
\end{array}\right](0,0) \ 
\vartheta \left[\begin{array}{cc} 0 \ 0 \\  1 \ 1 \\
\end{array}\right](0,0)} .
\label{e5-12}
\end{eqnarray}
and $\tilde{A}=-\tilde{a}+\tilde{b}$ and $\tilde{B}=(\tilde{a} k_2^2- \tilde{b} k_1^2)$.\\
We can check that $a=\tilde{a}$ by using Eq.(\ref{e4-21}) and 
we can check $b=\tilde{b}$ by using Eq.(\ref{e4-20}). Then we have
$\tilde{A}=A$ and $\tilde{B}=B$.\\
Then we have 
\begin{eqnarray}
&& \frac{\partial}{\partial u} \sqrt{x_1 x_2}
=\frac{1}{2}\sqrt{\frac{x_2}{x_1}}\frac{\partial x_1}{\partial u}
 +\frac{1}{2}\sqrt{\frac{x_1}{x_2}}\frac{\partial x_2}{\partial u}
\nonumber\\
&&=\frac{1}{2} \left(
\sqrt{\frac{x_2}{x_1}}\frac{(A+Bx_2) \sqrt{f_5(x_1)}}{(x_2-x_1)}
-\sqrt{\frac{x_1}{x_2}}\frac{(A+Bx_1) \sqrt{f_5(x_2)}}{(x_2-x_1)}\right)  , 
\label{e5-13}\\
&& \frac{\partial}{\partial u} \sqrt{(1-x_1)(1- x_2)}
=-\frac{1}{2}\sqrt{\frac{1-x_2}{1-x_1}}\frac{\partial x_1}{\partial u}
 -\frac{1}{2}\sqrt{\frac{1-x_1}{1-x_2}}\frac{\partial x_2}{\partial u}
\nonumber\\
&&=-\frac{1}{2} \left(
\sqrt{\frac{1-x_2}{1-x_1}}\frac{(A+Bx_2) \sqrt{f_5(x_1)}}{(x_2-x_1)}
-\sqrt{\frac{1-x_1}{1-x_2}}\frac{(A+Bx_1) \sqrt{f_5(x_2)}}{(x_2-x_1)}\right) , 
\label{e5-14}
\end{eqnarray}
which gives\\
\begin{eqnarray}
\frac{\partial x_1}{\partial u}=\frac{(A+Bx_2) \sqrt{f_5(x_1)}}{(x_2-x_1)} , \ 
\frac{\partial x_2}{\partial u}=-\frac{(A+Bx_1) \sqrt{f_5(x_2)}}{(x_2-x_1)} . 
\label{e5-15}
\end{eqnarray}\\
Similarly we have
\begin{eqnarray}
\frac{\partial x_1}{\partial v}=\frac{(C+Dx_2) \sqrt{f_5(x_1)}}{(x_2-x_1)} , \ 
\frac{\partial x_2}{\partial v}=-\frac{(C+Dx_1) \sqrt{f_5(x_2)}}{(x_2-x_1)}
\label{e5-16}
\end{eqnarray}
where
\begin{eqnarray}
&& c=\frac{2 }{k'_2 k_{02}k_{12}} 
\frac{\vartheta \left[\begin{array}{cc} 0 \ 0 \\  1 \ 0 \\
\end{array}\right](0,0) \ 
\partial_v \vartheta \left[\begin{array}{cc} 1 \ 0 \\  1 \ 0 \\
\end{array}\right](0,v)\Big|_{0}}
{\vartheta \left[\begin{array}{cc} 1 \ 0 \\  0 \ 1 \\
\end{array}\right](0,0) \ 
\vartheta \left[\begin{array}{cc} 0 \ 0 \\  0 \ 1 \\
\end{array}\right](0,0)} , 
\label{e5-17}\\
&&d=\frac{2 }{k'_1 k_{01} k_{12}} 
\frac{\vartheta \left[\begin{array}{cc} 0 \ 1 \\  1 \ 0 \\
\end{array}\right](0,0) \ 
\partial_v \vartheta \left[\begin{array}{cc} 1 \ 1 \\  1 \ 0 \\
\end{array}\right](0,v)\Big|_{0} }
{\vartheta \left[\begin{array}{cc} 1 \ 0 \\  0 \ 1 \\
\end{array}\right](0,0) \ 
\vartheta \left[\begin{array}{cc} 0 \ 0 \\  0 \ 1 \\
\end{array}\right](0,0) }
\label{e5-18}
\end{eqnarray}
and $C=-c+d$ and $D=(ck_2^2-d k_1^2)$\\
This gives the differential equation
\begin{eqnarray}
\left(\begin{array}{c} dx_1 \\  dx_2 \\
\end{array}\right)=
\left(\begin{array}{cc} \alpha & \beta \\  \gamma & \delta \\
\end{array}\right)
\left(\begin{array}{c} du \\  dv \\
\end{array}\right)
\label{e5-19}
\end{eqnarray}
where
\begin{eqnarray}
&&\alpha=\frac{(A+Bx_2) \sqrt{f_5(x_1)}}{(x_2-x_1)} ,\quad
\beta=\frac{(C+Dx_2) \sqrt{f_5(x_1)}}{(x_2-x_1)}
\label{e5-20}\\
&&\gamma=-\frac{(A+Bx_1) \sqrt{f_5(x_2)}}{(x_2-x_1)}, \quad
\delta=-\frac{(C+Dx_1) \sqrt{f_5(x_2)}}{(x_2-x_1)} 
\label{e5-21}
\end{eqnarray}\\
Then we have 
\begin{eqnarray}
\left(\begin{array}{c} du \\  dv \\
\end{array}\right)=\frac{1}{\alpha \delta-\beta \gamma} 
\left(\begin{array}{cc} \delta & -\beta \\  -\gamma & \alpha \\
\end{array}\right)
\left(\begin{array}{c} dx_1 \\  dx_2 \\
\end{array}\right)
\label{e5-22}
\end{eqnarray}
where we have
$\alpha \delta-\beta \gamma=(AD-BC)\sqrt{f_5(x_1)}\sqrt{f_5(x_2)}/(x_2-x_1)$. \\
Then we finally find \\
\begin{eqnarray}
&&du=-\frac{1}{(AD-BC)}\left(
\frac{(C+Dx_1) }{\sqrt{f_5(x_1)}} dx_1
+\frac{(C+Dx_2)} {\sqrt{f_5(x_2)}} dx_2 \right)
\label{e5-23}\\
&&dv=\frac{1}{(AD-BC)}\left(
\frac{(A+Bx_1) }{\sqrt{f_5(x_1)}} dx_1
+\frac{(A+Bx_2)} {\sqrt{f_5(x_2)}} dx_2 \right)
\label{e5-24}
\end{eqnarray}\\
Then we can write in the form\\
\begin{eqnarray}
du=\frac{(P+Qx_1)dx_1}{\sqrt{f_5(x_1)}}
+\frac{(P+Qx_2) dx_2} {\sqrt{f_5(x_2)}} 
\label{e5-25}\\
dv=\frac{(R+Sx_1)dx_1}{\sqrt{f_5(x_1)}} 
+\frac{(R+Sx_2)dx_2} {\sqrt{f_5(x_2)}}  
\label{e5-26}
\end{eqnarray}
where we denote $P=-C/(AD-BC)$, $Q=-D/(AD-BC)$, $R=A/(AD-BC)$, $S=B/(AD-BC)$.\\
In this way, using the expression of the symmetric combination of $x_1$ and $x_2$ 
as the ratio of the hyperelliptic theta function with two variable, we derive 
the above differential equation  Eq.(\ref{e5-25}) and Eq.(\ref{e5-25}).
Then we have solved the Jacobi's inversion problem for genus two case, that is, 
$i)$ the single-valued function is the symmetric combination of $x_1$ and $x_2$, \ 
$ii)$ symmetric combination of $x_1$ and $x_2$ is expressed as the ratio of the 
hyperelliptic theta function with two variables $u$ and $v$.\\

\setcounter{equation}{0}
\section{Differential equation for $\tau_{12}=0$ case}

Here we consider the case $\tau_{12}=0$, which gives \\
\begin{eqnarray}
&&\vartheta\left[\begin{array}{cc} a \ c \\ b \ d \\
\end{array}\right](u,v) 
=\vartheta \left[\begin{array}{c} a \\ b \\ \end{array}\right](u) \ 
\vartheta \left[\begin{array}{c} c \\ d \\ \end{array}\right](v)  . 
\label{e6-1}
\end{eqnarray} \\
Using the expression 
\begin{eqnarray}
&&\sn(u,k)=
-\vartheta\left[\begin{array}{c} 0\\ 0\\ \end{array}\right](0)
\vartheta\left[\begin{array}{c} 1\\ 1\\ \end{array}\right](z)
/\vartheta\left[\begin{array}{c} 1\\ 0\\ \end{array}\right](0)
\vartheta\left[\begin{array}{c} 0\\ 1\\ \end{array}\right](z) ,
\label{e6-2}\\
&&\cn(u,k)=
\vartheta\left[\begin{array}{c} 0\\ 1\\ \end{array}\right](0)
\vartheta\left[\begin{array}{c} 1\\ 0\\ \end{array}\right](z)
/\vartheta\left[\begin{array}{c} 1\\ 0\\ \end{array}\right](0)
\vartheta\left[\begin{array}{c} 0\\ 1\\ \end{array}\right](z) ,
\label{e6-3}\\
&&\dn(u,k)=
\vartheta\left[\begin{array}{c} 0\\ 1\\ \end{array}\right](0)
\vartheta\left[\begin{array}{c} 0\\ 0\\ \end{array}\right](z)
/\vartheta\left[\begin{array}{c} 0\\ 0\\ \end{array}\right](0)
\vartheta\left[\begin{array}{c} 0\\ 1\\ \end{array}\right](z) ,
\label{e6-4}\\
&&k=
\vartheta^2\left[\begin{array}{c} 1\\ 0\\ \end{array}\right](0)/
\vartheta^2\left[\begin{array}{c} 0\\ 0\\ \end{array}\right](0) ,
\quad k'=\vartheta^2\left[\begin{array}{c} 0\\ 1\\ \end{array}\right](0)/
\vartheta^2\left[\begin{array}{c} 0\\ 0\\ \end{array}\right](0) ,
\label{e6-5}
\end{eqnarray}\\
From the relation 
\begin{eqnarray}
&&\cn^2(u,k)+\sn^2(u,k)=1, \quad \dn^2(u,k)+k^2 \sn^2(u,k)=1,
\label{e6-6}
\end{eqnarray}
we have three identities
\begin{eqnarray}
&&\vartheta^2\left[\begin{array}{c} 0\\ 0\\ \end{array}\right](0)
\vartheta^2\left[\begin{array}{c} 0\\ 0\\ \end{array}\right](z)
=\vartheta^2\left[\begin{array}{c} 0\\ 1\\ \end{array}\right](0)
\vartheta^2\left[\begin{array}{c} 0\\ 1\\ \end{array}\right](z)
+\vartheta^2\left[\begin{array}{c} 1\\ 0\\ \end{array}\right](0)
\vartheta^2\left[\begin{array}{c} 1\\ 0\\ \end{array}\right](z),
\label{e6-7}\\
&&\vartheta^2\left[\begin{array}{c} 0\\ 0\\ \end{array}\right](0)
\vartheta^2\left[\begin{array}{c} 1\\ 1\\ \end{array}\right](z)
=\vartheta^2\left[\begin{array}{c} 1\\ 0\\ \end{array}\right](0)
\vartheta^2\left[\begin{array}{c} 0\\ 1\\ \end{array}\right](z)
-\vartheta^2\left[\begin{array}{c} 0\\ 1\\ \end{array}\right](0)
\vartheta^2\left[\begin{array}{c} 1\\ 0\\ \end{array}\right](z),
\label{e6-8}\\
&&\vartheta^2\left[\begin{array}{c} 0\\ 0\\ \end{array}\right](0)
\vartheta^2\left[\begin{array}{c} 0\\ 1\\ \end{array}\right](z)
=\vartheta^2\left[\begin{array}{c} 0\\ 1\\ \end{array}\right](0)
\vartheta^2\left[\begin{array}{c} 0\\ 0\\ \end{array}\right](z)
+\vartheta^2\left[\begin{array}{c} 1\\ 0\\ \end{array}\right](0)
\vartheta^2\left[\begin{array}{c} 1\\ 1\\ \end{array}\right](z), 
\label{e6-9}\\
&&{\rm with}\nonumber\\
&&\vartheta^4\left[\begin{array}{c} 0\\ 0\\ \end{array}\right](0)
=\vartheta^4\left[\begin{array}{c} 0\\ 1\\ \end{array}\right](0)
+\vartheta^4\left[\begin{array}{c} 1\\ 0\\ \end{array}\right](0) .
\label{e6-10}
\end{eqnarray}
The fundamental theta identity of Eq.(\ref{B1-1}), Eq.(\ref{B1-2})
and Eq.(\ref{B1-3}) reduced to Eq.(\ref{e6-7}).\\

We define 
\begin{eqnarray}
&&x(=-\sn(u,k))=\vartheta\left[\begin{array}{c} 0\\ 0\\ \end{array}\right](0)
\vartheta\left[\begin{array}{c} 1\\ 1\\ \end{array}\right](u)
/\vartheta\left[\begin{array}{c} 1\\ 0\\ \end{array}\right](0)
\vartheta\left[\begin{array}{c} 0\\ 1\\ \end{array}\right](u) ,
\label{e6-11}\\
&&y(=-\sn(v,k))=\vartheta\left[\begin{array}{c} 0\\ 0\\ \end{array}\right](0)
\vartheta\left[\begin{array}{c} 1\\ 1\\ \end{array}\right](v)
/\vartheta\left[\begin{array}{c} 1\\ 0\\ \end{array}\right](0)
\vartheta\left[\begin{array}{c} 0\\ 1\\ \end{array}\right](v) ,
\label{e6-12}
\end{eqnarray}
and we express $x_1$ and $x_2$ in Eq.(\ref{e4-3}), Eq.(\ref{e4-4}) and 
Eq.(\ref{e4-5}) with $x$ and $y$.
Straightforward calculation, we have 
\begin{eqnarray}
&&\hskip -10mm k_0^2 x_1 x_2=x^2 ,\quad \frac{k_0^2}{1-k_0^2} (1-x_1)(1-x_2)=-(1-x^2) , \quad 
\frac{(1-k_0^2 x_1)(1-k_0^2 x_2)}{1-k_0^2} =0.
\label{e6-13}
\end{eqnarray}
Then we have two solutions
\begin{eqnarray}
&&i) x_1=x^2 , \quad x_2=\frac{1}{k_0^2} \quad ii) x_1=\frac{1}{k_0^2}, \quad x_2=x^2, 
\label{e6-14}
\end{eqnarray}
by cancel out the $v$-dependence.
Similarly, from  Eq.(\ref{e4-8}), Eq.(\ref{e4-9}) and Eq.(\ref{e4-10}), we have two solutions
\begin{eqnarray}
&& i) x_1=x^2 , \quad x_2=\frac{1}{k_1^2} \quad ii) x_1=\frac{1}{k_1^2}, \quad x_2=x^2, \quad  
\label{e6-15}
\end{eqnarray}
and from  Eq.(\ref{e4-13}), Eq.(\ref{e4-14}) and Eq.(\ref{e4-15}), we have two solutions
\begin{eqnarray}
&&i) x_1=x^2 , \ x_2=\frac{1}{k_2^2}, \quad ii) x_1=\frac{1}{k_2^2}, \ x_2=x^2, \quad  
\label{e6-16}
\end{eqnarray}
Combining these relations, we have
\begin{eqnarray}
&&i) x_1=x^2 , \ x_2=\frac{1}{k_0^2}=\frac{1}{k_1^2}=\frac{1}{k_2^2}=({\rm const.}),
\label{e6-17}\\ 
&&ii) x_1=\frac{1}{k_0^2}=\frac{1}{k_1^2}=\frac{1}{k_2^2}=({\rm const.}),\ x_2=x^2.
\label{e6-18}
\end{eqnarray}
We consider the case $i)$. Then, by using $x_1=x^2$ and 
$x_2=1/k_0^2=1/k_1^2=1/k_2^2$=const., the differential 
equations Eq.(\ref{e5-25}) and  Eq.(\ref{e5-26}) reduces
to only one differential equation
\begin{eqnarray}
&&du=\frac{(P+Qx_1) dx_1 }{\sqrt{\tilde{f}_5(x_1)}}
= \frac{(1-k_0^2 x_1)  dx_1}{2\sqrt{\tilde{f}_5(x_1)}}
\nonumber\\
&&=\frac{(1-k_0^2 x^2)  x dx}{\sqrt{x^2(1-x^2)(1-k_0^2 x^2)^3}}
=\frac{dx}{\sqrt{(1-x^2)(1-k_0^2 x^2)}} .
\label{e6-19}
\end{eqnarray}
where we use  $\tilde{f}_5(x)=x_1(1-x_1)(1-k_0^2 x_1)^3$. Then we have the 
differential equation of the elliptic function.

\setcounter{equation}{0}
\section{Summary and discussion}
\noindent
The Ising model, which is parameterized by the elliptic function, satisfies 
the two dimensional integrability equation with difference property.
We form the addition formula of the elliptic function into the integrable $SU(2)$ 
group structure in the previous paper.\\
We expect that the addition formula of the Abelian function with any genus 
will form some integrable Lie group structure.
For that purpose, we first study the Jacobi's inversion problem 
of the hyperelliptic integral with genus two case in this paper, to have the hint
to understand the structure of the addition formula of the Abelian 
function with genus two.
Even for the genus one or the genus two case, the essence of the addition 
formula of the elliptic function comes from the Riemann theta identity, so that 
we expect some integrable $SU(2)$ structure for the addition formula
of the hyperelliptic theta function with two varibles. \\


\newpage


\newpage

\appendix
\setcounter{equation}{0}
\section{\large \bf Property of theta function}

\subsection{Even odd property}
\begin{eqnarray}
\vartheta\left[\begin{array}{cc} a \ c \\ b \ d \\
\end{array}\right](-u,-v)
=(-1)^{ab+cd}\vartheta\left[\begin{array}{cc} a \ c \\ b \ d \\
\end{array}\right](u,v)
\label{A1-1}
\end{eqnarray}
Then the following 6 theta functions
$\vartheta\left[\begin{array}{cc} 1 \ 0 \\ 1 \ 0 \\ \end{array}\right](u,v)$, 
$\vartheta\left[\begin{array}{cc} 1 \ 1 \\ 1 \ 0 \\ \end{array}\right](u,v)$, 
$\vartheta\left[\begin{array}{cc} 1 \ 0 \\ 1 \ 1 \\ \end{array}\right](u,v)$,\\ 
$\vartheta\left[\begin{array}{cc} 0 \ 1 \\ 0 \ 1 \\ \end{array}\right](u,v)$, 
$\vartheta\left[\begin{array}{cc} 1 \ 1 \\ 0 \ 1 \\ \end{array}\right](u,v)$, 
$\vartheta\left[\begin{array}{cc} 0 \ 1 \\ 1 \ 1 \\ \end{array}\right](u,v)$ 
are odd function, so that we have \\
$\vartheta\left[\begin{array}{cc} 1 \ 0 \\ 1 \ 0 \\ \end{array}\right](0,0)=0$, 
$\vartheta\left[\begin{array}{cc} 1 \ 1 \\ 1 \ 0 \\ \end{array}\right](0,0)=0$, 
$\vartheta\left[\begin{array}{cc} 1 \ 0 \\ 1 \ 1 \\ \end{array}\right](0,0)=0$, 
$\vartheta\left[\begin{array}{cc} 0 \ 1 \\ 0 \ 1 \\ \end{array}\right](0,0)=0$, \\
$\vartheta\left[\begin{array}{cc} 1 \ 1 \\ 0 \ 1 \\ \end{array}\right](u,v)=0$, 
$\vartheta\left[\begin{array}{cc} 0 \ 1 \\ 1 \ 1 \\ \end{array}\right](u,v)=0$, 
and the rest 10 theta functions are even function.\\

\subsection{Half periodic property I}
\begin{eqnarray}
&&\vartheta\left[\begin{array}{cc} 0 & c \\  0  & d \\
\end{array}\right](u+\frac{1}{2},v)
=\vartheta\left[\begin{array}{cc} 0 & c \\  1 & d \\
\end{array}\right](u,v) , 
\label{A1-2}\\
&&\vartheta\left[\begin{array}{cc} 1 & c \\ 0 & d \\
\end{array}\right](u+\frac{1}{2},v)
=\vartheta\left[\begin{array}{cc} 1 & c \\ 1 & d \\
\end{array}\right](u,v) , 
\label{A1-3}\\
&&\vartheta\left[\begin{array}{cc} 0 & c \\ 1 & d \\
\end{array}\right](u+\frac{1}{2},v)
=\vartheta\left[\begin{array}{cc} 0 & c \\ 0 & d \\
\end{array}\right](u,v) , 
\label{A1-4}\\
&&\vartheta\left[\begin{array}{cc} 1 & c \\ 1 & d \\
\end{array}\right](u+\frac{1}{2},v)
=(-1)\vartheta\left[\begin{array}{cc} 1 & c \\ 0 & d \\
\end{array}\right](u,v) . 
\label{A1-5}
\end{eqnarray}

\subsection{Half periodic property II}
\begin{eqnarray}
&&\vartheta\left[\begin{array}{cc} 0 & c \\  0  & d \\
\end{array}\right](u+\frac{\tau_1}{2},v+\frac{\tau_{12}}{2})
=e^{-i \pi \tau_1 /4 -i \pi u} \ 
\vartheta\left[\begin{array}{cc} 1 & c \\  0 & d \\
\end{array}\right](u,v) , 
\label{A1-6}\\
&&\vartheta\left[\begin{array}{cc} 1 & c \\  0  & d \\
\end{array}\right](u+\frac{\tau_1}{2},v+\frac{\tau_{12}}{2})
=e^{-i \pi \tau_1 /4 -i \pi u} \ 
\vartheta\left[\begin{array}{cc} 0 & c \\  0 & d \\
\end{array}\right](u,v) , 
\label{A1-7}\\
&&\vartheta\left[\begin{array}{cc} 0 & c \\  1  & d \\
\end{array}\right](u+\frac{\tau_1}{2},v+\frac{\tau_{12}}{2})
=(-\sqrt{-1}) e^{-i \pi \tau_1 /4 -i \pi u} \ 
\vartheta\left[\begin{array}{cc} 1 & c \\  1 & d \\
\end{array}\right](u,v) , 
\label{A1-8}\\
&&\vartheta\left[\begin{array}{cc} 1 & c \\  1  & d \\
\end{array}\right](u+\frac{\tau_1}{2},v+\frac{\tau_{12}}{2})
=(-1\sqrt{-1}) e^{-i \pi \tau_1 /4 -i \pi u} \ 
\vartheta\left[\begin{array}{cc} 0 & c \\  1 & d \\
\end{array}\right](u,v) .
\label{A1-9}
\end{eqnarray}

\subsection{Half periodic property III}

\begin{eqnarray}
&&\vartheta\left[\begin{array}{cc} 0 & c \\  0  & d \\
\end{array}\right](u+\frac{\tau_1}{2}+\frac{1}{2},v+\frac{\tau_{12}}{2})
=(-\sqrt{-1} )e^{-i \pi \tau_1 /4 -i \pi u} \ 
\vartheta\left[\begin{array}{cc} 1 & c \\  1 & d \\
\end{array}\right](u,v) , 
\label{A1-10}\\
&&\vartheta\left[\begin{array}{cc} 1 & c \\  0  & d \\
\end{array}\right](u+\frac{\tau_1}{2}+\frac{1}{2},v+\frac{\tau_{12}}{2})
=(-\sqrt{-1}) e^{-i \pi \tau_1 /4 -i \pi u} \ 
\vartheta\left[\begin{array}{cc} 0 & c \\  1 & d \\
\end{array}\right](u,v) , 
\label{A1-11}\\
&&\vartheta\left[\begin{array}{cc} 0 & c \\  1  & d \\
\end{array}\right](u+\frac{\tau_1}{2}+\frac{1}{2},v+\frac{\tau_{12}}{2})
=e^{-i \pi \tau_1 /4 -i \pi u} \ 
\vartheta\left[\begin{array}{cc} 1 & c \\  0 & d \\
\end{array}\right](u,v) , 
\label{A1-12}\\
&&\vartheta\left[\begin{array}{cc} 1 & c \\  1  & d \\
\end{array}\right](u+\frac{\tau_1}{2}+\frac{1}{2},v+\frac{\tau_{12}}{2})
=(-1)e^{-i \pi \tau_1 /4 -i \pi u} \ 
\vartheta\left[\begin{array}{cc} 0 & c \\  0 & d \\
\end{array}\right](u,v) .
\label{A1-13}
\end{eqnarray}

\subsection{Periodic property I}
\begin{eqnarray}
&&\vartheta\left[\begin{array}{cc} 0 & c \\ b & d \\
\end{array}\right](u+1,v)
=\vartheta\left[\begin{array}{cc} 0 & c \\ b & d \\
\end{array}\right](u,v)  , 
\label{A1-14}\\
&&\vartheta\left[\begin{array}{cc} 1 & c \\ b & d \\
\end{array}\right](u+1,v)
=(-1)\vartheta\left[\begin{array}{cc} 1 & c \\  b & d \\
\end{array}\right](u,v) . 
\label{A1-15}
\end{eqnarray}

\subsection{Periodic property II}
\begin{eqnarray}
&&\vartheta\left[\begin{array}{cc} a& c \\  0 & d \\
\end{array}\right](u+\tau_1,v+\tau_{12})
=e^{-i \pi \tau_1 -2 i \pi u} \ 
\vartheta\left[\begin{array}{cc} a & c \\  0 & d \\
\end{array}\right](u,v) , 
\label{A1-16}\\
&&\vartheta\left[\begin{array}{cc} a & c \\  1 & d \\
\end{array}\right](u+\tau_1,v+\tau_{12})
=(-1)e^{-i \pi \tau_1 -2 i \pi u} \ 
\vartheta\left[\begin{array}{cc} a & c \\  1 & d \\
\end{array}\right](u,v)  . 
\label{A1-17}
\end{eqnarray}

\vskip 10mm
\setcounter{equation}{0}
\section{\large \bf Various theta identity}

\begin{eqnarray}
&&\vartheta^2 \left[\begin{array}{cc} 0 \ 0 \\  0 \ 0 \\
\end{array}\right](0,0) \ 
\vartheta^2 \left[\begin{array}{cc} 0 \ 0 \\  0 \ 0 \\
\end{array}\right](u,v)
=\vartheta^2 \left[\begin{array}{cc} 0 \ 0 \\  1 \ 0 \\
\end{array}\right](0,0) \ 
\vartheta^2 \left[\begin{array}{cc} 0 \ 0 \\  1 \ 0 \\
\end{array}\right](u,v)  \nonumber\\
&&+\vartheta^2 \left[\begin{array}{cc} 1 \ 0 \\  0 \ 0 \\
\end{array}\right](0,0) \ 
\vartheta^2 \left[\begin{array}{cc} 1 \ 0 \\  0 \ 0 \\
\end{array}\right](u,v)
+\vartheta^2 \left[\begin{array}{cc} 1 \ 1 \\  1 \ 1 \\
\end{array}\right](0,0) \ 
\vartheta^2 \left[\begin{array}{cc} 1 \ 1 \\  1 \ 1 \\
\end{array}\right](u,v) ,
\label{B1-1}\\
\nonumber\\
&&\vartheta^2 \left[\begin{array}{cc} 0 \ 1 \\  0 \ 0 \\
\end{array}\right](0,0) \ 
\vartheta^2 \left[\begin{array}{cc} 0 \ 0 \\  0 \ 0 \\
\end{array}\right](u,v)
=\vartheta^2 \left[\begin{array}{cc} 0 \ 1 \\  1 \ 0 \\
\end{array}\right](0,0) \ 
\vartheta^2 \left[\begin{array}{cc} 0 \ 0 \\  1 \ 0 \\
\end{array}\right](u,v)  \nonumber\\
&&+\vartheta^2 \left[\begin{array}{cc} 1 \ 1 \\  0 \ 0 \\
\end{array}\right](0,0) \ 
\vartheta^2 \left[\begin{array}{cc} 1 \ 0 \\  0 \ 0 \\
\end{array}\right](u,v)
+\vartheta^2 \left[\begin{array}{cc} 1 \ 1 \\  1 \ 1 \\
\end{array}\right](0,0) \ 
\vartheta^2 \left[\begin{array}{cc} 1 \ 0 \\  1 \ 1 \\
\end{array}\right](u,v) , 
\label{B1-2}\\
\nonumber\\
&&\vartheta^2 \left[\begin{array}{cc} 0 \ 0 \\  0 \ 1 \\
\end{array}\right](0,0) \ 
\vartheta^2 \left[\begin{array}{cc} 0 \ 0 \\  0 \ 0 \\
\end{array}\right](u,v)
=\vartheta^2 \left[\begin{array}{cc} 0 \ 0 \\  1 \ 1 \\
\end{array}\right](0,0) \ 
\vartheta^2 \left[\begin{array}{cc} 0 \ 0 \\  1 \ 0 \\
\end{array}\right](u,v)  \nonumber\\
&&+\vartheta^2 \left[\begin{array}{cc} 1 \ 0 \\  0 \ 1 \\
\end{array}\right](0,0) \ 
\vartheta^2 \left[\begin{array}{cc} 1 \ 0 \\  0 \ 0 \\
\end{array}\right](u,v)
-\vartheta^2 \left[\begin{array}{cc} 1 \ 1 \\  1 \ 1 \\
\end{array}\right](0,0) \ 
\vartheta^2 \left[\begin{array}{cc} 1 \ 1 \\  1 \ 0 \\
\end{array}\right](u,v) .  
\label{B1-3}
\end{eqnarray}

\subsection{Derivation of Eq.(\ref{B1-1}) and Eq.(\ref{B1-2})}
Substituting $u_1=u_2=u$, $u_3=u_4=0$, $v_1=v_2=v$, $v_3=v_4=0$, which gives 
$\tilde{u}_1=\tilde{u}_2=u$, $\tilde{u}_3=\tilde{u}_4=0$, $\tilde{v}_1=v_2=v$, 
$\tilde{v}_3=\tilde{v}_4=0$ in Eq.(\ref{e3-26}), 
that is,  $2 \tilde{M}=M+M'+M''+M'''$, we have 

\begin{eqnarray}
&&\left\{
\vartheta^2 \left[\begin{array}{cc} 0 \ 0 \\  0 \ 0 \\
\end{array}\right](0,0) \ 
\vartheta^2 \left[\begin{array}{cc} 0 \ 0 \\  0 \ 0 \\
\end{array}\right](u,v) 
-\left(
\vartheta^2 \left[\begin{array}{cc} 0 \ 0 \\  1 \ 0 \\
\end{array}\right](0,0) \ 
\vartheta^2 \left[\begin{array}{cc} 0 \ 0 \\  1 \ 0 \\
\end{array}\right](u,v)   \right. \right.
\nonumber\\ 
&&\left. \left. +\vartheta^2 \left[\begin{array}{cc} 1 \ 0 \\  0 \ 0 \\
\end{array}\right](0,0) \ 
\vartheta^2 \left[\begin{array}{cc} 1 \ 0 \\  0 \ 0 \\
\end{array}\right](u,v)
+\vartheta^2 \left[\begin{array}{cc} 1 \ 1 \\  1 \ 1 \\
\end{array}\right](0,0) \ 
\vartheta^2 \left[\begin{array}{cc} 1 \ 1 \\  1 \ 1 \\
\end{array}\right](u,v)
\right)\right\}
\nonumber\\
&&+\left\{
\vartheta^2 \left[\begin{array}{cc} 0 \ 1 \\  0 \ 0 \\
\end{array}\right](0,0) \ 
\vartheta^2 \left[\begin{array}{cc} 0 \ 1 \\  0 \ 0 \\
\end{array}\right](u,v) 
-\left(
\vartheta^2 \left[\begin{array}{cc} 0 \ 1 \\  1 \ 0 \\
\end{array}\right](0,0) \ 
\vartheta^2 \left[\begin{array}{cc} 0 \ 1 \\  1 \ 0 \\
\end{array}\right](u,v)   \right. \right.
\nonumber\\ 
&&\left. \left. +\vartheta^2 \left[\begin{array}{cc} 1 \ 1 \\  0 \ 0 \\
\end{array}\right](0,0) \ 
\vartheta^2 \left[\begin{array}{cc} 1 \ 1 \\  0 \ 0 \\
\end{array}\right](u,v)
-\vartheta^2 \left[\begin{array}{cc} 1 \ 1 \\  1 \ 1 \\
\end{array}\right](0,0) \ 
\vartheta^2 \left[\begin{array}{cc} 1 \ 1 \\  1 \ 1 \\
\end{array}\right](u,v)
\right)\right\}=0   . 
\label{B1-4}
\end{eqnarray}\\
Next, substituting $u_1=u_2=u$, $u_3=u_4=0$, $v_1=v+1$, $v_2=v$, $v_3=v_4=0$, which gives 
$\tilde{u}_1=\tilde{u}_2=u$, $\tilde{u}_3=\tilde{u}_4=0$, $\tilde{v}_1=\tilde{v}_2=v+1/2$, 
$\tilde{v}_3=\tilde{v}_4=1/2$ in Eq.(\ref{e3-28}), 
that is,  $2 \tilde{M}''=M-M'+M''-M'''$, we have 

\begin{eqnarray}
&&\left\{
\vartheta^2 \left[\begin{array}{cc} 0 \ 0 \\  0 \ 0 \\
\end{array}\right](0,0) \ 
\vartheta^2 \left[\begin{array}{cc} 0 \ 0 \\  0 \ 0 \\
\end{array}\right](u,v) 
-\left(
\vartheta^2 \left[\begin{array}{cc} 0 \ 0 \\  1 \ 0 \\
\end{array}\right](0,0) \ 
\vartheta^2 \left[\begin{array}{cc} 0 \ 0 \\  1 \ 0 \\
\end{array}\right](u,v)   \right. \right.
\nonumber\\ 
&&\left. \left. +\vartheta^2 \left[\begin{array}{cc} 1 \ 0 \\  0 \ 0 \\
\end{array}\right](0,0) \ 
\vartheta^2 \left[\begin{array}{cc} 1 \ 0 \\  0 \ 0 \\
\end{array}\right](u,v)
+\vartheta^2 \left[\begin{array}{cc} 1 \ 1 \\  1 \ 1 \\
\end{array}\right](0,0) \ 
\vartheta^2 \left[\begin{array}{cc} 1 \ 1 \\  1 \ 1 \\
\end{array}\right](u,v)
\right)\right\}
\nonumber\\
&&-\left\{
\vartheta^2 \left[\begin{array}{cc} 0 \ 1 \\  0 \ 0 \\
\end{array}\right](0,0) \ 
\vartheta^2 \left[\begin{array}{cc} 0 \ 1 \\  0 \ 0 \\
\end{array}\right](u,v) 
-\left(
\vartheta^2 \left[\begin{array}{cc} 0 \ 1 \\  1 \ 0 \\
\end{array}\right](0,0) \ 
\vartheta^2 \left[\begin{array}{cc} 0 \ 1 \\  1 \ 0 \\
\end{array}\right](u,v)   \right. \right.
\nonumber\\ 
&&\left. \left. +\vartheta^2 \left[\begin{array}{cc} 1 \ 1 \\  0 \ 0 \\
\end{array}\right](0,0) \ 
\vartheta^2 \left[\begin{array}{cc} 1 \ 1 \\  0 \ 0 \\
\end{array}\right](u,v)
-\vartheta^2 \left[\begin{array}{cc} 1 \ 1 \\  1 \ 1 \\
\end{array}\right](0,0) \ 
\vartheta^2 \left[\begin{array}{cc} 1 \ 1 \\  1 \ 1 \\
\end{array}\right](u,v)
\right)\right\} =0  . 
\label{B1-5}
\end{eqnarray}\\
Combining Eq.(\ref{B1-4}) and Eq.(\ref{B1-5}), we have 
 Eq.(\ref{B1-1}) and Eq.(\ref{B1-2}).

\subsection{Derivation of Eq.(\ref{B1-3})}
Substituting $u_1=u+1$, $u_2=u$, $u_3=1/2$, $u_4=1/2$, $v_1=v_2=v+1/2$, $v_3=1/2$, 
$v_4=-1/2$, which gives 
$\tilde{u}_1=u+1$, $\tilde{u}_2=u$, $\tilde{u}_3=1/2$, $\tilde{u}_4=1/2$, 
$\tilde{v}_1=\tilde{v}_2=v+1/2$, $\tilde{v}_3=1/2$, $\tilde{v}_4=-1/2$
 in Eq.(\ref{e3-26}), 
that is,  $2 \tilde{M}=M+M'+M''+M'''$, we have \\
\begin{eqnarray}
&&\vartheta^2 \left[\begin{array}{cc} 0 \ 0 \\  0 \ 1 \\
\end{array}\right](0,0) \ 
\vartheta^2 \left[\begin{array}{cc} 0 \ 0 \\  0 \ 1 \\
\end{array}\right](u,v)
=\vartheta^2 \left[\begin{array}{cc} 0 \ 0 \\  1 \ 1 \\
\end{array}\right](0,0) \ 
\vartheta^2 \left[\begin{array}{cc} 0 \ 0 \\  1 \ 1 \\
\end{array}\right](u,v)
\nonumber\\
&&+\vartheta^2 \left[\begin{array}{cc} 1 \ 0 \\  0 \ 1 \\
\end{array}\right](0,0) \ 
\vartheta^2 \left[\begin{array}{cc} 1 \ 0 \\  0 \ 1 \\
\end{array}\right](u,v)
-\vartheta^2 \left[\begin{array}{cc} 1 \ 1 \\  1 \ 1 \\
\end{array}\right](0,0) \ 
\vartheta^2 \left[\begin{array}{cc} 1 \ 1 \\  1 \ 1 \\
\end{array}\right](u,v)  .
\label{B1-6}
\end{eqnarray}\\
We further replace $u\rightarrow u$, $v \rightarrow v+1/2$, we have 
\begin{eqnarray}
&&\vartheta^2 \left[\begin{array}{cc} 0 \ 0 \\  0 \ 1 \\
\end{array}\right](0,0) \ 
\vartheta^2 \left[\begin{array}{cc} 0 \ 0 \\  0 \ 0 \\
\end{array}\right](u,v)
=\vartheta^2 \left[\begin{array}{cc} 0 \ 0 \\  1 \ 1 \\
\end{array}\right](0,0) \ 
\vartheta^2 \left[\begin{array}{cc} 0 \ 0 \\  1 \ 0 \\
\end{array}\right](u,v)
\nonumber\\
&&+\vartheta^2 \left[\begin{array}{cc} 1 \ 0 \\  0 \ 1 \\
\end{array}\right](0,0) \ 
\vartheta^2 \left[\begin{array}{cc} 1 \ 0 \\  0 \ 0 \\
\end{array}\right](u,v)
-\vartheta^2 \left[\begin{array}{cc} 1 \ 1 \\  1 \ 1 \\
\end{array}\right](0,0) \ 
\vartheta^2 \left[\begin{array}{cc} 1 \ 1 \\  1 \ 0 \\
\end{array}\right](u,v)  . 
\label{B1-7}
\end{eqnarray}
which gives Eq.(\ref{B1-3}).

\vskip 10mm
\setcounter{equation}{0}
\section{\large \bf Parameterization of constant}

In order that these parameterization is consistent, we must have \\
\begin{eqnarray}
&&\frac{\vartheta^2 \left[\begin{array}{cc} 1 \ 0 \\ 0 \ 1 \\
\end{array}\right](0,0)} 
{\vartheta^2 \left[\begin{array}{cc} 0 \ 0 \\ 0 \ 1 \\
\end{array}\right](0,0)}k_0^2
=\frac{\vartheta^2 \left[\begin{array}{cc} 1 \ 0 \\ 0 \ 0 \\
\end{array}\right](0,0)} 
{\vartheta^2 \left[\begin{array}{cc} 0 \ 0 \\ 0 \ 0 \\
\end{array}\right](0,0)}k_1^2
=\frac{\vartheta^2 \left[\begin{array}{cc} 1 \ 1 \\ 0 \ 0 \\
\end{array}\right](0,0)} 
{\vartheta^2 \left[\begin{array}{cc} 0 \ 1 \\ 0 \ 0 \\
\end{array}\right](0,0)}k_2^2  , 
\label{C1-1}\\
&&\frac{\vartheta^2 \left[\begin{array}{cc} 1 \ 0 \\ 0 \ 1 \\
\end{array}\right](0,0)} 
{\vartheta^2 \left[\begin{array}{cc} 0 \ 0 \\ 1 \ 1 \\
\end{array}\right](0,0)}\frac{k_0^2}{1-k_0^2}
=\frac{\vartheta^2 \left[\begin{array}{cc} 1 \ 0 \\ 0 \ 0 \\
\end{array}\right](0,0)} 
{\vartheta^2 \left[\begin{array}{cc} 0 \ 0 \\ 1 \ 0 \\
\end{array}\right](0,0)}\frac{k_1^2}{1-k_1^2}
=\frac{\vartheta^2 \left[\begin{array}{cc} 1 \ 1 \\ 0 \ 0 \\
\end{array}\right](0,0)} 
{\vartheta^2 \left[\begin{array}{cc} 0 \ 1 \\ 1 \ 0 \\
\end{array}\right](0,0)}\frac{k_2^2}{1-k_2^2}  , 
\label{C1-2}
\end{eqnarray}
which gives \\
\begin{eqnarray}
&&\hspace{-10mm}k_0^2=\frac{\vartheta^2 \left[\begin{array}{cc} 1 \ 0 \\ 0 \ 0 \\
\end{array}\right](0,0) \ 
\vartheta^2 \left[\begin{array}{cc} 1 \ 1 \\ 0 \ 0 \\
\end{array}\right](0,0)} 
{\vartheta^2 \left[\begin{array}{cc} 0 \ 0 \\ 0 \ 0 \\
\end{array}\right](0,0) \ 
\vartheta^2 \left[\begin{array}{cc} 0 \ 1 \\ 0 \ 0 \\
\end{array}\right](u,v)} , 
\frac{k_0^2}{1-k_0^2}=\frac{\vartheta^2 \left[\begin{array}{cc} 1 \ 0 \\ 0 \ 0 \\
\end{array}\right](0,0) \ 
\vartheta^2 \left[\begin{array}{cc} 1 \ 1 \\ 0 \ 0 \\
\end{array}\right](0,0)} 
{\vartheta^2 \left[\begin{array}{cc} 0 \ 0 \\ 1 \ 0 \\
\end{array}\right](0,0) \ 
\vartheta^2 \left[\begin{array}{cc} 0 \ 1 \\ 1 \ 0 \\
\end{array}\right](0,0)} , 
\label{C1-3}\\
&&\hspace{-10mm}k_1^2=\frac{\vartheta^2 \left[\begin{array}{cc} 1 \ 0 \\ 0 \ 1 \\
\end{array}\right](0,0) \ 
\vartheta^2 \left[\begin{array}{cc} 1 \ 1 \\ 0 \ 0 \\
\end{array}\right](0,0)} 
{\vartheta^2 \left[\begin{array}{cc} 0 \ 0 \\ 0 \ 1 \\
\end{array}\right](0,0) \ 
\vartheta^2 \left[\begin{array}{cc} 0 \ 1 \\ 0 \ 0 \\
\end{array}\right](0,0)} , 
\frac{k_1^2}{1-k_1^2}=\frac{\vartheta^2 \left[\begin{array}{cc} 1 \ 0 \\ 0 \ 1 \\
\end{array}\right](0,0) \ 
\vartheta^2 \left[\begin{array}{cc} 1 \ 1 \\ 0 \ 0 \\
\end{array}\right](0,0)} 
{\vartheta^2 \left[\begin{array}{cc} 0 \ 0 \\ 1 \ 1 \\
\end{array}\right](0,0) \ 
\vartheta^2 \left[\begin{array}{cc} 0 \ 1 \\ 1 \ 0 \\
\end{array}\right](0,0)}  , 
\label{C1-4}\\
&&\hspace{-10mm} k_2^2=\frac{\vartheta^2 \left[\begin{array}{cc} 1 \ 0 \\ 0 \ 1 \\
\end{array}\right](0,0) \ 
\vartheta^2 \left[\begin{array}{cc} 1 \ 0 \\ 0 \ 0 \\
\end{array}\right](0,0)} 
{\vartheta^2 \left[\begin{array}{cc} 0 \ 0 \\ 0 \ 1 \\
\end{array}\right](0,0) \ 
\vartheta^2 \left[\begin{array}{cc} 0 \ 0 \\ 0 \ 0 \\
\end{array}\right](0,0)}  , 
\frac{k_2^2}{1-k_2^2}=\frac{\vartheta^2 \left[\begin{array}{cc} 1 \ 0 \\ 0 \ 1 \\
\end{array}\right](0,0) \ 
\vartheta^2 \left[\begin{array}{cc} 1 \ 0 \\ 0 \ 0 \\
\end{array}\right](0,0)} 
{\vartheta^2 \left[\begin{array}{cc} 0 \ 0 \\ 1 \ 1 \\
\end{array}\right](0,0) \ 
\vartheta^2 \left[\begin{array}{cc} 0 \ 0 \\ 1 \ 0 \\
\end{array}\right](0,0)}  . 
\label{C1-5}
\end{eqnarray}\\
Consistency of $k_0^2$ in Eq.(\ref{C1-3}), consistency of  $k_1^2$ in Eq.(\ref{C1-4}),
consistency of $k_2^2$ in Eq.(\ref{C1-5}), give the following relation\\
\begin{eqnarray}
&&1=\frac{\vartheta^2 \left[\begin{array}{cc} 0 \ 0 \\ 0 \ 0 \\
\end{array}\right](0,0) \ 
\vartheta^2 \left[\begin{array}{cc} 0 \ 1 \\ 0 \ 0 \\
\end{array}\right](0,0)} 
{\vartheta^2 \left[\begin{array}{cc} 1 \ 0 \\ 0 \ 0 \\
\end{array}\right](0,0) \ 
\vartheta^2 \left[\begin{array}{cc} 1 \ 1 \\ 0 \ 0 \\
\end{array}\right](0,0)}
-\frac{\vartheta^2 \left[\begin{array}{cc} 0 \ 0 \\ 1 \ 0 \\
\end{array}\right](0,0) \ 
\vartheta^2 \left[\begin{array}{cc} 0 \ 1 \\ 1 \ 0 \\
\end{array}\right](0,0)} 
{\vartheta^2 \left[\begin{array}{cc} 1 \ 0 \\ 0 \ 0 \\
\end{array}\right](0,0) \ 
\vartheta^2 \left[\begin{array}{cc} 1 \ 1 \\ 0 \ 0 \\
\end{array}\right](0,0)}, 
\label{C1-6}\\
&&1=\frac{\vartheta^2 \left[\begin{array}{cc} 0 \ 0 \\ 0 \ 1 \\
\end{array}\right](0,0) \ 
\vartheta^2 \left[\begin{array}{cc} 0 \ 1 \\ 0 \ 0 \\
\end{array}\right](0,0)} 
{\vartheta^2 \left[\begin{array}{cc} 1 \ 0 \\ 0 \ 1 \\
\end{array}\right](0,0) \ 
\vartheta^2 \left[\begin{array}{cc} 1 \ 1 \\ 0 \ 0 \\
\end{array}\right](0,0)}
-\frac{\vartheta^2 \left[\begin{array}{cc} 0 \ 0 \\ 1 \ 1 \\
\end{array}\right](0,0) \ 
\vartheta^2 \left[\begin{array}{cc} 0 \ 1 \\ 1 \ 0 \\
\end{array}\right](0,0)} 
{\vartheta^2 \left[\begin{array}{cc} 1 \ 0 \\ 0 \ 1 \\
\end{array}\right](0,0) \ 
\vartheta^2 \left[\begin{array}{cc} 1 \ 1 \\ 0 \ 0 \\
\end{array}\right](0,0)}, 
\label{C1-7}\\
&&1=\frac{\vartheta^2 \left[\begin{array}{cc} 0 \ 0 \\ 0 \ 1 \\
\end{array}\right](0,0) \ 
\vartheta^2 \left[\begin{array}{cc} 0 \ 0 \\ 0 \ 0 \\
\end{array}\right](0,0)} 
{\vartheta^2 \left[\begin{array}{cc} 1 \ 0 \\ 0 \ 1 \\
\end{array}\right](0,0) \ 
\vartheta^2 \left[\begin{array}{cc} 1 \ 0 \\ 0 \ 0 \\
\end{array}\right](0,0)}
-\frac{\vartheta^2 \left[\begin{array}{cc} 0 \ 0 \\ 1 \ 1 \\
\end{array}\right](0,0) \ 
\vartheta^2 \left[\begin{array}{cc} 0 \ 0 \\ 1 \ 0 \\
\end{array}\right](0,0)} 
{\vartheta^2 \left[\begin{array}{cc} 1 \ 0 \\ 0 \ 1 \\
\end{array}\right](0,0) \ 
\vartheta^2 \left[\begin{array}{cc} 1 \ 0 \\ 0 \ 0 \\
\end{array}\right](0,0)} . 
\label{C1-8}
\end{eqnarray}\\
Eq.(\ref{C1-6}) is derived from Eq.(\ref{B1-1}) by putting $u=\tau_{12}/2$,
$v= \tau_2/2$. Eq.(\ref{C1-7}) is derived from Eq.(\ref{B1-3}) by 
putting $u=\tau_{12}/2$, $v= \tau_2/2$. Eq.(\ref{C1-8}) is derived from 
Eq.(\ref{B1-3}) by putting $u=0$, $v=0$. \\
As we use the expression ${k'_{0}}^2=1-k_{0}^2$, ${k'_{1}}^2=1-k_{1}^2$, 
${k'_{2}}^2=1-k_{2}^2$, 
$k_{01}^2=k_{0}^2-k_{1}^2$, $k_{02}^2=k_{0}^2-k_{2}^2$, $k_{12}^2=k_{1}^2-k_{2}^2$,
expressed with the ration of theta function with zero argument.\\
From Eq.(\ref{e4-21}),  Eq.(\ref{e4-22}), Eq.(\ref{e4-23}), we have 
\begin{eqnarray}
&&{k'_{0}}^2=1-k_0^2=
\frac{\vartheta^2 \left[\begin{array}{cc} 0 \ 0 \\ 1 \ 0 \\
\end{array}\right](0,0) \ 
\vartheta^2 \left[\begin{array}{cc} 0 \ 1 \\ 1 \ 0 \\
\end{array}\right](0,0)} 
{\vartheta^2 \left[\begin{array}{cc} 0 \ 0 \\ 0 \ 0 \\
\end{array}\right](0,0) \ 
\vartheta^2 \left[\begin{array}{cc} 0 \ 1 \\ 0 \ 0 \\
\end{array}\right](0,0)}  ,
\label{C1-9}\\
&&{k'_{1}}^2=1-k_1^2=
\frac{\vartheta^2 \left[\begin{array}{cc} 0 \ 0 \\ 1 \ 1 \\
\end{array}\right](0,0) \ 
\vartheta^2 \left[\begin{array}{cc} 0 \ 1 \\ 1 \ 0 \\
\end{array}\right](0,0)} 
{\vartheta^2 \left[\begin{array}{cc} 0 \ 0 \\ 0 \ 1 \\
\end{array}\right](0,0) \ 
\vartheta^2 \left[\begin{array}{cc} 0 \ 1 \\ 0 \ 0 \\
\end{array}\right](0,0)}  , 
\label{C1-10}\\
&&{k'_{2}}^2=1-k_2^2=
\frac{\vartheta^2 \left[\begin{array}{cc} 0 \ 0 \\ 1 \ 1 \\
\end{array}\right](0,0) \ 
\vartheta^2 \left[\begin{array}{cc} 0 \ 0 \\ 1 \ 0 \\
\end{array}\right](0,0)} 
{\vartheta^2 \left[\begin{array}{cc} 0 \ 0 \\ 0 \ 1 \\
\end{array}\right](0,0) \ 
\vartheta^2 \left[\begin{array}{cc} 0 \ 0 \\ 0 \ 0 \\
\end{array}\right](0,0)}  , 
\label{C1-11}
\end{eqnarray}
and
$k_{01}^2=k_{0}^2-k_{1}^2$, $k_{02}^2=k_{0}^2-k_{2}^2$, $k_{12}^2=k_{1}^2-k_{2}^2$,
are expressed with the ration of theta function with zero argument.\\
We have 
\begin{eqnarray}
&&k_{01}^2=k_{0}^2-k_{1}^2=
\frac{\vartheta^2 \left[\begin{array}{cc} 1 \ 1 \\ 0 \ 0 \\
\end{array}\right](0,0) }
{\vartheta^2 \left[\begin{array}{cc} 0 \ 1 \\ 0 \ 0 \\
\end{array}\right](0,0)} 
\nonumber\\
&&\times
\frac{\left(\vartheta^2 \left[\begin{array}{cc} 1 \ 0 \\ 0 \ 0 \\
\end{array}\right](0,0) \ 
\vartheta^2 \left[\begin{array}{cc} 0 \ 0 \\ 0 \ 1 \\
\end{array}\right](0,0) 
-\vartheta^2 \left[\begin{array}{cc} 0 \ 0 \\ 0 \ 0 \\
\end{array}\right](0,0) \ 
\vartheta^2 \left[\begin{array}{cc} 1 \ 0 \\ 0 \ 1 \\
\end{array}\right](0,0)\right) }
{\vartheta^2 \left[\begin{array}{cc} 0 \ 0 \\ 0 \ 0 \\
\end{array}\right](0,0) \ 
\vartheta^2 \left[\begin{array}{cc} 0 \ 0 \\ 0 \ 1 \\
\end{array}\right](0,0) }
\nonumber\\
&&=\frac{\vartheta^2 \left[\begin{array}{cc} 1 \ 1 \\ 0 \ 0 \\
\end{array}\right](0,0) \ 
\vartheta^2 \left[\begin{array}{cc} 1 \ 1 \\ 1 \ 1 \\
\end{array}\right](0,0) \ 
\vartheta^2 \left[\begin{array}{cc} 0 \ 1 \\ 1 \ 0 \\
\end{array}\right](0,0) }
{\vartheta^2 \left[\begin{array}{cc} 0 \ 1 \\ 0 \ 0 \\
\end{array}\right](0,0) \ 
\vartheta^2 \left[\begin{array}{cc} 0 \ 0 \\ 0 \ 0 \\
\end{array}\right](0,0) \ 
\vartheta^2 \left[\begin{array}{cc} 0 \ 0 \\ 0 \ 1 \\
\end{array}\right](0,0) }   , 
\label{C1-12}
\end{eqnarray}
by using Eq.(\ref{B1-1}) with putting $u=\tau_{1}/2$, $v=\tau_{12}/2+1/2$.\\
Similarly, we have  
We have 
\begin{eqnarray}
&&k_{02}^2=k_{0}^2-k_{2}^2=
\frac{\vartheta^2 \left[\begin{array}{cc} 1 \ 0 \\ 0 \ 0 \\
\end{array}\right](0,0) }
{\vartheta^2 \left[\begin{array}{cc} 0 \ 0 \\ 0 \ 0 \\
\end{array}\right](0,0)} 
\nonumber\\
&&\times
\frac{\left(\vartheta^2 \left[\begin{array}{cc} 1 \ 1 \\ 0 \ 0 \\
\end{array}\right](0,0) \ 
\vartheta^2 \left[\begin{array}{cc} 0 \ 0 \\ 0 \ 1 \\
\end{array}\right](0,0) 
-\vartheta^2 \left[\begin{array}{cc} 0 \ 1 \\ 0 \ 0 \\
\end{array}\right](0,0) \ 
\vartheta^2 \left[\begin{array}{cc} 1 \ 0 \\ 0 \ 1 \\
\end{array}\right](0,0)\right) }
{\vartheta^2 \left[\begin{array}{cc} 0 \ 1 \\ 0 \ 0 \\
\end{array}\right](0,0) \ 
\vartheta^2 \left[\begin{array}{cc} 0 \ 0 \\ 0 \ 1 \\
\end{array}\right](0,0) }
\nonumber\\
&&=\frac{\vartheta^2 \left[\begin{array}{cc} 1 \ 0 \\ 0 \ 0 \\
\end{array}\right](0,0) \ 
\vartheta^2 \left[\begin{array}{cc} 1 \ 1 \\ 1 \ 1 \\
\end{array}\right](0,0) \ 
\vartheta^2 \left[\begin{array}{cc} 0 \ 0 \\ 1 \ 0 \\
\end{array}\right](0,0) }
{\vartheta^2 \left[\begin{array}{cc} 0 \ 0 \\ 0 \ 0 \\
\end{array}\right](0,0) \ 
\vartheta^2 \left[\begin{array}{cc} 0 \ 1 \\ 0 \ 0 \\
\end{array}\right](0,0) \ 
\vartheta^2 \left[\begin{array}{cc} 0 \ 0 \\ 0 \ 1 \\
\end{array}\right](0,0) } , 
\label{C1-13}
\end{eqnarray}
by using Eq.(\ref{B1-2}) with putting $u=\tau_{1}/2$, $v=\tau_{12}/2+1/2$.\\
Similarly, we have  
\begin{eqnarray}
&&k_{12}^2=k_{1}^2-k_{2}^2=
\frac{\vartheta^2 \left[\begin{array}{cc} 1 \ 0 \\ 0 \ 1 \\
\end{array}\right](0,0) }
{\vartheta^2 \left[\begin{array}{cc} 0 \ 0 \\ 0 \ 1 \\
\end{array}\right](0,0)} 
\nonumber\\
&&\times
\frac{\left(\vartheta^2 \left[\begin{array}{cc} 0 \ 0 \\ 0 \ 0 \\
\end{array}\right](0,0) \ 
\vartheta^2 \left[\begin{array}{cc} 1 \ 1 \\ 0 \ 0 \\
\end{array}\right](0,0) 
-\vartheta^2 \left[\begin{array}{cc} 1 \ 0 \\ 0 \ 0 \\
\end{array}\right](0,0) \ 
\vartheta^2 \left[\begin{array}{cc} 0 \ 1 \\ 0 \ 0 \\
\end{array}\right](0,0)\right) }
{\vartheta^2 \left[\begin{array}{cc} 0 \ 1 \\ 0 \ 0 \\
\end{array}\right](0,0) \ 
\vartheta^2 \left[\begin{array}{cc} 0 \ 0 \\ 0 \ 0 \\
\end{array}\right](0,0) }
\nonumber\\
&&=\frac{\vartheta^2 \left[\begin{array}{cc} 1 \ 0 \\ 0 \ 1 \\
\end{array}\right](0,0) \ 
\vartheta^2 \left[\begin{array}{cc} 1 \ 1 \\ 1 \ 1 \\
\end{array}\right](0,0) \ 
\vartheta^2 \left[\begin{array}{cc} 0 \ 0 \\ 1 \ 1 \\
\end{array}\right](0,0) }
{\vartheta^2 \left[\begin{array}{cc} 0 \ 0 \\ 0 \ 1 \\
\end{array}\right](0,0) \ 
\vartheta^2 \left[\begin{array}{cc} 0 \ 1 \\ 0 \ 0 \\
\end{array}\right](0,0) \ 
\vartheta^2 \left[\begin{array}{cc} 0 \ 0 \\ 0 \ 0 \\
\end{array}\right](0,0) } , 
\label{C1-14}
\end{eqnarray}
by using Eq.(\ref{B1-1}) with putting $u=\tau_{1}/2+\tau_{12}/2$, 
$v=\tau_{2}/2+\tau_{12}/2$.

\vskip 10mm
\setcounter{equation}{0}
\section{\large \bf Parameterization of the ratio of other theta function}

\subsection{Derivation of Eq.(\ref{e4-35})[6-th parameterization]}
In this appendix, we will first show \\
\begin{eqnarray}
&&\frac{\vartheta^2 \left[\begin{array}{cc} 0 \ 0 \\  0 \ 1 \\
\end{array}\right](u,v) }
{\vartheta^2 \left[\begin{array}{cc} 0 \ 0 \\  1 \ 1 \\
\end{array}\right](u,v)}
=-\frac{x_1 x_2 (1-x_1) (1-x_2) }{k'_0 k'_1 k'_2 (x_2-x_1)^2}
\left\{ \frac{\sqrt{f_5(x_1)}}{x_1 (1-x_1)} \mp \frac{\sqrt{f_5(x_2)}}{x_2 (1-x_2)}
\right\}^2   .
\label{D-1}
\end{eqnarray}\\
When we take the square root, in order that the right hand side of 
the above is symmetric for $ x_1 \leftrightarrow x_2$, we  take minus sign 
of $\mp$ in the above, and we have \\
\begin{eqnarray}
&&\frac{\vartheta \left[\begin{array}{cc} 0 \ 0 \\  0 \ 1 \\
\end{array}\right](u,v) }
{\vartheta \left[\begin{array}{cc} 0 \ 0 \\  1 \ 1 \\
\end{array}\right](u,v)}
=\sqrt{-1}{\frac{\sqrt{x_1 x_2 (1-x_1) (1-x_2)} }{\sqrt{k'_0 k'_1 k'_2} (x_2-x_1)}}
\left\{ \frac{\sqrt{f_5(x_1)}}{x_1 (1-x_1)} -\frac{\sqrt{f_5(x_2)}}{x_2 (1-x_2)}
\right\}  .
\label{D-2}
\end{eqnarray}\\
In that purpose, we want to derive the algebraic equation of 
$\vartheta \left[\begin{array}{cc} 0 \ 0 \\  0 \ 1 \\
\end{array}\right](u,v)$ and 
$\vartheta \left[\begin{array}{cc} 0 \ 0 \\  1 \ 1 \\
\end{array}\right](u,v)$. For that purpose, we use the identity\\
\begin{eqnarray}
&&\vartheta \left[\begin{array}{cc} 0 \ 0 \\  0 \ 0 \\
\end{array}\right](0,0) \ 
\vartheta \left[\begin{array}{cc} 0 \ 1 \\  0 \ 0 \\
\end{array}\right](0,0) \ 
\vartheta \left[\begin{array}{cc} 0 \ 1 \\  0 \ 1 \\
\end{array}\right](u,v) \ 
\vartheta \left[\begin{array}{cc} 0 \ 0 \\  0 \ 1 \\
\end{array}\right](u,v) \ 
\nonumber\\
&&-\vartheta \left[\begin{array}{cc} 1 \ 0 \\  0 \ 0 \\
\end{array}\right](0,0) \ 
\vartheta \left[\begin{array}{cc} 1 \ 1 \\  0 \ 0 \\
\end{array}\right](0,0) \ 
\vartheta \left[\begin{array}{cc} 1 \ 0 \\  0 \ 1 \\
\end{array}\right](u,v) \ 
\vartheta \left[\begin{array}{cc} 1 \ 1 \\  0 \ 1 \\
\end{array}\right](u,v) \ 
\nonumber\\
&&-\vartheta \left[\begin{array}{cc} 0 \ 0 \\  1 \ 0 \\
\end{array}\right](0,0) \ 
\vartheta \left[\begin{array}{cc} 0 \ 1 \\  1 \ 0 \\
\end{array}\right](0,0) \ 
\vartheta \left[\begin{array}{cc} 0 \ 0 \\  1 \ 1 \\
\end{array}\right](u,v) \ 
\vartheta \left[\begin{array}{cc} 0 \ 1 \\  1 \ 1 \\
\end{array}\right](u,v)=0  , 
\label{D-3}
\end{eqnarray}
where Eq.(\ref{D-3}) is obtained by replacing 
$u_1 \rightarrow u+\tau_{12}/2$, $u_2 \rightarrow u$, 
$u_3 \rightarrow \tau_{12}/2$, $u_4 \rightarrow 0$, 
$v_1 \rightarrow v+\tau_2/2+1/2$, $v_2 \rightarrow v+1/2$, 
$v_3 \rightarrow \tau_2/2$, $v_4 \rightarrow 0$, which gives
$\tilde{u}_1 \rightarrow u+\tau_{12}/2$, 
$\tilde{u}_2 \rightarrow u$, $\tilde{u}_3 \rightarrow \tau_{12}/2$, 
$\tilde{u}_4 \rightarrow 0$, 
$\tilde{v}_1 \rightarrow v+\tau_2/2+1/2$, $\tilde{v}_2 \rightarrow v+1/2$, 
$\tilde{v}_3 \rightarrow \tau_2/2$, $\tilde{v}_4 \rightarrow 0$ in 
Eq.(\ref{e3-26}), that is, $2\tilde{M}=M+M'+M''+M'''$.\\
In this identity, we use Eq.(\ref{e4-32}) for 
$\vartheta^2 \left[\begin{array}{cc} 0 \ 1 \\  0 \ 1 \\
\end{array}\right](u,v)$, and Eq.(\ref{e4-31}) for 
$\vartheta^2 \left[\begin{array}{cc} 1 \ 0 \\  0 \ 1 \\
\end{array}\right](u,v)$, so that it is necessary to use 
the identity to connect 
$\vartheta^2 \left[\begin{array}{cc} 0 \ 1 \\  1 \ 1 \\
\end{array}\right](u,v)$ and 
$\vartheta^2 \left[\begin{array}{cc} 1 \ 1 \\  0 \ 1 \\
\end{array}\right](u,v)$
with 
$\vartheta^2 \left[\begin{array}{cc} 0 \ 0 \\  0 \ 1 \\
\end{array}\right](u,v)$,
$\vartheta^2 \left[\begin{array}{cc} 0 \ 0 \\  1 \ 1 \\
\end{array}\right](u,v)$,
$\vartheta^2 \left[\begin{array}{cc} 0 \ 1 \\  0 \ 1 \\
\end{array}\right](u,v)$,  
$\vartheta^2 \left[\begin{array}{cc} 1 \ 0 \\  0 \ 1 \\
\end{array}\right](u,v)$.\\
Then we use the following such identity
\begin{eqnarray}
&&\vartheta^2 \left[\begin{array}{cc} 1 \ 1 \\  1 \ 1 \\
\end{array}\right](0,0) \ 
\vartheta^2 \left[\begin{array}{cc} 0 \ 1 \\  1 \ 1 \\
\end{array}\right](u,v) \ 
=\vartheta^2 \left[\begin{array}{cc} 0 \ 0 \\  1 \ 1 \\
\end{array}\right](0,0) \ 
\vartheta^2 \left[\begin{array}{cc} 1 \ 0 \\  1 \ 1 \\
\end{array}\right](u,v)
\nonumber\\
&&+\vartheta^2 \left[\begin{array}{cc} 0 \ 0 \\  0 \ 1 \\
\end{array}\right](0,0) \ 
\vartheta^2 \left[\begin{array}{cc} 1 \ 0 \\  0 \ 1 \\
\end{array}\right](u,v) \ 
-\vartheta^2 \left[\begin{array}{cc} 1 \ 0 \\  0 \ 1 \\
\end{array}\right](0,0) \ 
\vartheta^2 \left[\begin{array}{cc} 0 \ 0 \\  0 \ 1 \\
\end{array}\right](u,v),
\label{D-4}\\
&&\vartheta^2 \left[\begin{array}{cc} 1 \ 1 \\  1 \ 1 \\
\end{array}\right](0,0) \ 
\vartheta^2 \left[\begin{array}{cc} 1 \ 1 \\  0 \ 1 \\
\end{array}\right](u,v) \ 
=\vartheta^2 \left[\begin{array}{cc} 1 \ 0 \\  0 \ 1 \\
\end{array}\right](0,0) \ 
\vartheta^2 \left[\begin{array}{cc} 1 \ 0 \\  1 \ 1 \\
\end{array}\right](u,v)
\nonumber\\
&&-\vartheta^2 \left[\begin{array}{cc} 0 \ 0 \\  0 \ 1 \\
\end{array}\right](0,0) \ 
\vartheta^2 \left[\begin{array}{cc} 0 \ 0 \\  1 \ 1 \\
\end{array}\right](u,v) \ 
+\vartheta^2 \left[\begin{array}{cc} 0 \ 0 \\  1 \ 1 \\
\end{array}\right](0,0) \ 
\vartheta^2 \left[\begin{array}{cc} 0 \ 0 \\  0 \ 1 \\
\end{array}\right](u,v),
\label{D-5}
\end{eqnarray}
where Eq.(\ref{D-4}) is obtained by replacing $u \rightarrow u+\tau_1/2$, 
$v \rightarrow v+\tau_{12}/2+1/2$ in Eq.(\ref{B1-3}), and
Eq.(\ref{D-5}) is obtained by replacing $u \rightarrow u+1/2$, 
$v \rightarrow v+1/2$ in Eq.(\ref{B1-3}).\\
After the straightforward but quite tedious calculation, we have 
\begin{eqnarray}
&&\frac{\vartheta^4 \left[\begin{array}{cc} 0 \ 0 \\  0 \ 1 \\
\end{array}\right](u,v)}
{\vartheta^4 \left[\begin{array}{cc} 0 \ 0 \\  1 \ 1 \\
\end{array}\right](u,v)}
+2\left\{ \frac{F_{01}(x_2)F_{234}(x_1)+F_{01}(x_1)F_{234}(x_2)}
{k'_0 k'_1 k'_2 (x_2-x_1)^2} \right\}
\frac{\vartheta^2 \left[\begin{array}{cc} 0 \ 0 \\  0 \ 1 \\
\end{array}\right](u,v)}
{\vartheta^2 \left[\begin{array}{cc} 0 \ 0 \\  1 \ 1 \\
\end{array}\right](u,v)} \nonumber\\
&&+\left\{ \frac{F_{01}(x_2)F_{234}(x_1)-F_{01}(x_1)F_{234}(x_2))}
{k'_0 k'_1 k'_2 (x_2-x_1)^2} \right\}^2 =0  , 
\label{D-6}
\end{eqnarray}
where we use 
$F_{01}(x)=x(1-x)$ and $F_{234}(x)=(1-k_0^2 x) (1-k_1^2 x) (1-k_2^2 x)=f_5(x)/F_{01}(x)$. \\
From this we have  
\begin{eqnarray}
&&\frac{\vartheta^2 \left[\begin{array}{cc} 0 \ 0 \\  0 \ 1 \\
\end{array}\right](u,v) }
{\vartheta^2 \left[\begin{array}{cc} 0 \ 0 \\  1 \ 1 \\
\end{array}\right](u,v)}
=-\frac{x_1 x_2 (1-x_1) (1-x_2) }{k'_0 k'_1 k'_2 (x_2-x_1)^2}
\left\{ \frac{\sqrt{f_5(x_1)}}{x_1 (1-x_1)} \mp \frac{\sqrt{f_5(x_2)}}{x_2 (1-x_2)}
\right\}^2  , 
\label{D-7}
\end{eqnarray}
which gives Eq.(\ref{e4-35}).

\subsection{Check of Eq.(\ref{e4-36})[7-th parameterization]}
We sketch how to check Eq.(\ref{e4-36}) $\sim$  Eq.(\ref{e4-44}) by using Eq.(\ref{e4-35})
and various theta identities.\\
Replacing $u \rightarrow u+\tau_1/2$ and 
$v  \rightarrow v+\tau_{12}/2+1/2$ in Eq.(\ref{B1-3}),
we have 
\begin{eqnarray}
&&\vartheta^2 \left[\begin{array}{cc} 0 \ 0 \\  0 \ 1 \\
\end{array}\right](0,0) \ 
\vartheta^2 \left[\begin{array}{cc} 1 \ 0 \\  0 \ 1 \\
\end{array}\right](u,v)
=-\vartheta^2 \left[\begin{array}{cc} 0 \ 0 \\  1 \ 1 \\
\end{array}\right](0,0) \ 
\vartheta^2 \left[\begin{array}{cc} 1 \ 0 \\  1 \ 1 \\
\end{array}\right](u,v)
\nonumber\\
&&+\vartheta^2 \left[\begin{array}{cc} 1 \ 0 \\  0 \ 1 \\
\end{array}\right](0,0) \ 
\vartheta^2 \left[\begin{array}{cc} 0 \ 0 \\  0 \ 1 \\
\end{array}\right](u,v)
+\vartheta^2 \left[\begin{array}{cc} 1 \ 1 \\  1 \ 1 \\
\end{array}\right](0,0) \ 
\vartheta^2 \left[\begin{array}{cc} 0 \ 1 \\  1 \ 1 \\
\end{array}\right](u,v)  , 
\label{D-8}
\end{eqnarray}
which gives \\
\begin{eqnarray}
&& \frac{\vartheta^2 \left[\begin{array}{cc} 1 \ 1 \\ 1 \ 1 \\
\end{array}\right](0,0)}
{\vartheta^2 \left[\begin{array}{cc} 0 \ 0 \\  0 \ 0 \\
\end{array}\right](0,0)}
\frac{\vartheta^2 \left[\begin{array}{cc} 0 \ 1 \\  1 \ 1 \\
\end{array}\right](u,v)}
{\vartheta^2 \left[\begin{array}{cc} 0 \ 0 \\  1 \ 1 \\
\end{array}\right](u,v)}
+\frac{\vartheta^2 \left[\begin{array}{cc} 1 \ 0 \\ 0 \ 1 \\
\end{array}\right](0,0)}
{\vartheta^2 \left[\begin{array}{cc} 0 \ 0 \\  0 \ 0 \\
\end{array}\right](0,0)}
\frac{\vartheta^2 \left[\begin{array}{cc} 0 \ 0 \\ 0 \ 1 \\
\end{array}\right](u,v)}
{\vartheta^2 \left[\begin{array}{cc} 0 \ 0 \\  1 \ 1 \\
\end{array}\right](u,v)}
\nonumber\\
&&=\frac{\vartheta^2 \left[\begin{array}{cc} 0 \ 0 \\ 1 \ 1 \\
\end{array}\right](0,0)}
{\vartheta^2 \left[\begin{array}{cc} 0 \ 0 \\  0 \ 0 \\
\end{array}\right](0,0)}
\frac{\vartheta^2 \left[\begin{array}{cc} 1 \ 0 \\  1 \ 1 \\
\end{array}\right](u,v)}
{\vartheta^2 \left[\begin{array}{cc} 0 \ 0 \\  1 \ 1 \\
\end{array}\right](u,v)}
+\frac{\vartheta^2 \left[\begin{array}{cc} 0 \ 0 \\ 0 \ 1 \\
\end{array}\right](0,0)}
{\vartheta^2 \left[\begin{array}{cc} 0 \ 0 \\  0 \ 0 \\
\end{array}\right](0,0)}
\frac{\vartheta^2 \left[\begin{array}{cc} 1 \ 0 \\ 0 \ 1 \\
\end{array}\right](u,v)}
{\vartheta^2 \left[\begin{array}{cc} 0 \ 0 \\  1 \ 1 \\
\end{array}\right](u,v)}  . 
\label{D-9}
\end{eqnarray}
By using Eq.(\ref{e4-30}), Eq.(\ref{e4-31}), Eq.(\ref{e4-35}),  
Eq.(\ref{e4-36}), after the 
straightforward but tedious calculation, we have \\ 
\begin{eqnarray}
&&{\rm \left(Left-hand\ side\ of\ Eq.(\ref{D-9})\right)
=\left(Right-hand\ side\ of\ Eq.(\ref{D-9})\right)}
\nonumber\\
&&=\frac{k_0^2 k_2 k_{12} } { {k'_1}^2 k'_2 k_{02} }  
\left\{(k_1^2 k_2^2-k_1^2-k_2^2)x_1x_2+x_1+x_2-1\right\}  ,
\nonumber
\end{eqnarray}
which gives Eq.(\ref{e4-36}).

\subsection{Check of Eq.(\ref{e4-37})[8-th parameterization]}
Similarly, by replacing $u \rightarrow u+\tau_1/2$ and 
$v  \rightarrow v+\tau_{12}/2+1/2$ in Eq.(\ref{B1-1}),
we have \\
\begin{eqnarray}
&& \frac{\vartheta^2 \left[\begin{array}{cc} 1 \ 1 \\ 1 \ 1 \\
\end{array}\right](0,0)}
{\vartheta^2 \left[\begin{array}{cc} 0 \ 0 \\  0 \ 0 \\
\end{array}\right](0,0)}
\frac{\vartheta^2 \left[\begin{array}{cc} 0 \ 1 \\  1 \ 0 \\
\end{array}\right](u,v)}
{\vartheta^2 \left[\begin{array}{cc} 0 \ 0 \\  1 \ 1 \\
\end{array}\right](u,v)}
-\frac{\vartheta^2 \left[\begin{array}{cc} 1 \ 0 \\ 0 \ 0 \\
\end{array}\right](0,0)}
{\vartheta^2 \left[\begin{array}{cc} 0 \ 0 \\  0 \ 0 \\
\end{array}\right](0,0)}
\frac{\vartheta^2 \left[\begin{array}{cc} 0 \ 0 \\ 0 \ 1 \\
\end{array}\right](u,v)}
{\vartheta^2 \left[\begin{array}{cc} 0 \ 0 \\  1 \ 1 \\
\end{array}\right](u,v)}
\nonumber\\
&&=-\frac{\vartheta^2 \left[\begin{array}{cc} 0 \ 0 \\ 1 \ 0 \\
\end{array}\right](0,0)}
{\vartheta^2 \left[\begin{array}{cc} 0 \ 0 \\  0 \ 0 \\
\end{array}\right](0,0)}
\frac{\vartheta^2 \left[\begin{array}{cc} 1 \ 0 \\  1 \ 1 \\
\end{array}\right](u,v)}
{\vartheta^2 \left[\begin{array}{cc} 0 \ 0 \\  1 \ 1 \\
\end{array}\right](u,v)}
-\frac{\vartheta^2 \left[\begin{array}{cc} 1 \ 0 \\ 0 \ 1 \\
\end{array}\right](u,v)}
{\vartheta^2 \left[\begin{array}{cc} 0 \ 0 \\  1 \ 1 \\
\end{array}\right](u,v)}  .
\label{D-10}
\end{eqnarray}
By using Eq.(\ref{e4-30}), Eq.(\ref{e4-31}), 
Eq.(\ref{e4-35}), Eq.(\ref{e4-37}), after the 
straightforward but tedious calculation, we have \\
\begin{eqnarray}
&&{\rm \left(Left-hand\ side\ of\ Eq.(\ref{D-10})\right)
=\left(Right-hand\ side\ of\ Eq.(\ref{D-10})\right)}
\nonumber\\
&&=-\frac{ k_0 k_1 k_2 } { k'_0 k'_1 k'_2 }
\left\{(k_0^2 k_2^2-k_0^2 -k_2^2)x_1x_2+x_1+x_2-1\right\}  ,
\nonumber
\end{eqnarray}
which gives Eq.(\ref{e4-37}).

\subsection{Check of Eq.(\ref{e4-38})[9-th parameterization]}
Similarly, by replacing $u \rightarrow u+\tau_1/2$ and 
$v  \rightarrow v+\tau_{12}/2+1/2$ in Eq.(\ref{B1-2}), we have \\
\begin{eqnarray}
&& \frac{\vartheta^2 \left[\begin{array}{cc} 1 \ 1 \\ 1 \ 1 \\
\end{array}\right](0,0)}
{\vartheta^2 \left[\begin{array}{cc} 0 \ 0 \\  0 \ 0 \\
\end{array}\right](0,0)}
\frac{\vartheta^2 \left[\begin{array}{cc} 0 \ 0 \\  1 \ 0 \\
\end{array}\right](u,v)}
{\vartheta^2 \left[\begin{array}{cc} 0 \ 0 \\  1 \ 1 \\
\end{array}\right](u,v)}
-\frac{\vartheta^2 \left[\begin{array}{cc} 1 \ 1 \\ 0 \ 0 \\
\end{array}\right](0,0)}
{\vartheta^2 \left[\begin{array}{cc} 0 \ 0 \\  0 \ 0 \\
\end{array}\right](0,0)}
\frac{\vartheta^2 \left[\begin{array}{cc} 0 \ 0 \\ 0 \ 1 \\
\end{array}\right](u,v)}
{\vartheta^2 \left[\begin{array}{cc} 0 \ 0 \\  1 \ 1 \\
\end{array}\right](u,v)}
\nonumber\\
&&=-\frac{\vartheta^2 \left[\begin{array}{cc} 0 \ 1 \\ 1 \ 0 \\
\end{array}\right](0,0)}
{\vartheta^2 \left[\begin{array}{cc} 0 \ 0 \\  0 \ 0 \\
\end{array}\right](0,0)}
\frac{\vartheta^2 \left[\begin{array}{cc} 1 \ 0 \\  1 \ 1 \\
\end{array}\right](u,v)}
{\vartheta^2 \left[\begin{array}{cc} 0 \ 0 \\  1 \ 1 \\
\end{array}\right](u,v)}
-\frac{\vartheta^2 \left[\begin{array}{cc} 0 \ 1 \\ 0 \ 0 \\
\end{array}\right](0,0)}
{\vartheta^2 \left[\begin{array}{cc} 0 \ 0 \\  0 \ 0 \\
\end{array}\right](0,0)}
\frac{\vartheta^2 \left[\begin{array}{cc} 1 \ 0 \\ 0 \ 1 \\
\end{array}\right](u,v)}
{\vartheta^2 \left[\begin{array}{cc} 0 \ 0 \\  1 \ 1 \\
\end{array}\right](u,v)} .
\label{D-11}
\end{eqnarray}
By using Eq.(\ref{e4-30}), Eq.(\ref{e4-31}), Eq.(\ref{e4-35}), Eq.(\ref{e4-38}), 
after the straightforward but tedious calculation, we have \\
\begin{eqnarray}
&&{\rm \left(Left-hand\ side\ of\ Eq.(\ref{D-11})\right)
=\left(Right-hand\ side\ of\ Eq.(\ref{D-11})\right)}
\nonumber\\
&&=-\frac{ k_0 k_2^2 k_{01} } { k'_0 {k'_1}^2 k_{02} }
\left\{(k_0^2 k_1^2-k_0^2 -k_1^2)x_1x_2+x_1+x_2-1\right\} , 
\nonumber
\end{eqnarray}
which gives Eq.(\ref{e4-38}).

\subsection{Check of Eq.(\ref{e4-39})[10-th parameterization]}
Similarly, by replacing $u \rightarrow u$ and $v  \rightarrow v+1/2$ 
in Eq.(\ref{B1-3}), we have \\
\begin{eqnarray}
&& \frac{\vartheta^2 \left[\begin{array}{cc} 1 \ 1 \\ 1 \ 1 \\
\end{array}\right](0,0)}
{\vartheta^2 \left[\begin{array}{cc} 0 \ 0 \\  0 \ 0 \\
\end{array}\right](0,0)}
\frac{\vartheta^2 \left[\begin{array}{cc} 1 \ 1 \\  1 \ 1 \\
\end{array}\right](u,v)}
{\vartheta^2 \left[\begin{array}{cc} 0 \ 0 \\  1 \ 1 \\
\end{array}\right](u,v)}
+\frac{\vartheta^2 \left[\begin{array}{cc} 0 \ 0 \\ 0 \ 1 \\
\end{array}\right](0,0)}
{\vartheta^2 \left[\begin{array}{cc} 0 \ 0 \\  0 \ 0 \\
\end{array}\right](0,0)}
\frac{\vartheta^2 \left[\begin{array}{cc} 0 \ 0 \\ 0 \ 1 \\
\end{array}\right](u,v)}
{\vartheta^2 \left[\begin{array}{cc} 0 \ 0 \\  1 \ 1 \\
\end{array}\right](u,v)}
\nonumber\\
&&=\frac{\vartheta^2 \left[\begin{array}{cc} 1 \ 0 \\ 0 \ 1 \\
\end{array}\right](0,0)}
{\vartheta^2 \left[\begin{array}{cc} 0 \ 0 \\  0 \ 0 \\
\end{array}\right](0,0)}
\frac{\vartheta^2 \left[\begin{array}{cc} 1 \ 0 \\  0 \ 1 \\
\end{array}\right](u,v)}
{\vartheta^2 \left[\begin{array}{cc} 0 \ 0 \\  1 \ 1 \\
\end{array}\right](u,v)}
+\frac{\vartheta^2 \left[\begin{array}{cc} 0 \ 0 \\ 1 \ 1 \\
\end{array}\right](0,0)}
{\vartheta^2 \left[\begin{array}{cc} 0 \ 0 \\  0 \ 0 \\
\end{array}\right](0,0)}  .
\label{D-12}
\end{eqnarray}
By using Eq.(\ref{e4-31}), Eq.(\ref{e4-35}), Eq.(\ref{e4-39}), after the 
straightforward but tedious calculation, we have \\
\begin{eqnarray}
&&{\rm \left(Left-hand\ side\ of\ Eq.(\ref{D-12})\right)
=\left(Right-hand\ side\ of\ Eq.(\ref{D-12})\right)}
\nonumber\\
&&=-\frac{ k_0 k_{12} } { k_1 {k'_1}^2 k'_2 k_{02} }  
\left\{k_1^2 k_2^2 x_1x_2-k_1^2 k_2^2(x_1+x_2)+k_1^2+ k_2^2-1 \right\} , 
\nonumber
\end{eqnarray}
which gives Eq.(\ref{e4-39}).

\subsection{Check of Eq.(\ref{e4-40})[11-th parameterization]}
Similarly, by replacing $u \rightarrow u$ and $v  \rightarrow v+1/2$ 
in Eq.(\ref{B1-1}), we have \\
\begin{eqnarray}
&& -\frac{\vartheta^2 \left[\begin{array}{cc} 1 \ 1 \\ 1 \ 1 \\
\end{array}\right](0,0)}
{\vartheta^2 \left[\begin{array}{cc} 0 \ 0 \\  0 \ 0 \\
\end{array}\right](0,0)}
\frac{\vartheta^2 \left[\begin{array}{cc} 1 \ 1 \\  1 \ 0 \\
\end{array}\right](u,v)}
{\vartheta^2 \left[\begin{array}{cc} 0 \ 0 \\  1 \ 1 \\
\end{array}\right](u,v)}
+\frac{\vartheta^2 \left[\begin{array}{cc} 0 \ 0 \\ 0 \ 1 \\
\end{array}\right](u,v)}
{\vartheta^2 \left[\begin{array}{cc} 0 \ 0 \\  1 \ 1 \\
\end{array}\right](u,v)}
\nonumber\\
&&=\frac{\vartheta^2 \left[\begin{array}{cc} 1 \ 0 \\ 0 \ 0 \\
\end{array}\right](0,0)}
{\vartheta^2 \left[\begin{array}{cc} 0 \ 0 \\  0 \ 0 \\
\end{array}\right](0,0)}
\frac{\vartheta^2 \left[\begin{array}{cc} 1 \ 0 \\  0 \ 1 \\
\end{array}\right](u,v)}
{\vartheta^2 \left[\begin{array}{cc} 0 \ 0 \\  1 \ 1 \\
\end{array}\right](u,v)}
+\frac{\vartheta^2 \left[\begin{array}{cc} 0 \ 0 \\ 1 \ 0 \\
\end{array}\right](0,0)}
{\vartheta^2 \left[\begin{array}{cc} 0 \ 0 \\  0 \ 0 \\
\end{array}\right](0,0)} .
\label{D-13}
\end{eqnarray}
By using Eq.(\ref{e4-31}), Eq.(\ref{e4-35}), Eq.(\ref{e4-40}), after the 
straightforward but tedious calculation, we have \\
\begin{eqnarray}
&&{\rm \left(Left-hand\ side\ of\ Eq.(\ref{D-13})\right)
=\left(Right-hand\ side\ of\ Eq.(\ref{D-13})\right)}
\nonumber\\
&&=-\frac{1} {k'_0 k'_1 k'_2 }  
\left\{k_0^2 k_2^2 x_1x_2-k_0^2 k_2^2(x_1+x_2)+k_0^2+ k_2^2-1 \right\} , 
\nonumber
\end{eqnarray}
which gives Eq.(\ref{e4-40}).

\subsection{Check of Eq.(\ref{e4-41})[12-th parameterization]}
Similarly, by replacing $u \rightarrow u$ and $v  \rightarrow v+1/2$ 
in Eq.(\ref{B1-2}), we have \\
\begin{eqnarray}
&& -\frac{\vartheta^2 \left[\begin{array}{cc} 1 \ 1 \\ 1 \ 1 \\
\end{array}\right](0,0)}
{\vartheta^2 \left[\begin{array}{cc} 0 \ 0 \\  0 \ 0 \\
\end{array}\right](0,0)}
\frac{\vartheta^2 \left[\begin{array}{cc} 1 \ 0 \\  1 \ 0 \\
\end{array}\right](u,v)}
{\vartheta^2 \left[\begin{array}{cc} 0 \ 0 \\  1 \ 1 \\
\end{array}\right](u,v)}
+\frac{\vartheta^2 \left[\begin{array}{cc} 0 \ 1 \\ 0 \ 0 \\
\end{array}\right](0,0)}
{\vartheta^2 \left[\begin{array}{cc} 0 \ 0 \\  0 \ 0 \\
\end{array}\right](0,0)}
\frac{\vartheta^2 \left[\begin{array}{cc} 0 \ 0 \\ 0 \ 1 \\
\end{array}\right](u,v)}
{\vartheta^2 \left[\begin{array}{cc} 0 \ 0 \\  1 \ 1 \\
\end{array}\right](u,v)}
\nonumber\\
&&=\frac{\vartheta^2 \left[\begin{array}{cc} 1 \ 1 \\ 0 \ 0 \\
\end{array}\right](0,0)}
{\vartheta^2 \left[\begin{array}{cc} 0 \ 0 \\  0 \ 0 \\
\end{array}\right](0,0)}
\frac{\vartheta^2 \left[\begin{array}{cc} 1 \ 0 \\  0 \ 1 \\
\end{array}\right](u,v)}
{\vartheta^2 \left[\begin{array}{cc} 0 \ 0 \\  1 \ 1 \\
\end{array}\right](u,v)}
+\frac{\vartheta^2 \left[\begin{array}{cc} 0 \ 1 \\ 1 \ 0 \\
\end{array}\right](0,0)}
{\vartheta^2 \left[\begin{array}{cc} 0 \ 0 \\  0 \ 0 \\
\end{array}\right](0,0)} . 
\label{D-14}
\end{eqnarray}
By using Eq.(\ref{e4-31}), Eq.(\ref{e4-35}), Eq.(\ref{e4-41}), after the 
straightforward but tedious calculation, we have \\
\begin{eqnarray}
&&{\rm \left(Left-hand\ side\ of\ Eq.(\ref{D-14})\right)
=\left(Right-hand\ side\ of\ Eq.(\ref{D-14})\right)}
\nonumber\\
&&=-\frac{k_2 k_{01}} {k'_0 k_1 {k'_1}^2 k_{02} }  
\left\{k_0^2 k_1^2 x_1x_2-k_0^2 k_1^2(x_1+x_2)+k_0^2+ k_1^2-1 \right\} , 
\nonumber
\end{eqnarray}
which gives of Eq.(\ref{e4-41}).

\subsection{Check of Eq.(\ref{e4-42})[13-th parameterization]}
Similarly, by replacing $u \rightarrow u+1/2$ and $v  \rightarrow v+1/2$ 
in Eq.(\ref{B1-3}), we have \\
\begin{eqnarray}
&& -\frac{\vartheta^2 \left[\begin{array}{cc} 1 \ 1 \\ 1 \ 1 \\
\end{array}\right](0,0)}
{\vartheta^2 \left[\begin{array}{cc} 0 \ 0 \\  0 \ 0 \\
\end{array}\right](0,0)}
\frac{\vartheta^2 \left[\begin{array}{cc} 1 \ 1 \\  0 \ 1 \\
\end{array}\right](u,v)}
{\vartheta^2 \left[\begin{array}{cc} 0 \ 0 \\  1 \ 1 \\
\end{array}\right](u,v)}
+\frac{\vartheta^2 \left[\begin{array}{cc} 0 \ 0 \\ 1 \ 1 \\
\end{array}\right](0,0)}
{\vartheta^2 \left[\begin{array}{cc} 0 \ 0 \\  0 \ 0 \\
\end{array}\right](0,0)}
\frac{\vartheta^2 \left[\begin{array}{cc} 0 \ 0 \\ 0 \ 1 \\
\end{array}\right](u,v)}
{\vartheta^2 \left[\begin{array}{cc} 0 \ 0 \\  1 \ 1 \\
\end{array}\right](u,v)}
\nonumber\\
&&=-\frac{\vartheta^2 \left[\begin{array}{cc} 1 \ 0 \\ 0 \ 1 \\
\end{array}\right](0,0)}
{\vartheta^2 \left[\begin{array}{cc} 0 \ 0 \\  0 \ 0 \\
\end{array}\right](0,0)}
\frac{\vartheta^2 \left[\begin{array}{cc} 1 \ 0 \\  1 \ 1 \\
\end{array}\right](u,v)}
{\vartheta^2 \left[\begin{array}{cc} 0 \ 0 \\  1 \ 1 \\
\end{array}\right](u,v)}
+\frac{\vartheta^2 \left[\begin{array}{cc} 0 \ 0 \\ 0 \ 1 \\
\end{array}\right](0,0)}
{\vartheta^2 \left[\begin{array}{cc} 0 \ 0 \\  0 \ 0 \\
\end{array}\right](0,0)}  . 
\label{D-15}
\end{eqnarray}
By using Eq.(\ref{e4-30}), Eq.(\ref{e4-35}), Eq.(\ref{e4-42}), after the 
straightforward but tedious calculation, we have \\
\begin{eqnarray}
&&{\rm \left(Left-hand\ side\ of\ Eq.(\ref{D-15})\right)
=\left(Right-hand\ side\ of\ Eq.(\ref{D-15})\right)}
\nonumber\\
&&=-\frac{k_0 k'_0 k_{12}} {k_1 k'_1 k_{02} }  
\left\{k_1^2 k_2^2 x_1x_2-1 \right\}  ,  
\nonumber
\end{eqnarray}
which gives Eq.(\ref{e4-42}).

\subsection{Check of Eq.(\ref{e4-43})[14-th parameterization]}
Similarly, by replacing $u \rightarrow u+1/2$ and $v  \rightarrow v+1/2$ 
in Eq.(\ref{B1-3}), we have \\
\begin{eqnarray}
&& \frac{\vartheta^2 \left[\begin{array}{cc} 1 \ 1 \\ 1 \ 1 \\
\end{array}\right](0,0)}
{\vartheta^2 \left[\begin{array}{cc} 0 \ 0 \\  0 \ 0 \\
\end{array}\right](0,0)}
\frac{\vartheta^2 \left[\begin{array}{cc} 1 \ 1 \\  0 \ 0 \\
\end{array}\right](u,v)}
{\vartheta^2 \left[\begin{array}{cc} 0 \ 0 \\  1 \ 1 \\
\end{array}\right](u,v)}
+\frac{\vartheta^2 \left[\begin{array}{cc} 0 \ 0 \\ 1 \ 0 \\
\end{array}\right](0,0)}
{\vartheta^2 \left[\begin{array}{cc} 0 \ 0 \\  0 \ 0 \\
\end{array}\right](0,0)}
\frac{\vartheta^2 \left[\begin{array}{cc} 0 \ 0 \\ 0 \ 1 \\
\end{array}\right](u,v)}
{\vartheta^2 \left[\begin{array}{cc} 0 \ 0 \\  1 \ 1 \\
\end{array}\right](u,v)}
\nonumber\\
&&=-\frac{\vartheta^2 \left[\begin{array}{cc} 1 \ 0 \\ 0 \ 0 \\
\end{array}\right](0,0)}
{\vartheta^2 \left[\begin{array}{cc} 0 \ 0 \\  0 \ 0 \\
\end{array}\right](0,0)}
\frac{\vartheta^2 \left[\begin{array}{cc} 1 \ 0 \\  1 \ 1 \\
\end{array}\right](u,v)}
{\vartheta^2 \left[\begin{array}{cc} 0 \ 0 \\  1 \ 1 \\
\end{array}\right](u,v)}+1   .   
\label{D-16}
\end{eqnarray}
By using Eq.(\ref{e4-30}), Eq.(\ref{e4-35}), Eq.(\ref{e4-43}), after the 
straightforward but tedious calculation, we have \\
\begin{eqnarray}
&&{\rm \left(Left-hand\ side\ of\ Eq.(\ref{D-16})\right)
=\left(Right-hand\ side\ of\ Eq.(\ref{D-16})\right)}
\nonumber\\
&&=-\left\{k_0^2 k_2^2 x_1x_2-1 \right\}   ,  
\nonumber
\end{eqnarray}
which gives Eq.(\ref{e4-43}).

\subsection{Check of Eq.(\ref{e4-44})[15-th parameterization]}
Similarly, by replacing $u \rightarrow u+1/2$ and $v  \rightarrow v+1/2$ 
in Eq.(\ref{B1-2}), we have \\
\begin{eqnarray}
&& \frac{\vartheta^2 \left[\begin{array}{cc} 1 \ 1 \\ 1 \ 1 \\
\end{array}\right](0,0)}
{\vartheta^2 \left[\begin{array}{cc} 0 \ 0 \\  0 \ 0 \\
\end{array}\right](0,0)}
\frac{\vartheta^2 \left[\begin{array}{cc} 1 \ 0 \\  0 \ 0 \\
\end{array}\right](u,v)}
{\vartheta^2 \left[\begin{array}{cc} 0 \ 0 \\  1 \ 1 \\
\end{array}\right](u,v)}
+\frac{\vartheta^2 \left[\begin{array}{cc} 0 \ 1 \\ 1 \ 0 \\
\end{array}\right](0,0)}
{\vartheta^2 \left[\begin{array}{cc} 0 \ 0 \\  0 \ 0 \\
\end{array}\right](0,0)}
\frac{\vartheta^2 \left[\begin{array}{cc} 0 \ 0 \\ 0 \ 1 \\
\end{array}\right](u,v)}
{\vartheta^2 \left[\begin{array}{cc} 0 \ 0 \\  1 \ 1 \\
\end{array}\right](u,v)}
\nonumber\\
&&=-\frac{\vartheta^2 \left[\begin{array}{cc} 1 \ 1 \\ 0 \ 0 \\
\end{array}\right](0,0)}
{\vartheta^2 \left[\begin{array}{cc} 0 \ 0 \\  0 \ 0 \\
\end{array}\right](0,0)}
\frac{\vartheta^2 \left[\begin{array}{cc} 1 \ 0 \\  1 \ 1 \\
\end{array}\right](u,v)}
{\vartheta^2 \left[\begin{array}{cc} 0 \ 0 \\  1 \ 1 \\
\end{array}\right](u,v)}
+\frac{\vartheta^2 \left[\begin{array}{cc} 0 \ 1 \\ 0 \ 0 \\
\end{array}\right](0,0)}
{\vartheta^2 \left[\begin{array}{cc} 0 \ 0 \\  0 \ 0 \\
\end{array}\right](0,0)}  . 
\label{D-17}
\end{eqnarray}
By using Eq.(\ref{e4-30}), Eq.(\ref{e4-35}), Eq.(\ref{e4-44}), after the 
straightforward but tedious calculation, we have \\
\begin{eqnarray}
&&{\rm \left(Left-hand\ side\ of\ Eq.(\ref{D-17})\right)
=\left(Right-hand\ side\ of\ Eq.(\ref{D-17})\right)}
\nonumber\\
&&=-\frac{k_2 k'_2 k_{01}} {k_1 k'_1 k_{02} }\left\{k_0^2 k_1^2 x_1x_2-1 \right\}  , 
\nonumber
\end{eqnarray}
which gives Eq.(\ref{e4-44}).

\vskip 10mm
\setcounter{equation}{0}
\section{\large \bf Addition formula of theta function and differential formula}

\subsection{Addition theorem of theta function}
We will show the following addition theorem\\
\begin{eqnarray}
&&1) \ \vartheta \left[\begin{array}{cc} 1 \ 0 \\  0 \ 1 \\
\end{array}\right](0,0) \ 
\vartheta \left[\begin{array}{cc} 0 \ 0 \\  0 \ 1 \\
\end{array}\right](0,0)
\left\{
\vartheta \left[\begin{array}{cc} 1 \ 0 \\  1 \ 1 \\
\end{array}\right](u+u',v+v') \ 
\vartheta \left[\begin{array}{cc} 0 \ 0 \\  1 \ 1 \\
\end{array}\right](u-u',v-v')  \right.
\nonumber\\ 
&& \left. -\vartheta \left[\begin{array}{cc} 0 \ 0 \\  1 \ 1 \\
\end{array}\right](u+u',v+v') \ 
\vartheta \left[\begin{array}{cc} 1 \ 0 \\  1 \ 1 \\
\end{array}\right](u-u',v-v') \right\}
\nonumber\\
&&=2\left\{
\vartheta \left[\begin{array}{cc} 1 \ 0 \\  0 \ 0 \\
\end{array}\right](u,v) \ 
\vartheta \left[\begin{array}{cc} 0 \ 0 \\  0 \ 0 \\
\end{array}\right](u,v) \ 
\vartheta \left[\begin{array}{cc} 0 \ 0 \\  1 \ 0 \\
\end{array}\right](u',v') \
\vartheta \left[\begin{array}{cc} 1 \ 0 \\  1 \ 0 \\
\end{array}\right](u',v') \right. 
\nonumber\\
&&\left. -\vartheta \left[\begin{array}{cc} 1 \ 1 \\  0 \ 0 \\
\end{array}\right](u,v) \ 
\vartheta \left[\begin{array}{cc} 0 \ 1 \\  0 \ 0 \\
\end{array}\right](u,v) \ 
\vartheta \left[\begin{array}{cc} 1 \ 1 \\  1 \ 0 \\
\end{array}\right](u',v') \ 
\vartheta \left[\begin{array}{cc} 0 \ 1 \\  1 \ 0 \\
\end{array}\right](u',v') \right\} ,  
\label{E-1}\\
&&\nonumber\\
&&2) \ \vartheta \left[\begin{array}{cc} 1 \ 0 \\  0 \ 1 \\
\end{array}\right](0,0) \ 
\vartheta \left[\begin{array}{cc} 0 \ 0 \\  1 \ 1 \\
\end{array}\right](0,0)
\left\{
\vartheta \left[\begin{array}{cc} 1 \ 0 \\  0 \ 1 \\
\end{array}\right](u+u',v+v') \ 
\vartheta \left[\begin{array}{cc} 0 \ 0 \\  1 \ 1 \\
\end{array}\right](u-u',v-v')  \right.
\nonumber\\ 
&& \left. -\vartheta \left[\begin{array}{cc} 0 \ 0 \\  1 \ 1 \\
\end{array}\right](u+u',v+v') \ 
\vartheta \left[\begin{array}{cc} 1 \ 0 \\  0 \ 1 \\
\end{array}\right](u-u',v-v') \right\}
\nonumber\\
&&=2\left\{
-\vartheta \left[\begin{array}{cc} 0 \ 0 \\  0 \ 0 \\
\end{array}\right](u,v) \ 
\vartheta \left[\begin{array}{cc} 1 \ 0 \\  1 \ 0 \\
\end{array}\right](u,v) \ 
\vartheta \left[\begin{array}{cc} 0 \ 0 \\  0 \ 0 \\
\end{array}\right](u',v') \
\vartheta \left[\begin{array}{cc} 1 \ 0 \\  1 \ 0 \\
\end{array}\right](u',v') \right. 
\nonumber\\
&&\left. +\vartheta \left[\begin{array}{cc} 0 \ 1 \\  0 \ 0 \\
\end{array}\right](u,v) \ 
\vartheta \left[\begin{array}{cc} 1 \ 1 \\  1 \ 0 \\
\end{array}\right](u,v) \ 
\vartheta \left[\begin{array}{cc} 0 \ 1 \\  0 \ 0 \\
\end{array}\right](u',v') \ 
\vartheta \left[\begin{array}{cc} 1 \ 1 \\  1 \ 0 \\
\end{array}\right](u',v') \right\}  . 
\label{E-2}
\end{eqnarray} \\
Replacing 
$u_1 \rightarrow u +\tau_1/2 +1/2$, $v_1 \rightarrow v +\tau_{12}/2 +1$, 
$u_2 \rightarrow u +1/2$, $v_2 \rightarrow v$, 
$u_3 \rightarrow u' +\tau_1/2$, $v_3 \rightarrow v' +\tau_{12}/2$, 
$u_4 \rightarrow u'$, $v_4 \rightarrow v'$, 
which gives
$\tilde{u}_1 \rightarrow u+u' +\tau_1/2 +1/2$, 
$\tilde{v}_1 \rightarrow v+v' +\tau_{12}/2 +1/2$, 
$\tilde{u}_2 \rightarrow u-u' +1/2$, 
$\tilde{v}_2 \rightarrow v-v'+1/2$, 
$\tilde{u}_3 \rightarrow \tau_1/2$, $v_3 \rightarrow \tau_{12}/2+1/2$, 
$u_4 \rightarrow 0$, $v_4 \rightarrow 1/2$, 
in Eq.(\ref{e3-25}), that is, 
$2M'''=\tilde{M}-\tilde{M}'-\tilde{M}''+\tilde{M}'''$. In this case,
we have $\tilde{M}''=0$, $\tilde{M}'''=0$, so that we have
$2M'''=\tilde{M}-\tilde{M}'$. 
After using the transformation property, we have Eq.(\ref{E-1}).\\
Similarly, replacing 
$u_1 \rightarrow u $, $v_1 \rightarrow v +1$, 
$u_2 \rightarrow u +\tau_1/2+1/2$, $v_2 \rightarrow v+\tau_{12}/2$, 
$u_3 \rightarrow u' $, $v_3 \rightarrow v' $, 
$u_4 \rightarrow u' +\tau_1/2 -1/2$, $v_4 \rightarrow v' +\tau_{12}/2$, 
which gives
$\tilde{u}_1 \rightarrow u+u' +\tau_1/2 $, 
$\tilde{v}_1 \rightarrow v+v' +\tau_{12}/2 +1/2$, 
$\tilde{u}_2 \rightarrow u-u' +1/2$, 
$\tilde{v}_2 \rightarrow v-v' +1/2$, 
$\tilde{u}_3 \rightarrow -\tau_1/2$, $v_3 \rightarrow -\tau_{12}/2+1/2$, 
$u_4 \rightarrow -1/2$, $v_4 \rightarrow 1/2$, 
in Eq.(\ref{e3-22}), that is, 
$2M'''=\tilde{M}+\tilde{M}'+\tilde{M}''+\tilde{M}'''$. In this case,
we have $\tilde{M}'=0$, $\tilde{M}'''=0$, so that we have
$2M=\tilde{M}+\tilde{M}''$. 
After using the transformation property, we have Eq.(\ref{E-2}).\\

\subsection{Derivative formula I}
We calculate 
\begin{eqnarray}
&&\frac{\partial}{\partial u} \left(
\vartheta \left[\begin{array}{cc} 1 \ 0 \\  1 \ 1 \\
\end{array}\right](u,v) /
\vartheta \left[\begin{array}{cc} 0 \ 0 \\  1 \ 1 \\
\end{array}\right](u,v)\right)
\nonumber\\
&&=\frac{\partial_u 
\vartheta \left[\begin{array}{cc} 1 \ 0 \\  1 \ 1 \\
\end{array}\right](u,v) \ 
\vartheta \left[\begin{array}{cc} 0 \ 0 \\  1 \ 1 \\
\end{array}\right](u,v) 
-\vartheta \left[\begin{array}{cc} 1 \ 0 \\  1 \ 1 \\
\end{array}\right](u,v) \ 
\partial_u  
\vartheta \left[\begin{array}{cc} 0 \ 0 \\  1 \ 1 \\
\end{array}\right](u,v)}  
{\vartheta^2 \left[\begin{array}{cc} 0 \ 0 \\  1 \ 1 \\
\end{array}\right](u,v)}  , 
\label{E-3}
\end{eqnarray}
by using the addition theorem.\\
For this purpose, we use the following addition formula
Eq.(\ref{E-1}) and put $u'=du$, $v'=0$,\\ 
\begin{eqnarray}
&&\ \vartheta \left[\begin{array}{cc} 1 \ 0 \\  0 \ 1 \\
\end{array}\right](0,0) \ 
\vartheta \left[\begin{array}{cc} 0 \ 0 \\  0 \ 1 \\
\end{array}\right](0,0)
\left\{
\left(\vartheta \left[\begin{array}{cc} 1 \ 0 \\  1 \ 1 \\
\end{array}\right](u,v)
+\frac{\partial}{\partial u} 
\vartheta \left[\begin{array}{cc} 1 \ 0 \\  1 \ 1 \\
\end{array}\right](u,v) \ du \right) \right.
\nonumber\\
&&\times \left(
\vartheta \left[\begin{array}{cc} 0 \ 0 \\  1 \ 1 \\
\end{array}\right](u,v)  
- \frac{\partial}{\partial u} 
\vartheta \left[\begin{array}{cc} 0 \ 0 \\  1 \ 1 \\
\end{array}\right](u,v) du \right)
-\left(\vartheta \left[\begin{array}{cc} 0 \ 0 \\  1 \ 1 \\
\end{array}\right](u,v)
+\frac{\partial}{\partial u} 
\vartheta \left[\begin{array}{cc} 0 \ 0 \\  1 \ 1 \\
\end{array}\right](u,v) \ du \right) 
\nonumber\\ 
&&\left. \times \left(
\vartheta \left[\begin{array}{cc} 1 \ 0 \\  1 \ 1 \\
\end{array}\right](u,v)  
- \frac{\partial}{\partial u} 
\vartheta \left[\begin{array}{cc} 1 \ 0 \\  1 \ 1 \\
\end{array}\right](u,v) du \right) \right\}
\nonumber\\ 
&&=2\left\{
\vartheta \left[\begin{array}{cc} 1 \ 0 \\  0 \ 0 \\
\end{array}\right](u,v) \ 
\vartheta \left[\begin{array}{cc} 0 \ 0 \\  0 \ 0 \\
\end{array}\right](u,v) \ 
\vartheta \left[\begin{array}{cc} 0 \ 0 \\  1 \ 0 \\
\end{array}\right](0,0) \
\partial_u
\vartheta \left[\begin{array}{cc} 1 \ 0 \\  1 \ 0 \\
\end{array}\right](u,0)\Big|_{0}\ du \right. 
\nonumber\\
&&\left. -\vartheta \left[\begin{array}{cc} 1 \ 1 \\  0 \ 0 \\
\end{array}\right](u,v) \ 
\vartheta \left[\begin{array}{cc} 0 \ 1 \\  0 \ 0 \\
\end{array}\right](u,v) \ 
\partial_u
\vartheta \left[\begin{array}{cc} 1 \ 1 \\  1 \ 0 \\
\end{array}\right](u,0)\Big|_{0} \ du \ 
\vartheta \left[\begin{array}{cc} 0 \ 1 \\  1 \ 0 \\
\end{array}\right](0,0) \right\}   , 
\label{E-4}
\end{eqnarray}
which gives the expression \\
\begin{eqnarray}
&&\ \vartheta \left[\begin{array}{cc} 1 \ 0 \\  0 \ 1 \\
\end{array}\right](0,0) \ 
\vartheta \left[\begin{array}{cc} 0 \ 0 \\  0 \ 1 \\
\end{array}\right](0,0) \ 
\vartheta^2 \left[\begin{array}{cc} 0 \ 0 \\  1 \ 1 \\
\end{array}\right](u,v) \ 
\frac{\partial}{\partial u} 
\left(\vartheta \left[\begin{array}{cc} 1 \ 0 \\  1 \ 1 \\
\end{array}\right](u,v)  /
\vartheta \left[\begin{array}{cc} 0 \ 0 \\  1 \ 1 \\
\end{array}\right](u,v) \right)
\nonumber\\
&&=
\left(\vartheta \left[\begin{array}{cc} 0 \ 0 \\  1 \ 0 \\
\end{array}\right](0,0) \
\partial_u
\vartheta \left[\begin{array}{cc} 1 \ 0 \\  1 \ 0 \\
\end{array}\right](u,0)_0  \ 
\vartheta \left[\begin{array}{cc} 1 \ 0 \\  0 \ 0 \\
\end{array}\right](u,v) \ 
\vartheta \left[\begin{array}{cc} 0 \ 0 \\  0 \ 0 \\
\end{array}\right](u,v) \  \right.
\nonumber\\ 
&&-\left.
\vartheta \left[\begin{array}{cc} 0 \ 1 \\  1 \ 0 \\
\end{array}\right](0,0) \ 
\partial_u
\vartheta \left[\begin{array}{cc} 1 \ 1 \\  1 \ 0 \\
\end{array}\right](u,0)_0 \ 
\vartheta \left[\begin{array}{cc} 1 \ 1 \\  0 \ 0 \\
\end{array}\right](u,v) \ 
\vartheta \left[\begin{array}{cc} 0 \ 1 \\  0 \ 0 \\
\end{array}\right](u,v) \ \right)   , 
\label{E-5}
\end{eqnarray}
which is used in Eq.(\ref{e5-9}) where we denote
$\partial_u
\vartheta \left[\begin{array}{cc} 1 \ 0 \\  1 \ 0 \\
\end{array}\right](u,0)\Big|_0$ for 
$\frac{\partial}{\partial u}
\vartheta \left[\begin{array}{cc} 1 \ 0 \\  1 \ 0 \\
\end{array}\right](u,0)\Big|_{u=0}$ etc.\\
Similarly, by putting $u'=0$, $v'=dv$, we have
the expression 
\begin{eqnarray}
&&\ \vartheta \left[\begin{array}{cc} 1 \ 0 \\  0 \ 1 \\
\end{array}\right](0,0) \ 
\vartheta \left[\begin{array}{cc} 0 \ 0 \\  0 \ 1 \\
\end{array}\right](0,0) \ 
\vartheta^2 \left[\begin{array}{cc} 0 \ 0 \\  1 \ 1 \\
\end{array}\right](u,v) \ 
\frac{\partial}{\partial v} 
\left(\vartheta \left[\begin{array}{cc} 1 \ 0 \\  1 \ 1 \\
\end{array}\right](u,v)  /
\vartheta \left[\begin{array}{cc} 0 \ 0 \\  1 \ 1 \\
\end{array}\right](u,v) \right)
\nonumber\\
&&=
\left(\vartheta \left[\begin{array}{cc} 0 \ 0 \\  1 \ 0 \\
\end{array}\right](0,0) \
\partial_v
\vartheta \left[\begin{array}{cc} 1 \ 0 \\  1 \ 0 \\
\end{array}\right](0,v)_0  \ 
\vartheta \left[\begin{array}{cc} 1 \ 0 \\  0 \ 0 \\
\end{array}\right](u,v) \ 
\vartheta \left[\begin{array}{cc} 0 \ 0 \\  0 \ 0 \\
\end{array}\right](u,v) \  \right.
\nonumber\\ 
&&-\left.
\vartheta \left[\begin{array}{cc} 0 \ 1 \\  1 \ 0 \\
\end{array}\right](0,0) \ 
\partial_v
\vartheta \left[\begin{array}{cc} 1 \ 1 \\  1 \ 0 \\
\end{array}\right](0,v)_0 \ 
\vartheta \left[\begin{array}{cc} 1 \ 1 \\  0 \ 0 \\
\end{array}\right](u,v) \ 
\vartheta \left[\begin{array}{cc} 0 \ 1 \\  0 \ 0 \\
\end{array}\right](u,v) \ \right) . 
\label{E-6}
\end{eqnarray}

\subsection{Derivative formula II}
We calculate 
\begin{eqnarray}
&&\frac{\partial}{\partial u} \left(
\vartheta \left[\begin{array}{cc} 1 \ 0 \\  0 \ 1 \\
\end{array}\right](u,v) /
\vartheta \left[\begin{array}{cc} 0 \ 0 \\  1 \ 1 \\
\end{array}\right](u,v)\right)
\nonumber\\
&&=\frac{\partial_u 
\vartheta \left[\begin{array}{cc} 1 \ 0 \\  0 \ 1 \\
\end{array}\right](u,v) \ 
\vartheta \left[\begin{array}{cc} 0 \ 0 \\  1 \ 1 \\
\end{array}\right](u,v) 
-\vartheta \left[\begin{array}{cc} 1 \ 0 \\  0 \ 1 \\
\end{array}\right](u,v) \ 
\partial_u  
\vartheta \left[\begin{array}{cc} 0 \ 0 \\  1 \ 1 \\
\end{array}\right](u,v)}  
{\vartheta^2 \left[\begin{array}{cc} 0 \ 0 \\  1 \ 1 \\
\end{array}\right](u,v)} , 
\label{E-7}
\end{eqnarray}
by using the addition theorem.\\
For this purpose, we use the following addition formula
Eq.(\ref{E-2}) and put $u'=du$, $v'=0$, 
\begin{eqnarray}
&&\ \vartheta \left[\begin{array}{cc} 1 \ 0 \\  0 \ 1 \\
\end{array}\right](0,0) \ 
\vartheta \left[\begin{array}{cc} 0 \ 0 \\  1 \ 1 \\
\end{array}\right](0,0)
\left\{
\left(\vartheta \left[\begin{array}{cc} 1 \ 0 \\  0 \ 1 \\
\end{array}\right](u,v)
+\frac{\partial}{\partial u} 
\vartheta \left[\begin{array}{cc} 1 \ 0 \\  0 \ 1 \\
\end{array}\right](u,v) \ du \right) \right.
\nonumber\\
&&\times \left(
\vartheta \left[\begin{array}{cc} 0 \ 0 \\  1 \ 1 \\
\end{array}\right](u,v)  
- \frac{\partial}{\partial u} 
\vartheta \left[\begin{array}{cc} 0 \ 0 \\  1 \ 1 \\
\end{array}\right](u,v) du \right)
-\left(\vartheta \left[\begin{array}{cc} 0 \ 0 \\  1 \ 1 \\
\end{array}\right](u,v)
+\frac{\partial}{\partial u} 
\vartheta \left[\begin{array}{cc} 0 \ 0 \\  1 \ 1 \\
\end{array}\right](u,v) \ du \right) 
\nonumber\\ 
&&\left. \times \left(
\vartheta \left[\begin{array}{cc} 1 \ 0 \\  0 \ 1 \\
\end{array}\right](u,v)  
- \frac{\partial}{\partial u} 
\vartheta \left[\begin{array}{cc} 1 \ 0 \\  0 \ 1 \\
\end{array}\right](u,v) du \right) \right\}
\nonumber\\ 
&&=2\left\{
-\vartheta \left[\begin{array}{cc} 0 \ 0 \\  0 \ 0 \\
\end{array}\right](u,v) \ 
\vartheta \left[\begin{array}{cc} 1 \ 0 \\  1 \ 0 \\
\end{array}\right](u,v) \ 
\vartheta \left[\begin{array}{cc} 0 \ 0 \\  0 \ 0 \\
\end{array}\right](0,0) \
\partial_u
\vartheta \left[\begin{array}{cc} 1 \ 0 \\  1 \ 0 \\
\end{array}\right](u,0)\Big|_{0}\ du \right. 
\nonumber\\
&&\left. +\vartheta \left[\begin{array}{cc} 0 \ 1 \\  0 \ 0 \\
\end{array}\right](u,v) \ 
\vartheta \left[\begin{array}{cc} 1 \ 1 \\  1 \ 0 \\
\end{array}\right](u,v) \ 
\vartheta \left[\begin{array}{cc} 0 \ 1 \\  0 \ 0 \\
\end{array}\right](0,0) \ 
\partial_u
\vartheta \left[\begin{array}{cc} 1 \ 1 \\  1 \ 0 \\
\end{array}\right](u,0)\Big|_{0} \ du \ \right\} , 
\label{E-8}
\end{eqnarray}
which gives the expression
\begin{eqnarray}
&&\ \vartheta \left[\begin{array}{cc} 1 \ 0 \\  0 \ 1 \\
\end{array}\right](0,0) \ 
\vartheta \left[\begin{array}{cc} 0 \ 0 \\  1 \ 1 \\
\end{array}\right](0,0) \ 
\vartheta^2 \left[\begin{array}{cc} 0 \ 0 \\  1 \ 1 \\
\end{array}\right](u,v) \ 
\frac{\partial}{\partial u} 
\left(\vartheta \left[\begin{array}{cc} 1 \ 0 \\  0 \ 1 \\
\end{array}\right](u,v)  /
\vartheta \left[\begin{array}{cc} 0 \ 0 \\  1 \ 1 \\
\end{array}\right](u,v) \right)
\nonumber\\
&&=
\left(-\vartheta \left[\begin{array}{cc} 0 \ 0 \\  0 \ 0 \\
\end{array}\right](0,0) \
\partial_u
\vartheta \left[\begin{array}{cc} 1 \ 0 \\  1 \ 0 \\
\end{array}\right](u,0)_0  \ 
\vartheta \left[\begin{array}{cc} 0 \ 0 \\  0 \ 0 \\
\end{array}\right](u,v) \ 
\vartheta \left[\begin{array}{cc} 1 \ 0 \\  1 \ 0 \\
\end{array}\right](u,v) \  \right.
\nonumber\\ 
&&+\left.
\vartheta \left[\begin{array}{cc} 0 \ 1 \\  0 \ 0 \\
\end{array}\right](0,0) \ 
\partial_u
\vartheta \left[\begin{array}{cc} 1 \ 1 \\  1 \ 0 \\
\end{array}\right](u,0)_0 \ 
\vartheta \left[\begin{array}{cc} 0 \ 1 \\  0 \ 0 \\
\end{array}\right](u,v) \ 
\vartheta \left[\begin{array}{cc} 1 \ 1 \\  1 \ 0 \\
\end{array}\right](u,v) \ \right) , 
\label{E-9}
\end{eqnarray}
which is used in Eq.(\ref{e5-10}).\\
Similarly, by putting $u'=0$, $v'=dv$, we have
the expression 
\begin{eqnarray}
&&\ \vartheta \left[\begin{array}{cc} 1 \ 0 \\  0 \ 1 \\
\end{array}\right](0,0) \ 
\vartheta \left[\begin{array}{cc} 0 \ 0 \\  1 \ 1 \\
\end{array}\right](0,0) \ 
\vartheta^2 \left[\begin{array}{cc} 0 \ 0 \\  1 \ 1 \\
\end{array}\right](u,v) \ 
\frac{\partial}{\partial v} 
\left(\vartheta \left[\begin{array}{cc} 1 \ 0 \\  0 \ 1 \\
\end{array}\right](u,v)  /
\vartheta \left[\begin{array}{cc} 0 \ 0 \\  1 \ 1 \\
\end{array}\right](u,v) \right)
\nonumber\\
&&=
\left(-\vartheta \left[\begin{array}{cc} 0 \ 0 \\  0 \ 0 \\
\end{array}\right](0,0) \
\partial_v
\vartheta \left[\begin{array}{cc} 1 \ 0 \\  1 \ 0 \\
\end{array}\right](0,v)_0  \ 
\vartheta \left[\begin{array}{cc} 0 \ 0 \\  0 \ 0 \\
\end{array}\right](u,v) \ 
\vartheta \left[\begin{array}{cc} 1 \ 0 \\  1 \ 0 \\
\end{array}\right](u,v) \  \right.
\nonumber\\ 
&&+\left.
\vartheta \left[\begin{array}{cc} 0 \ 1 \\  0 \ 0 \\
\end{array}\right](0,0) \ 
\partial_v
\vartheta \left[\begin{array}{cc} 1 \ 1 \\  1 \ 0 \\
\end{array}\right](0,v)_0 \ 
\vartheta \left[\begin{array}{cc} 0 \ 1 \\  0 \ 0 \\
\end{array}\right](u,v) \ 
\vartheta \left[\begin{array}{cc} 1 \ 1 \\  1 \ 0 \\
\end{array}\right](u,v) \ \right) .
\label{E-10}
\end{eqnarray}

\newpage
\end{document}